\documentclass[12pt]{puthesis}

\input epsf.sty

\usepackage{amsfonts}
\usepackage{amssymb}
\usepackage{amsthm}
\usepackage{latexsym}

\def\pa{{\partial}}

\def\ket{\rangle}
\def\bra{\langle}

\def\CF{{\cal F}}

\def\CN{{\cal N}}
\def\CO{{\cal O}}

\def\CQ{{\cal Q}}

\def\CV{{\cal V}}
\def\CW{{\cal W}}

\def\CZ{{\cal Z}}

\newcommand{\labell}[1]{\label{#1}}

\newcommand{\dslash}[0]{\slash{\hspace{-0.23cm}}\partial}

\newcommand{\ads}[1]{{\rm AdS}_{#1}}

\newcommand{\sph}[1]{{\rm S}^{#1}}

\newcommand{\be}{\begin{equation}}
\newcommand{\ee}{\end{equation}}
\newcommand{\bea}{\begin{eqnarray}}
\newcommand{\eea}{\end{eqnarray}}

\def\tr{\hbox{tr}}

\def\half{\frac{1}{2}}

\def\CF{{\cal F}}

\def\CN{{\cal N}}
\def\CO{{\cal O}}

\def\CQ{{\cal Q}}

\def\CV{{\cal V}}
\def\CW{{\cal W}}

\def\CZ{{\cal Z}}
\makeatletter
\def\overbracket#1{\mathop{\vbox{\ialign{##\crcr\noalign{\kern3\p@}
\downbracketfill\crcr\noalign{\kern3\p@\nointerlineskip}
$\hfil\displaystyle{#1}\hfil$\crcr}}}\limits}
\def\underbracket#1{\mathop{\vtop{\ialign{##\crcr
$\hfil\displaystyle{#1}\hfil$\crcr\noalign{\kern3\p@\nointerlineskip}
\upbracketfill\crcr\noalign{\kern3\p@}}}}\limits}
\def\overparenthesis#1{\mathop{\vbox{\ialign{##\crcr\noalign{\kern3\p@}
\downparenthfill\crcr\noalign{\kern3\p@\nointerlineskip}
$\hfil\displaystyle{#1}\hfil$\crcr}}}\limits}
\def\underparenthesis#1{\mathop{\vtop{\ialign{##\crcr
$\hfil\displaystyle{#1}\hfil$\crcr\noalign{\kern3\p@\nointerlineskip}
\upparenthfill\crcr\noalign{\kern3\p@}}}}\limits}
\def\downparenthfill{$\m@th\braceld\leaders\vrule\hfill\bracerd$}
\def\upparenthfill{$\m@th\bracelu\leaders\vrule\hfill\braceru$}
\def\upbracketfill{$\m@th\makesm@sh{\llap{\vrule\@height3\p@\@width.7\p@}}%
\leaders\vrule\@height.7\p@\hfill
\makesm@sh{\rlap{\vrule\@height3\p@\@width.7\p@}}$}
\def\downbracketfill{$\m@th
\makesm@sh{\llap{\vrule\@height.7\p@\@depth2.3\p@\@width.7\p@}}%
\leaders\vrule\@height.7\p@\hfill
\makesm@sh{\rlap{\vrule\@height.7\p@\@depth2.3\p@\@width.7\p@}}$}
\makeatother

\def\Tr{{\rm Tr}}

\newcommand{\eref}[1]{(\ref{#1})}
\newcommand{\fref}[1]{Figure~\ref{#1}}
\newcommand{\beq}{\begin{equation}}
\newcommand{\eeq}{\end{equation}}
\def\IR{\mathbb{R}}

\title{Aspects of Supersymmetric Gauge Theory and String Theory}
\submitted{2004}  %graduation date
\author{Minxin Huang}

\abstract{
\begin{center}
Aspects of Supersymmetric Gauge Theory and String Theory\\
Minxin Huang\\
Supervisor: Vijay Balasubramanian
\end{center}
\vspace{.3in}

This thesis consists of two parts. In the first part we study some
topics in $\CN=1$ supersymmeric gauge theory and the relation to
matrix models. We review the relevant non-perturbative techniques
for computing effective superpotential, such as Seiberg-Witten
curve. Then we review the proposal of Dijkgraaf and Vafa that
relates the glueball superpotentials to the computation in matrix
models. We then consider a case of multi-trace superpotential. We
perform the perturbative computation of glueball superpotential in
this case and explain the subtlety in identifying the glueball
superfield. We also use these techniques to study phases of
$\CN=1$ gauge theory with flavors.

In the second part we study topics in AdS/CFT correspondence and
its plane wave limit. We review the plane wave geometry and BMN
operators that corresponding to string modes. Then we study string
interactions in the case of a highly curved plane wave background,
and demonstrate the agreements between calculations of string
interaction amplitudes in the two dual theories. Finally we study
D3-brane giant gravitons and open string attached to them. Giant
gravitons are non-perturbative objects that have very large
R-charge.

}

\acknowledgements{
First I thank my advisor Vijay Balasubramanian. He taught me many
knowledge in physics and gave me many insights. He also gave me
numerous encouragement during the course of my graduate study. He
trained me with all kinds of other necessary skills. I am very
grateful to have such a great advisor.

I should thank Asad Naqvi, who was a postdoc here. He provided
many specific guidance when my understanding was still very naive.
He seemed to be able to answer most questions I had in a very
broad regime of subjects. Also I have learned tremendously from
his expertise in field theory. Besides my advisor, he has played
important role in the formation of my intellectual perspective.

I thank other faculty members of the math and physics department
such as Mirjam Cvetic, Ron Donagi, Paul Langacker, Burt Ovrut,
Gino Segre and Matt Strassler to provide wonderful courses and
stimulating discussions. I also thank Jan de Boer, Bo Feng,
Yang-Hui He, Tommy Levi, Vishnu Jejjala for collaborations and
discussions, especially Bo Feng for his insights. I also thank
other (ex)postdocs, graduate students and visitors such as David
Berenstein, Volker Braun, Evgeny Buchbinder, Can Gurses, Junhai
Kang, Klaus Larjo, Tianjun Li, Tao Liu, Oleg Lunin, Makoto
Natsuume, Brent Nelson, Rene Reinbacher, Gary Shiu, Joan Simon for
useful discussions.

Finally I thank my family and friends for moral support.

}

\begin{document}

\chapter{Introduction}  \label{chapter1}%insert chapter 1 title
Our understanding of the physical laws of the nature has
progressed tremendously during the past century. The foundation of
this progress rests on two pillars: general relativity and quantum
field theory. General relativity is a classical theory of gravity.
It explains the macroscopic phenomena of our world, such as the
motion of planets, galaxies and the whole cosmos. On the other
hand, the microscopic world is well governed by quantum field
theory. In particular, a paradigm known as ``standard model" has
emerged and successfully incorporated all three kind of
interactions of elementary particles, namely, the
electro-magnetic, weak and strong interactions. These two theories
have been verified by experimental tests in a very broad regime
that is available to the probe of current experimental devices.

Yet, there are many problems with general relativity and the
standard model, and it is generally believed that these two
theories can not be a complete description of nature. Firstly,
there are too many arbitrary parameters in the standard model.
Throught the study of a more fundamental theory, one might hope to
explain some of the patterns in standard model or even make a
quantitative derivation of some of these parameters. Also some
parameters in the standard model appear to have an unnatural
magnitude. One such example is the hierarchy problem. In quantum
field theory, the mass of a scalar particle is renormalized by
quantum effects and naturally has a mass scale comparable to the
fundamental cut-off scale of the theory. The Higgs is a scalar, so
its natural mass scale is the Planck scale, which is $10^{19}$
Gev. However, the mass of the Higgs is constrained by experiment
to be at about $10^2$ Gev, and therefore appears to be very
unnaturally small. Secondly, the general relativity Lagrangian is
non-renormalizable. This means that general relativity is only an
effective theory at low energy. The theory must break down and
needs to be modified above some high energy scale. Thirdly, recent
astronomical observations suggest our universe is dominated by a
kind of dark energy with negative pressure. A natural explanation
of the dark energy is the vacuum energy caused by a very small
positive cosmological constant. However, quantum field theory
predicts the vacuum energy to be roughly the cut-off scale of the
fundamental theory, which is on a order of $10^{120}$ larger than
the observational value of the cosmological constant. This
cosmological problem is generally regarded as a hint of new
physics beyond classical general relativity.

Supersymmetry provides a beautiful solution to the hierarchy
problem. In a supersymmetric theory bosons and fermions always
appear in pairs. The contributions of bosons and fermions to the
quantum correction of the mass of Higgs cancel, therefore the
Higgs can have a naturally light mass compared to Planck scale.
There are also other arguments for supersymmetry. For example, it
improves the predictions of Grand Unified Theory. Due to many of
these arguments, it is generally believed that ${\cal N}=1$
supersymmetric gauge theory might be relevant to the description
of our world. In Chapter \ref{chapter3} we focus on low energy
dynamics of asymptotically free supersymmetric gauge theories.
Asymptotically free gauge theories are strongly coupled in low
energy, therefore the usual method of perturbation theory breaks
down. It has been a long standing and important problem to solve
the low energy dynamics of non-supersymmetric QCD, which is at
present only accessible by large computer simulation of QCD on
lattice. Meanwhile, in the past years, it has been shown the low
energy dynamics of ${\cal N}=2 $ supersymmetric gauge theories are
solvable \cite{SW1, SW2}. Recently, Dijkgraaf and Vafa proposed
the superpotentials of some ${\cal N}=1$ supersymmetric gauge
theories can be computed from matrix models \cite{DV1, DV2, DV3}.
In Chapter \ref{chapter3} we review these developments and present
some results we obtained along these lines.

There are many reasons why these results are useful and
interesting. We have mentioned they may be useful to predict
effective low energy dynamics of baryons and mesons. Furthermore,
these studies might be relevant to understanding of phenomenology
of supersymmetry in the next generation of accelerator, namely the
LHC (large hadron collider). It is widely hoped that supersymmetry
will be discovered in LHC. We know supersymmetry, if exist, is
broken in the scale we currently observe. However, there are many
mechanisms to break supersymmetry. To maintain a small Higgs mass,
supersymmetry must be ``softly'' broken. One such appealing
mechanism is that supersymmetry is broken by non-perturbative
effects, i.e. dynamically broken. Studies of non-perturbative
superpotentials in Chapter \ref{chapter3} might serve as a
learning grounds for these subjects.

String theory is a leading candidate of quantum theory of gravity
that requires supersymmetry. For a general textbook introduction,
see \cite{GSW, Polchinski}.  String theory was originally invented
in 1960's to account for some empirical formulae in strong
interaction experiments. Later it was realized that string theory
also contains gravity and does not suffer the annoying divergence
associated with the quantization of gravity. In the First String
Revolution in the mid 1980's, it was realized that string theory
can cancel anomaly and obtain many appealing phenomenological
features of the standard model, such as chiral fermions. There
appeared to be five distinct perturbative formulations of string
theory. However, in the Second String Revolution in mid 1990's,
many evidences suggested that the five theories can be embedded in
a single theory known as ``M-theory'', whose low energy dynamics
is described by an eleven dimensional supergravity. The existence
of the five perturbative string formulations strongly suggests
that a non-perturbative definition of the M-theory should exist.
Finding this conjectured non-perturbative definition is very
desirable, since it will enable us to find the true vacuum of the
theory that may describe our world.

Perhaps the most compelling reason for believing the conjectured
``M-theory'' is the correct description of nature is that the
theory is unique and has the richest structure of all known
theories. There is no dimensionless free parameter in the theory.
The theory only contains a dimensionful constant that determine
the fundamental scale of the theory. \footnote{However, M theory
has a very large moduli space of vacua. Other parameters such as
string coupling constants are related to the expectation value of
dilaton, so correspond to different points of the moduli space.
Physics at different vacua are dramatically different. At present
we do not where we are in this space of vacua, so can not make
very precise phenomenological predictions. } Furthermore, We have
mentioned supersymmetry solves many problems. It is also true that
the more supersymmetries a theory has, the richer the theory is.
The maximal amount of supersymmetry that a consistent quantum
theory can have is  32 supersymmetries, i.e. the theory has 32
real supercharges. M-theory has 32 supersymmetries and has the
highest spacetime dimension, namely 11 dimension, of theories with
32 supersymmetries. The other five perturbative string theories
can be obtained by compactifying M-theory on some one-dimensional
spaces and using various string dualities. One may naively think a
theory in higher spacetime dimension is richer than a theory in
lower spacetime dimension. Thus we claim M-theory is the richest
theory.

There are not many theories with 32 supersymmetries. It turns out
many of them are dual to each other. This is in some sense not
surprising, due to the scarcity or uniqueness of rich structure .
A more radical claim would be all theories with 32 supersymmetries
are equivalent to each other. However, the dimension of spacetime
a theory lives in should also be relevant. We have been a little
too naive in saying a higher dimensional theory is richer in
structure than a lower dimensional theory. Indeed, it has been
long speculated that a quantum theory of gravity is
holographically dual to a quantum theory without gravity in one
lower dimension. This ``holography principle'' comes from the
studies of black hole information problem. The AdS/CFT
correspondence is such an example \cite{Maldacena, Gubser,
Witten}. For a general review see \cite{Aharony, Freedman}. The
AdS/CFT correspondence states that Type IIB string theory on the
$AdS_5\times S^5$ background is equivalent to ${\cal N}=4$
supersymmetric gauge theory in 4 dimension. \footnote{Both
theories are maximally supersymmetric. In four dimension $\CN=4$
supersymmetry has 16 supercharges. The conformal symmetry of the
theory will effectively double the amount of supersymmetry.}
Besides being conceptually astonishing, the AdS/CFT correspondence
is also a powerful technical tool for obtaining new results in
both theories. Some very difficult problems in one theory might
turn out to be easy to solve in the other one. For example, it has
been shown that the very hard problem of doing calculation in the
strongly coupled gauge theory can be done easily using classical
supergravity in $AdS_5\times S^5$. In this thesis, among other
things, we will provide another such example. We use the free
gauge theory to study type IIB string theory in a highly curved
plane wave background, where the seemingly intractable problem of
computing string interaction amplitudes to all loops becomes easy
and computable!

In Chapter \ref{chapter2} we first review the AdS/CFT conjecture
and discuss the tests of it in the regime where string theory can
be approximated by classical supergravtiy. Then we focus on the
more recent development of string theory on plane wave backgrounds
\cite{BMN}. The plane wave background can be obtained as Penrose
limit of $AdS_5\times S^5$ \cite{Blau}. The study of AdS/CFT in
this limit is very interesting,  since the gauge theory is weakly
coupled and computable, and we can go beyond the classical
supergravity approximation. Finally we will go beyond the plane
wave limit, where the strings blow up into some higher dimensional
objects known as giant gravitons \cite{McGreevy}.

Although there are some conceptual connections, Chapter
\ref{chapter3} and Chapter \ref{chapter2} are roughly technically
independent of each other. I have provided self-contain
introduction to each chapters. The readers can  read individual
chapter according to their own interests and needs. We hope this
writing might be of some values to beginning students as well as
active researchers in the field. However, this thesis focus mostly
on the author's own work, and is not a comprehensive review of
various subjects presented here. Therefore I apologize in advance
for those authors whose works are less emphasized here.
    %insert chapter 1 file name

\chapter{Topics in Supersymmetric Gauge Theories} \label{chapter3} %insert chapter 1 title
\section{Supersymmetric Lagrangian}
We start in this section by introducing the notation of  ${\cal
N}=1$ supersymmetry in $4$ dimensions. Then we review the
Seiberg-Witten solution of ${\cal N}=2$ supersymmetric gauge
theory \cite{SW1, SW2}, and how to use the Seiberg-Witten curve to
study ${\cal N}=1$ supersymmetric gauge theory. There are many
excellent reviews on this subject. For a general introduction to
supersymmetry in 4 dimension see e.g. \cite{WB, argyres}. For
review of ${\cal N}=1$ supersymmetric gauge theory see e.g.
\cite{IS, Peskin}.

There are two kinds of multiplets in ${\cal N}=1$ supersymmetric
gauge theory in 4 dimensions, the chiral multiplet $\Phi$ and the
gauge or vector multiplet $V$. We will consider theory with $U(N)$
or $SU(N)$ gauge group. The vector multiplet $V$ is in the adjoint
representation of the gauge group, and the chiral multiplet can be
in adjoint or (anti)fundamental representations. The gauge field
strength is
\begin{equation}
{\cal W}_\alpha =i\bar{D}^2 e^{-V}D_\alpha e^{V}
\end{equation}
and the glueball superfield is
\begin{equation}
S=\frac{1}{32\pi^2} \Tr({\cal W}_\alpha {\cal W}^\alpha )
\end{equation}
where $D_\alpha$ and $\bar{D}_{\dot\alpha}$ are the superspace
covariant derivatives. We use the notations of \cite{WB}. The
gluino condensate $S$ is a commuting field constructed out of a
pair of fermionic operators $\CW_\alpha$. We can write a general
${\cal N}=1$ Lagrangian with a chiral and a vector multiplet in
the adjoint representation
\begin{equation} \label{lagrangian}
{\cal L}=\int d^4\theta K(\Phi^{\dagger}, e^V \Phi)+2\pi i\int
d^2\theta\tau(\Phi)S+h.c.+\int d^2\theta W(\Phi)+h.c.
\end{equation}
Here $K(\Phi^{\dagger}, e^V \Phi)$ and $W(\Phi)$ are known as
Kahler potential and superpotential. We have used superfield
notation in writing the Lagrangian. After integrating over
$\theta$,  one can find the usual potential consisting of
``D-term'' and ``F-term'' coming from integrating out the
auxiliary fields $D$ and $F$.

The Lagrangian (\ref{lagrangian}) is usually referred to as tree
level or bare Lagragian. In asymptotically free gauge theory,
which will be our focus, the theory becomes strongly coupled and
non-perturbative below some energy scale usually referred to as
$\Lambda$. The scale $\Lambda$ is generated dynamically by
dimensional transmutation. We refer to the Lagrangian that
describe the full quantum dynamics below scale $\Lambda$ as
effective or Wilsonian Lagrangian. The effective Lagrangian is
constrained by many properties such as holomorphy. In a confining
gauge theory, if we assume the glueball $S$ is a good variable to
describe low energy dynamics, then the effective superpotential is
a function of $\Lambda$ and $S$ known as the glueball
superpotential $W(\Lambda, S)$. The glueball has a mass scale
comparable to $\Lambda$. If we only consider IR dynamics with
energy scale much less than $\Lambda$, then we can ``integrate
out'' the glueball in $W(\Lambda, S)$ and find the low energy
effective superpotential $W(\Lambda)$. In this thesis we are
primarily concerned with the computations of glueball
superpotentials in ${\cal N}=1$ supersymmetric gauge theory.

\subsubsection{ The Veneziano-Yankielowicz Superpotential}

The glueball superpotential of a $SU(N)$ gauge theory consists of
two parts: the Veneziano-Yankielowicz superpotential and a
perturbative part.
\begin{equation}
W_{eff}(\Lambda,S)=W_{VY}(\Lambda,S)+W_{per}(S)
\end{equation}
The Veneziano-Yankielowicz superpotential has been known for a
long time \cite{VY}
\begin{equation} \label{Veneziano}
W_{VY}(\Lambda,S)=S(-\ln(\frac{S^N}{\Lambda^b_0})+N)
\end{equation}
Here $b_0$ is the coefficient of one-loop $\beta$ function of the
gauge theory. In ${\cal N}=1$ gauge theory with an adjoint chiral
multiplet, $b_0=2N$; while in pure ${\cal N}=1$ gauge theory
$b_0=3N$. We can ``integrate out'' the glueball in
Veneziano-Yankielowicz superpotential (\ref{Veneziano}) by solve
the equation of motion for glueball $\frac{\partial
W_{VY}}{\partial S}=0$, and plug it back into equation
(\ref{Veneziano}). For example, for pure ${\cal N}=1$ gauge theory
the superpotential only consists of the Veneziabo-Yankielowicz
part, we find
\begin{equation}
W_{eff}=N\Lambda^3
\end{equation}
This is the known low energy superpotantial of pure guage theory
without the chiral superfield in (\ref{lagrangian}). It is
generated nonperturbatively by instanton.

\subsubsection{ Seiberg-Witten curve}

Now we consider ${\cal N}=2$ supersymmetric gauge theory. There
are two kinds of ${\cal N}=2$ multiplet usually referred to as
${\cal N}=2$ vector multiplet and ${\cal N}=2$ hypermultiplet. A
${\cal N}=2$ vector multiplet consists of a ${\cal N}=1$  vector
multiplet and a ${\cal N}=1$ chiral multiplet in the adjoint
representation; while a ${\cal N}=2$ hypermultiplet consists of
two ${\cal N}=1$ chiral multiplets. The Lagrangian ${\cal N}=2$
theory is determined by only one holomorphic function of the
chiral superfield known as ``prepotential''. The pure ${\cal N}=2$
super Yang-Mills has one ${\cal N}=2$ vector multiplet, and is a
special case of the Lagrangian (\ref{lagrangian})
\begin{equation} \label{N=2lagrangian}
{\cal L}=\int d^4\theta \Phi^{\dagger} e^V \Phi+2\pi i \tau\int
d^2\theta S+h.c.
\end{equation}
The modili space of vacua of the theory can be parameterized by
the expectation value of gauge invariant operators
$u_k=\frac{1}{k}\bra \Tr(\Phi^k)\ket$, $k=1, 2,\cdots, N$.
\footnote{Sometimes it is convenient to take the gauge group to be
$U(N)$ and keep the parameter $u_1$. The $U(1)$ factor decouples
in the inferred and will not affect our discussion.}

Seiberg and Witten gives a solution of low energy effective
prepotential of (\ref{N=2lagrangian}) \cite{SW1}. Here we will not
go into the details of the solution, but instead directly use the
Seiberg-Witten Curve. The Seiberg-Witten curve of the pure ${\cal
N}=2$ super Yang-Mills is
\begin{equation}
y^2=P_N(x)^2-4\Lambda^{2N}
\end{equation}
where $P_N(x,u_k)=\bra \det(xI-\Phi) \ket$ is degree $N$
polynomial parameterized by $u_k=\frac{1}{k}\bra \Tr(\Phi^k)\ket$,
and can be explicitly obtained by writing the determinant in terms
of traces.

We can break the ${\cal N}=2$ supersymmetry by adding a degree $n$
tree level superpotential to (\ref{N=2lagrangian}) with $n\leq N$
\begin{equation} \label{tree}
W_{tree}=\sum^{n+1}_{i=1}g_iu_i
\end{equation}
Classically, the vacuum structure is given by the F-term and
D-term equation. The D-term equation requires $\Phi$ to be
diagonal, and the F-term equation requires $W^{\prime}(\Phi)=0$.
In the IR we have a pure Yang-Mills with gauge group broken to
$U(N)\rightarrow U(N_1)\times U(N_2)\times\cdots \times U(N_n)$.
Quantum mechanically, the tree level superpotential lift the
points in the moduli space of vacua except those points where at
least $N-n$ mutually local monopoles becomes massless. The
presence of the superpotential then forces the monopoles to
condense and cause the confinement of ${\cal N}=1$ electric
charge. These points are characterized by $N-m$ ($m\leq n$) double
roots in the Seiberg-Witten curve.
\begin{equation} \label{factorization}
P_N(x,u_k)^2-4\Lambda^{2N}=F_{2m}(x)H_{N-m}^2(x)
\end{equation}
The original Coulomb moduli space is a $N$-dimensional space
parameterized by $u_k, k=1,2,\cdots N$. The constrain to satisfy
(\ref{factorization}) restricts the moduli space to a
$m$-dimensional sub-space. When $m=1$, the low energy degree of
freedom has only a decoupled $U(1)$ factor from the original
$U(N)$ gauge group and the remaining $SU(N)$ confines. In this
$m=1$ vacuum, known as the confining vacuum, the one-parameter
explicit solution of $P_N(x,u_k)$ is known to be given by the
Chebyshev polynomials \cite{DS}. The study of Dijkgraaf-Vafa
conjecture is rather simple in the confining vacuum. We will focus
on the confining vacuum in this chapter, until in the last
section, where we explore interpolation between various vacua.

To find the low energy effective superpotential, we minimize the
tree level superpotential (\ref{tree}) in the $m$-dimensional
sub-space of vacua constrained by the factorization
(\ref{factorization}). The resulting low energy effective
$W_{eff}(\Lambda, g_i)$ does not include the glueball superfield
$S$. We can reverse the ``integrate out'' procedure to ``integrate
in'' the glueball. To match the Veneziano-Yankielowicz
superpotential, the glueball superpotential must satisfy $\frac
{\partial W_{eff}(\Lambda, S, g_i)}{\partial
\ln(\Lambda^{2N})}=S$.

\subsubsection{ An example} \label{AE}

The above discussion seems a little abstract. We now consider an
explicit example. In confining vacuum, the solution of
Seiberg-Witten factorization can be derived from the Chebyshev
polynomials and is given by \cite{DS, Ferrari}
\begin{eqnarray} \label{che}
&& u_1=Nz \\
&&
u_p=\frac{N}{p}\sum_{q=0}^{[p/2]}C^{2q}_{p}C^q_{2p}\Lambda^{2q}z^{p-2q},
~~~~~~~~~~p\geq 2 \nonumber
\end{eqnarray}
Here $C^p_n\equiv \frac{n!}{p!(n-p)!}$ and we use $z$ to
parameterize the one-parameter space of vacua. It will be
integrated out shortly. We consider the following superpotential
\begin{equation} \label{tree34}
W_{tree}=u_2+4g_4u_4
\end{equation}
Using (\ref{che}) we find
\begin{equation}
u_2=\frac{N}{2}(z^2+2\Lambda^2)
\end{equation}
\begin{equation}
u_4=\frac{N}{4}(z^4+12\Lambda^2z^2+6\Lambda^4)
\end{equation}
We minimize the tree level superpotential (\ref{tree34}) over $z$
and find
\begin{equation} \label{low12}
W_{eff}=N\Lambda^2(1+6g_4\Lambda^2)
\end{equation}
We use the following procedure to ``integrate in'' the glueball.
We set $\Delta\equiv\Lambda^2$, and impose the following equation
in order to match the Veneziano-Yankielowicz superpotential
\begin{equation} \label{2.16}
S=\frac {\partial W_{eff}}{\partial
\ln(\Lambda^{2N})}=\Delta+12g_4\Delta^2
\end{equation}
Then the effective glueball superpotential is
\begin{eqnarray}\label{eff}
W_{eff}(\Lambda,
S)&=&-NS\ln(\frac{\Delta(S)}{\Lambda^2})+W_{tree}(\Lambda,
\Delta(S)) \\ \nonumber &=&
-NS\ln(\frac{\Delta(S)}{\Lambda^2})+N\Delta(S)(1+6g_4\Delta(S))
\end{eqnarray}
One can verify this procedure by integrate out $S$ and reproduce
(\ref{low12}). We can also eliminate the variable $\Delta$ by
solving $\Delta$ in terms of $S$ from (\ref{2.16})
\begin{equation}
\Delta=\frac{-1+\sqrt{1+48g_4S}}{24g_4}
\end{equation}
Plug this into (\ref{eff}) and expand in powers of the glueball
$S$ we find
\begin{equation} \label{SU}
W_{eff}(\Lambda,
S)=-NS(\log(\frac{S}{\Lambda^2})-1)+N(6g_4S^2-72g_4^2S^3+1440g_4^3S^4+\cdots)
\end{equation}
The leading term is indeed the Veneziano-Yankielowicz
superpotential.

\section{Matrix Models and Dijkgraaf-Vafa Conjecture}
\subsection{Matrix Models and saddle point method}
\label{MMSPA} In this subsection we review the solution of the
matrix model by saddle point approximation in planar limit. The
saddle point method is a standard technique of matrix model
reviewed in e.g \cite{DiFrancesco}. We consider a simple example
first studied in \cite{Brezin}. It is a Hermitian $U(M)$ matrix
model with a potential similar to the tree level superpotential
(\ref{tree34})
\begin{equation} \label{2.20}
V(\Phi) = \frac{1}{2} \Tr(\Phi^2) + g_4 \Tr(\Phi^4)
\end{equation}
In large $M$ limit only planar diagram survives. The free energy
of the matrix model can be computed by perturbative method by
doing the combinatorics of the planar Feymann diagram. However,
the matrix model can be solved by saddle point method
\cite{Brezin}. The solution will conveniently sum up all loop
planar amplitude for us. The large $M$ limit planar free energy
${\cal F}$ of the matrix model is defined as
\begin{eqnarray} \label{5.1}
\exp(-M^2 {\cal F}) &=& %\frac{1}{Vol(U(M))}
\int d^{M^2}(\Phi)\exp\{-M(\frac{1}{2}\Tr(\Phi^2)+
g_4\Tr(\Phi^4))\} \\
\nonumber
&=& %\frac{1}{Vol(U(M))}
\int \prod_i d\lambda_i \exp\{ M(-\frac12 \sum_i \lambda_i^2 - g_4
\sum_i \lambda_i^4) + \sum_{i \ne j} \log |\lambda_i - \lambda_j|
\}
\end{eqnarray}

Here $\lambda$ are the $M$ eigenvalues of $\Phi$ and $\cal F$ is
the free energy, which can be evaluated by saddle point
approximation at the planar limit. The $\log$ term comes from the
standard Vandermonde determinant \cite{DiFrancesco}. This matrix
model is Hermitian with rank $M$.

For a one-cut solution, the density of eigenvalues \beq
\rho(\lambda):=\frac{1}{M}\sum_{i=1}^{M}\delta(\lambda-\lambda_i)
\eeq becomes continuous in an interval $(-2a,2a)$ when $M$ goes to
infinity in the planar limit for some $a \in \IR^+$. Here the
interval is symmetric around zero since our model is an even
function. The normalization condition for eigenvalue density is
\begin{equation} \label{5.2}
\int_{-2a}^{2a}d\lambda \rho(\lambda)=1 \ .
\end {equation}
We can rewrite (\ref{5.1}) in terms of the eigenvalue density in
the continuum limit as
\begin{eqnarray} \label{5.11}
\exp(-M^2 {\cal F})&=&\int \prod _{i=1}^{M} d\lambda_i
\exp\{-M^2(\int_{-2a}^{2a} d\lambda \rho(\lambda)(
\frac{1}{2}\lambda^2+g_4\lambda^4) \\ \nonumber &-&
\int_{-2a}^{2a}\int_{-2a}^{2a} d\lambda d\mu
\rho(\lambda)\rho(\mu)\ln|\lambda-\mu|)\} \ .
\end {eqnarray}
In large $M$ limit, the free energy is dominated by the saddle
point distribution of eigenvalues $\rho(\lambda)$. The saddle
point equation that determines $\rho(\lambda)$ can be found by
standard variation of the above action. We find the saddle point
equation is
\begin{equation} \label{5.3}
\frac{1}{2}\lambda+2g_4\lambda^3=P
\int_{-2a}^{2a}d\mu\frac{\rho(\mu)}{\lambda-\mu},
\end{equation}
where $P$ means principal value integration.

The solution of $\rho(\lambda)$ to (\ref{5.3}) can be obtained by
standard matrix model techniques by introducing a resolvent. The
answer is
\begin{equation} \label{5.5}
\rho(\lambda)=\frac{1}{\pi}\left(\frac{1}{2}+
4g_4a^2+2g_4\lambda^2\right)\sqrt{4a^2-\lambda}.
\end {equation}
Plugging the solution into (\ref{5.2}) we obtain the equation that
determine the parameters $a$ :
\begin{equation} \label{5.7}
12g_4a^4+a^2-1=0,
\end{equation}

Substituting these expressions into (\ref{5.11}) gives us the free
energy in the planar limit $M \rightarrow \infty$ as:
\begin{eqnarray}
{\cal F}&=&\int_{-2a}^{2a} d\lambda \rho(\lambda)(
\frac{1}{2}\lambda^2+g_4\lambda^4) -\int\int_{-2a}^{2a} d\lambda
d\mu
\rho(\lambda)\rho(\mu)\ln|\lambda-\mu|.\nonumber \\
\end{eqnarray}
One obtains
\begin{equation} \label{freeenergy}
{\cal F}(g_4)-{\cal
F}(0)=\frac{1}{4}(a^2-1)+(6g_4a^4+a^2-2)g_4a^4-\frac{1}{2}\log
(a^2) \ .
\end{equation}
Equation (\ref{freeenergy}) together with (\ref{5.7}) give the
planar free energy. We can also expand the free energy in powers
of the couplings, by using (\ref{5.7}) to solve for $a^2$
perturbatively
\begin{eqnarray}
a^2&=&1-12g_4+288g_4^2- 8640g_4^3+\cdots.
\end{eqnarray}
Plugging this back into (\ref{freeenergy}) we find the free energy
as a perturbative series
\begin{equation} \label{free}
\CF_0 = {\cal F}(g_4)-{\cal F}(0)=2g_4-18g_4^2 +288g_4^3+\cdots.
\end{equation}
This result sums up all loop amplitude of the planar diagram
combinatorics as explicitly  demonstrated in \cite{Brezin}.

\subsection{Dijkgraaf-Vafa conjecture and a diagrammatic
derivation} \label{DVCDD}

Dijkgraaf and Vafa made a remarkable proposal that the glueball
superpotential and other holomorphic data of $\CN =1$
supersymmetric gauge theories in four dimensions can be computed
from an auxiliary matrix model \cite{DV1,DV2,DV3}. While the
original proposal arose from consideration of stringy dualities
arising in context of geometrically engineered field theories
\cite{GV, CIV}, two subsequent papers have suggested direct field
theory proofs of the proposal \cite{DGLVZ,CDSW}.   These works
considered $U(N)$ gauge theories with an adjoint chiral matter
multiplet $\Phi$ and a tree-level superpotential $W(\Phi) = \sum_k
\frac{g_k}{k} \Tr(\Phi^k)$. Using somewhat different techniques
(\cite{DGLVZ} uses properties of superspace perturbation theory,
while \cite{CDSW} relies on factorization of chiral correlation
functions, symmetries, and the Konishi anomaly) these papers
conclude that:
\begin{enumerate}
\item The computation of the effective superpotential as a function of the
    glueball superfield reduces to computing matrix integrals.
\item Because of holomorphy and symmetries (or properties of
    superspace perturbation theory),
    only planar Feynman diagrams contribute.
\item These diagrams can be summed up by the large-$N$ limit of an
    auxiliary Matrix model.  The field theory effective superpotential is
    obtained as a derivative of the Matrix model free energy.
\end{enumerate}
Various generalizations and extensions of these ideas ({\em e.g.},
$\CN=1^*$ theories \cite{Dorey}, fundamental matter
\cite{Argurio:2002xv},  non-supersymmetric cases
\cite{Dijkgraaf:2002wr}, other gauge groups \cite{Ita:2002kx},
baryonic matter \cite{Argurio:2002hk}, gravitational corrections
\cite{Klemm:2002pa}, Seiberg Duality \cite{Feng:2002zb}, some
scaling properties \cite{Matone}, and using holomorphy to solve
matrix model \cite{holomorphy} ) have been considered in the
literature. For a review see \cite{Argurio}. In what follows we
focus on the glueball superpotential in the confining vacuum,
although the Dijkgraaf-Vafa conjecture also concerns other
holomorphic data such as the low energy $U(1)$ gauge coupling
constant.

\subsubsection{ The statement of the conjecture and an
explicit check}

Condiser the following $\CN =1$ $U(N)$ gauge theory
\begin{equation}
{\cal L}=\int d^4\theta \Phi^{\dagger} e^V \Phi+2\pi i \tau\int
d^2\theta S+h.c.+ \int d^2\theta W(\Phi)+h.c
\end{equation}
with a single-trace tree level superpotential
\begin{equation} \label{tree56}
W_{tree}=\sum^{n+1}_{k=1}\frac{g_k}{k}\Tr(\Phi^k)
\end{equation}
In the confining vacuum, the low energy degree of freedom consists
a decoupled $U(1)$ and confining $SU(N)$. The low energy dynamics
of the confining $SU(N)$ can be described by a glueball superfield
$S$. The effective glueball superpotential is the
Veneziano-Yankielowicz part plus a perturbative part
\begin{equation}
W_{eff}(\Lambda,S)=NS(-\ln(\frac{S}{\Lambda^2})+1)+W_{per}(S)
\end{equation}
Dijkgraaf and Vafa considered the free energy of a corresponding
matrix model with the same potential
\begin{equation}
Z = \exp( \frac{\CF_0}{g_s^2}) = \frac{1}{{\rm Vol}(U(M))} \int
[D\Phi] \exp\left(-\frac{1}{g_s}\Tr\, W(\Phi)\right) .
\label{matmod}
\end{equation}
We take a large $M$ limit  with the 't Hooft coupling $S \equiv M
g_s$ fixed, so only planar diagrams contribute to the free energy.
The free energy of matrix model can be expressed as a power series
of the 't Hooft coupling constant $S$. A planar diagram with $h$
index loops (i.e. ``holes'' in the Feymann diagram) contributes
$S^h$ to the free energy.
\begin{eqnarray}
\CF_0(S) &=& \sum_h \CF_{0,h} \, S^h.
\end{eqnarray}
Dijkgraaf and Vafa identified the 't Hooft coupling $S \equiv M
g_s$ with the glueball in the gauge theory and conjectured the
perturbative part of the superpotential is given by the derivative
of the matrix model free energy
\begin{eqnarray}
W_{pert}(S) &=&  N \frac{\partial}{\partial S} \CF_0(S),
\label{deriv}
\end{eqnarray}
Dijkgraaf and Vafa also pointed out the Veneziano-Yankielowicz
term in $W_{eff}(S)$ arises from the volume factor in the
integration over matrices in (\ref{matmod}), but here we will not
explore this point. We should note the rank the field theory gauge
group $N$ is finite, but the matrix model is a $M\times M$ matrix
with $M\rightarrow \infty$ to select planar diagrams.

Now we can make a simple check of the Dijkgraaf-Vafa conjecture.
Consider the superpotential we studied in (\ref{tree34})
(\ref{2.20})
\begin{equation}
W(\Phi) = \frac{1}{2} \Tr(\Phi^2) + g_4 \Tr(\Phi^4)
\end{equation}
In subsection \ref{MMSPA} we solve the matrix model free energy
(\ref{free}) in planar limit. We have took the 't Hooft coupling
constant to be $1$ in subsection \ref{MMSPA}. It is not difficult
to recover the 't Hooft coupling constant in the free energy
expression (\ref{free}), and we find
\begin{equation}
\CF_0(S) =2g_4S^3-18g_4^2S^4 +288g_4^3S^5+\cdots.
\end{equation}
Then the Dijkgraaf-Vafa conjecture states that the effective
glueball superpotential is
\begin{eqnarray}
W_{eff}(\Lambda,S)&=&NS(-\ln(\frac{S}{\Lambda^2})+1)+N
\frac{\partial}{\partial S} \CF_0(S) \\ \nonumber &=&
NS(-\ln(\frac{S}{\Lambda^2})+1)+N(6g_4S^2-72g_4^2S^3+1440g_4^3S^4+\cdots)
\end{eqnarray}
This is in agreement with the glueball superpotential (\ref{SU})
we derived from Seiberg-Witten curve. Thus we provide a check of
Dijkgraaf-Vafa conjecture in this simple case. It is easy to show
the effective glueball superpotential agrees with matrix model
computation to all order in $S$, by keeping the exact expression
together with the constraining equation. For general arbitrary
polynomial single-trace tree level superpotential, the
Dijkgraaf-Vafa conjecture is checked  in confining vacuum in
\cite{Ferrari}. Some non-confining vacuum cases are studied in
\cite{Gukov}.

\subsubsection{ A diagrammatic derivation of Dijkgraaf-Vafa conjecture}

In section \ref{MTS} we will use the diagrammatic approach in
\cite{DGLVZ} to derive glueball superpotential of gauge theory
with a double-trace tree level superpotential. We give a schematic
review here.

\begin{description}

\item[1. The Power of Holomorphy: ]
We are interested in expressing the effective superpotential in
terms of the {\em chiral} glueball superfield $S$.   Holomorphy
tells us that it will be independent of the parameters of the
anti-holomorphic part of the tree-level superpotential.
% we need only concern ourselves with the
%{\em holomorphic} quantity $\vev{\Phi\Phi}$.
%Furthermore, holomorphy lets us {\em choose} a particularly simple form
Therefore, without loss of generality, we can choose a
particularly simple form for $\bar{W}(\bar{\Phi})$:
\begin{equation}
\bar{W}(\bar{\Phi}) = \frac{1}{2}\bar{m}\bar{\Phi}^2.
\end{equation}
Integrating out the anti-holomorphic fields and performing
standard superspace manipulations as discussed in Sec.\ 2 of
\cite{DGLVZ}, gives
\begin{equation}
\label{niceaction} S = \int d^4x d^2\theta
\left(-\frac{1}{2\bar{m}}\Phi\left(\Box- i{\cal W}^{\alpha}
D_{\alpha}\right)\Phi + W_{tree}(\Phi)\right)
\end{equation}
as the part of the action that is relevant for computing the
effective potential as a function of $S$. Here,
$\Box=\frac{1}{2}\pa_{\alpha\dot\alpha}\pa^{\alpha\dot\alpha}$ is
the d'Alembertian, and $W_{tree}$ is the tree-level
superpotential, expanded as $\frac{1}{2}m\Phi^2 +
\mbox{interactions}$.   (The reader may consult Sec.\ 2 of
\cite{DGLVZ} for a discussion of various subtleties such as why
the $\Box$ can be taken as the ordinary d'Alembertian as opposed
to a gauge covariantized $\Box_{{\rm cov}}$).
\item[2. The Propagator: ]
After reduction into the form (\ref{niceaction}), the quadratic
part gives the propagator.  We write the covariant derivative in
terms of Grassmann momentum variables
\begin{equation}
D_\alpha = \pa/\pa\theta^\alpha := -i\pi_\alpha,
\end{equation}
and it has been shown in \cite{DGLVZ} that by rescaling the
momenta we can put $\bar{m}=1$ since all $\bar{m}$ dependence
cancels out. Then the momentum space representation of the
propagator is simply
\begin{equation}
\label{schwinger} \int_0^\infty ds_i\, \exp\left(-s_i(p_i^2 +
{\cal W}^{\alpha} \pi_{i\alpha} + m) \right),
\end{equation}
where $s_i$ is the Schwinger time parameter of $i$-th Feynman
propagator. Here the precise form of the $\CW^\alpha \pi_\alpha$
depends on the representation of the gauge group that is carried
by the field propagating in the loop.
\item[3. Calculation of Feynman Diagrams: ]
The effective superpotential as a function of the glueball $S$ is
a sum of vacuum Feynman diagrams computed in the background of a
fixed constant $\CW_\alpha$ leading to insertions of this field
along propagators.  In general there will be $\ell$ momentum
loops, and the corresponding momenta must be integrated over
yielding the contribution
\begin{eqnarray}
I &=& \left(\int\prod_i ds_i e^{-s_i m}\right)
\left(\int\prod_{a,i} d^4p_a\,  e^{-s_i p_i^2}\right) \cdot
\left(\int\prod_{a,i} d^2\pi_a\,
e^{-s_i\CW^\alpha\pi_{i\alpha}}\right) \cdot
 \nonumber \\
&=& \left(\int\prod_i ds_i e^{-s_i m}\right) I_{boson} \cdot
I_{fermion} \cdot \nonumber \\
&=&  I_{boson} \cdot I_{fermion}  \frac{1}{m^P} \label{mominteg}
\end{eqnarray}
to the overall amplitude. We will show that th $s_i$ dependence in
$I_{boson} \cdot I_{fermion}$ cancels later and thus explain the
origin of the last line. Here $a$ labels momentum loops, while $i
= 1, \ldots, P$ labels propagators.  The momenta in the
propagators are linear combinations of the loop momenta because of
momentum conservation.
\item[4. Bosonic Momentum Integrations: ]
The bosonic contribution can be expressed as
\begin{equation}
I_{boson} = \int\prod_{a=1}^\ell \frac{d^4p_a}{(2\pi)^4}
            \exp\left[-\sum_{a,b} p_a M_{ab}(s) p_b\right]
          = \frac{1}{(4\pi)^{2\ell}} \frac{1}{(\det\, M(s))^2},
\end{equation}
where we have defined the momentum of the $i$-th propagator in
terms of the independent loop momenta $p_a$
\begin{equation}
p_i = \sum_{a} L_{ia} p_a
\end{equation}
via the matrix elements $L_{ia}\in \{0,\pm 1\}$ and
\begin{equation}
M_{ab}(s) = \sum_i s_i L_{ia} L_{ib}.
\end{equation}
\item[5. Which Diagrams Contribute: ]
Since each momentum loop comes with two fermionic $\pi_\alpha$
integrations (\ref{mominteg}), a non-zero amplitude will require
the insertion of $2\ell$ $\pi_\alpha$s.  From (\ref{schwinger}) we
see that that $\pi_\alpha$ insertions arise from the power series
expansion of the fermionic part of the propagator and that each
$\pi_\alpha$ is accompanied by a $\CW_\alpha$.  So in total we
expect an amplitude containing $2\ell$ factors of $\CW_\alpha$.
Furthermore, since we wish to compute the superpotential as a
function of $S \sim \Tr(\CW^\alpha \CW_\alpha)$ each index loop
can only have zero or two $\CW_\alpha$ insertions. These
considerations together imply that if a diagram contributes to the
effective superpotential as function of the $S$, then number of
index loops $h$ must be greater than or equal to the number of
momentum loops $\ell$, {\em i.e.},
\begin{equation}
h \geq \ell. \label{hbigl}
\end{equation}

\item[6. Planarity: ]
The above considerations are completely general. Now let us
specialize to $U(N)$ theories with single-trace operators. A
diagram with $\ell$ momentum loops has
\begin{equation}
h = \ell + 1 - 2g \label{eq:loops}
\end{equation}
index loops, where $g$ is the genus of the surface generated by 't
Hooft double line notation.  Combining this with (\ref{hbigl})
tell us that $g  = 0$, {\em i.e.}, only planar diagrams
contribute.

\item[7. Doing The Fermionic Integrations: ]
First let us discuss the combinatorial factors that arise from the
fermionic integrations.   Since the number of momentum loops is
one less than the number of index loops, we must choose which of
the latter to leave free of $\CW_\alpha$ insertions.  This gives a
combinatorial factor of $h$, and the empty index loop gives a
factor of $N$ from the sum over color.   For each loop with two
$\CW_\alpha$ insertions we get a factor of ${1 \over 2}
\Tr(\CW^\alpha \CW_\alpha) = 16 \pi^2 S$. Since we are dealing
with adjoint matter, the action of $\CW_\alpha$ is through a
commutator
\begin{equation}
\exp\left(-s_i[\CW_i^\alpha,-]\pi_{i\alpha}\right)
\label{commprop}
\end{equation}
in the Schwinger term. As in the bosonic integrals above, it is
convenient to express the fermionic propagator momenta as sums of
the independent loop momenta:
\begin{equation}
\pi_{i\alpha} = \sum_a L_{ia} \pi_{a\alpha},
\end{equation}
where the $L_{ia}$ are the same matrix elements as introduced
above. The authors of \cite{DGLVZ} also find it convenient to
introduce auxiliary fermionic variables via the equation
\begin{equation}
\CW_i^\alpha = \sum_a L_{ia} \CW_a^\alpha \, . \label{auxferm}
\end{equation}
Here, the $L_{ia}=\pm 1$ denotes the left- or right-action of the
commutator. In terms of the $\CW_a^\alpha$, the fermionic
contribution to the amplitude  can be written as
\begin{eqnarray}
I_{fermion} &=& N h(16\pi^2 S)^\ell \int \prod_a d^2\pi_a\,
d^2\CW_a\,
\exp\left[-\sum_{a,b} \CW_a^\alpha M_{ab}(s) \pi_{b\alpha}\right] \nonumber \\
&=& (4\pi)^{2\ell} N h S^\ell (\det\, M(s))^2. \label{planferm}
\end{eqnarray}

\item[8. Localization: ]
The Schwinger parameter dependence in the bosonic and fermionic
momentum integrations cancel exactly
\begin{equation}
\label{prefact} I_{boson} \cdot I_{fermion} = N h S^\ell,
\end{equation}
implying that the computation of the effective superpotential as a
function of the $S$ localizes to summing matrix integrals.   All
the four-dimensional spacetime dependence has washed out. The full
effective superpotential $W_{eff}(S)$ is thus a sum over planar
matrix graphs with the addition of the Veneziano-Yankielowicz term
for the pure Yang-Mills theory \cite{VY}.  The terms in the
effective action proportional to $S^\ell$ arise exclusively from
planar graphs with $\ell$ momentum loops giving a perturbative
computation of the {\it exact} superpotential.

%
%The dynamics are described by the
%glueball superfield $S$.  Note that terms in the effective action with an
%$S^\ell$ interaction arise exclusively from those planar graphs with $\ell$
%momentum loops. The {\em exact} superpotential is obtained perturbatively
%via the prescription we have just outlined.

%
\item[9. The Matrix Model: ]
The localization of the field theory computation to a set of
planar matrix diagrams suggests that the sum of diagrams can be
computed exactly by the large-$M$ limit of a bosonic Matrix model.
(We distinguish between $M$, the rank of the matrices in the
Matrix model and $N$, the rank of the gauge group.)  The
prescription of Dijkgraaf and Vafa does exactly this for
single-trace superpotentials.   Since the number of momentum loops
is one less than the number of index loops in a planar diagram,
i.e. $\ell=h-1$, the net result of the bosonic and fermionic
integrations in (\ref{prefact}) can be written as
\begin{equation}
    I_{boson} \cdot I_{fermion}  = N {\partial S^h \over \partial S}.
\end{equation}
%{\em A priori} there need not be relation between this result for the effective
%superpotential and a bosonic matrix model.  However, the Dijkgraaf-Vafa
%prescription equates these two.
Because of this, the perturbative part of the effective
superpotential, namely the sum over planar diagrams in the field
theory, can be written in terms of the genus zero free energy
$\CF_0(S)$ of the corresponding matrix model:
\begin{eqnarray}
W_{pert}(S) &=&  N \frac{\partial}{\partial S} \CF_0(S),  \label{deriv} \\
\CF_0(S) &=& \sum_h \CF_{0,h} \, S^h.
\end{eqnarray}

Thus we completed the schematic review of the derivation of the
Dijkgraaf-Vafa conjecture in \cite{DGLVZ}.

\end{description}

\section{Multi-trace Superpotentials} \label{MTS}
A stringent and simple test of the Dijkgraaf-Vafa proposal and of
the proofs presented in \cite{DGLVZ,CDSW} is to consider
superpotentials containing multi-trace terms such as \beq W(\Phi)
= \frac{1}{2}\Tr(\Phi^2) + g_4 \Tr(\Phi^4) + \widetilde{g}_2
(\Tr(\Phi^2))^2. \label{doubtrace} \eeq In \cite{Balasubramanian}
we find that for such multi-trace theories:
\begin{enumerate}
\item The computation of the effective superpotential as a function of the
glueball superfield still reduces to computing matrix integrals.
\item  Holomorphy and symmetries do not forbid non-planar
contributions; nevertheless only a certain subset of the planar
diagrams contributes to the effective superpotential.
\item  This subclass of planar graphs can be summed up by
the large-$N$ limit of an associated multi-trace Matrix model.
However, because of differences in combinatorial factors, the
field theory effective superpotential {\it cannot} be obtained
simply as a derivative of the multi-trace Matrix model free energy
as in \cite{DV3, DGLVZ}.
\item
Multi-trace theories can be linearized in traces by the addition
of auxiliary singlet fields $A_i$. The superpotentials for these
theories as a function of both the $A_i$ and the glueball can be
computed from an associated Matrix model. This shows that the
basic subtlety involves the correct identification of the field
theory glueball as a variable in a related Matrix model. This is
similar in spirit to the UV ambiguity of $Sp(N)$ gauge theory with
antisymmetric matter discussed in \cite{Kraus}
\end{enumerate}

It is worth mentioning several further reasons why multi-trace
superpotentials are interesting. First of all, the general
deformation of a pure $\CN = 2$ field theory to an $\CN = 1$
theory with adjoint matter involves multi-trace superpotentials,
and therefore these deformations are important to understand. What
is more, multi-trace superpotentials cannot be geometrically
engineered \cite{Reverse} in the usual manner for a simple reason:
in geometric engineering of gauge theories the tree-level
superpotential arises from a disc diagram for open strings on a
D-brane and these, having only one boundary, produce single-trace
terms.   In this context, even if multi-trace terms could be
produced by quantum corrections, their coefficients would be
determined by the tree-level couplings and would not be freely
tunable.  Hence comparison of the low-energy physics arising from
multi-trace superpotentials with the corresponding Matrix model
calculations is a useful probe of the extent to which the
Dijkgraaf-Vafa proposal is tied to its geometric and D-brane
origins.   In addition to these motivations, it is worth recalling
that the double  scaling limit of the $U(N)$ matrix model with a
double-trace potential is related to a theory of two-dimensional
gravity with a cosmological constant.   This matrix model also
displays phase transitions between smooth, branched polymer and
intermediate phases \cite{Double}.   It would be interesting to
understand whether and how these phenomena manifest themselves as
effects  in a four dimensional field theory. Finally, multi-trace
deformations of field theories have recently made an appearance in
the contexts of the $AdS$/CFT correspondence and a proposed
definition of string theories with a nonlocal worldsheet theory
\cite{AdS}.

Now we follow \cite{Balasubramanian} to compute the multi-trace
superpotential.

\subsection{A classification of multi-trace diagrams} \label{CMTD}
We have reviewed in subsection \ref{DVCDD} the field theory
calculation of the effective superpotential for a single-trace
theory localizes to a matrix model computation.  In this
subsection we show how the argument is modified when the
tree-level superpotential includes multi-trace terms. We consider
an $\CN=1$ theory with the tree-level superpotential
\begin{equation}
\label{tree} W_{tree}=\frac{1}{2}
\Tr(\Phi^2)+g_4\Tr(\Phi^4)+\widetilde {g}_2(\Tr(\Phi^2))^2 \ .
\end{equation}
To set the stage for our perturbative computation of the effective
superpotential we begin by analyzing the structure of the new
diagrams introduced by the double-trace term.  If $\widetilde{g}_2
= 0$, the connected diagrams we get are the familiar single-trace
ones; we will call these {\it primitive diagrams}. When
$\widetilde{g}_2 \neq 0$  propagators in primitive diagrams can be
spliced together by new double-trace vertices. It is useful to do
an explicit example to see how this splicing occurs.
\begin{figure}
  \begin{center}
 \epsfysize=5cm
   \mbox{\epsfbox{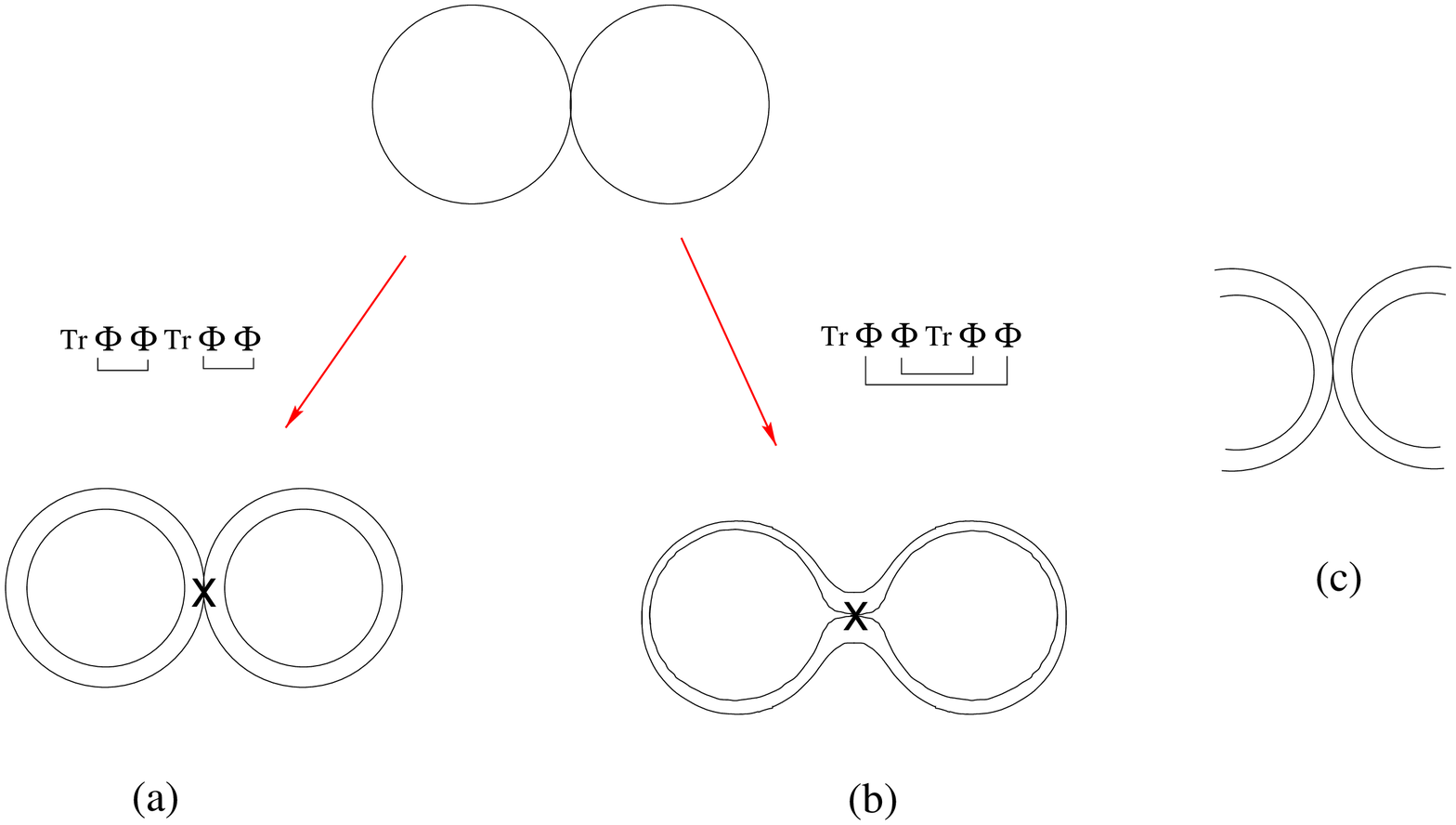}}
\end{center}
\caption{Two ways in which the double-trace operator: $\Tr(\Phi^2)
\Tr(\Phi^2)$ can be contracted using the vertex shown in (c).}
\label{f:contract}
\end{figure}

\begin{figure}
  \begin{center}
 \epsfysize=2.0in
   \mbox{\epsfbox{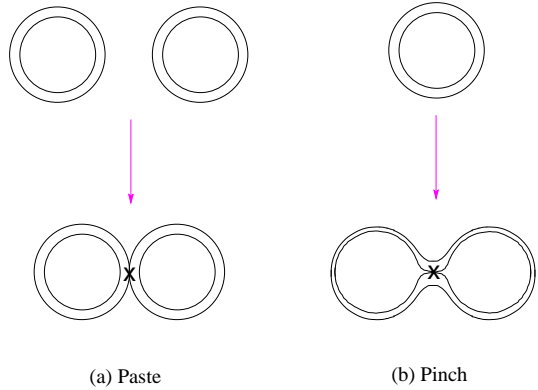}}
\end{center}
\caption{With the inclusion of the double-trace term we need new
types of vertices. These can be obtained from the ``primitive''
diagrams associated with the pure single-trace superpotential by
(a) pasting or (b) pinching. The vertices have been marked with a
cross.} \label{pastepinch}
\end{figure}

As an example, let us study the expectation value of the
double-trace operator: $\langle \Tr(\Phi^2) \Tr(\Phi^2) \rangle$.
To lowest order in couplings, the two ways to contract $\Phi$s
give rise to the two diagrams in \fref{f:contract}.  When we draw
these diagrams in double line notation, we find that
\fref{f:contract}a  corresponding to
 $\Tr (\overbracket{\Phi \Phi}) \Tr (\overbracket{\Phi \Phi})$ has
four index loops, while \fref{f:contract}b corresponding to $\Tr
(\overbracket{\Phi \overbracket{\Phi) \Tr (\Phi} \Phi})$ has only
two index loops.   Both these graphs have two momentum loops. For
our purposes both of these Feynman diagrams can also be generated
by a simple pictorial algorithm:  we splice together propagators
of primitive diagrams using the vertex in \fref{f:contract}c, as
displayed in \fref{pastepinch}a and b.   All graphs of the
double-trace theory can be generated from primitive diagrams by
this simple algorithm.  Note that the number of index loops never
changes when primitive diagrams are spliced by this pictorial
algorithm.

\begin{figure}
  \begin{center}
 \epsfysize=1.2in
   \mbox{\epsfbox{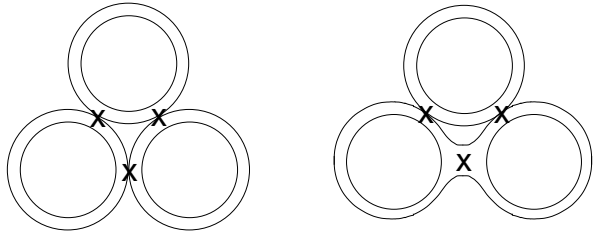}}
\end{center}
\caption{More examples of ``pinched'' diagrams.} \label{notpaste}
\end{figure}

If a splicing of diagrams does not create a new momentum loop we
say that the diagrams have been {\it pasted} together.   This
happens when the diagrams being spliced are originally
disconnected as, for example, in \fref{pastepinch}a.  In fact
because of momentum conservation, no momentum at all flows between
pasted diagrams.  If a new momentum loop is created we say that
that the diagrams have been {\it pinched}. This happens when two
propagators within an already connected diagram are spliced
together as, for example, in \fref{pastepinch}b.   In this example
one momentum loop becomes two because momentum can flow through
the double-trace vertex.  Further examples of pinched diagrams are
given in \fref{notpaste} where the new momentum loop arises from
momentum flowing between the primitive diagrams via double-trace
vertices.

To make the above statement more clear, let us provide some
calculations. First, according to our operation, the number of
double index loops never increases whether under pasting or
pinching. Second, we can calculate the total number of independent
momentum loops $\ell$ by $\ell= P-V+1$ where $P$ is the number of
propagators and $V$, the number of vertices. If we connect two
separate diagrams by {\it pasting}, we will have
$P_{tot}=(P_1-1)+(P_2-1)+4$, $V_{tot}=V_1+V_2+1$ and \beq
\ell_{tot}=P_{tot}-V_{tot}+1=\ell_1+\ell_2, \eeq which means that
the total number of momentum loops is just the sum of the
individual ones. If we insert the double-trace vertex in a single
connected diagram by {\it pinching}, we will have $P_{tot}=P-2+4$,
$V_{tot}=V+1$, and \beq \ell_{tot}= P_{tot}-V_{tot}+1 = \ell+1,
\eeq which indicates the creation of one new momentum loop.

Having understood the structure of double-trace diagrams in this
way, we can adapt the techniques of \cite{DGLVZ} to our case.  The
steps 1-4 as described in Sec. \ref{DVCDD} go through without
modification since they are independent of the details of the
tree-level superpotential. However the steps 5-9 are modified in
various ways.   First of all naive counting of  powers of
fermionic momenta as in step 5 leads to the selection rule
\begin{equation}
  h \geq \ell,
\label{selectno}
\end{equation}
where $h$ is the total number of index loops and $\ell$ is the
total
 number of momentum loops.  (The holomorphy and symmetry based
 arguments of \cite{CDSW} would lead to the same conclusion.)
 Since
no momentum flows between pasted primitive diagrams it is clear
that
 this selection rule would permit some of the primitive components to
 be non-planar.   Likewise, both planar and some non-planar pinching
 diagrams are admitted.   An example of a planar pinching diagram that
 can contribute according to this rule is \fref{pastepinch}b.
 However, we will show in the next subsection that more careful
 consideration of the structure of perturbative diagrams shows that
 only diagrams built by pasting planar primitive graphs give non-zero
 contributions to the effective superpotential.

\subsection{Which diagrams contribute: selection rules}

In order to explain which diagrams give non-zero contributions to
the multi-trace superpotential it is useful to first give another
perspective on the fermionic momentum integrations described in
steps 5-7 in Sec. \ref{DVCDD}.   A key step in the argument of
\cite{DGLVZ} was to split the glueball insertions up in terms of
auxiliary fermionic variables associated with each of the momentum
loops as in (\ref{auxferm}).   We will take a somewhat different
approach. In the end we want to attach zero or two fields ${\cal
W}^{\alpha}_{(p)}$ to each {\it index} loop, where $p$ labels the
index loop, and the total number of such fields must bring down
enough fermionic momenta to soak up the corresponding
integrations.   On each oriented propagator, with momentum
$\pi_{i\alpha}$, we have a left index line which we  label $p_L$
and a right index line which we label $p_R$. Because of the
commutator in (\ref{commprop}), the contribution of this
propagator will be \beq \label{e1} \exp( - s_i ( \pi_{i\alpha}
({\cal W}^{\alpha}_{(p_L)} -
 {\cal W}^{\alpha}_{(p_R)} )).
%\label{indexW}
\eeq Notice that we are omitting $U(N)$ indices,
which are simply replaced by the different index loop labels.  In
a standard
 planar diagram for
a single-trace theory, we have one more index loop than momentum
loop.
 So even in this case the choice of auxiliary variables in (\ref{e1})
 is not quite the same as in (\ref{auxferm}), since the number of
 $\CW_\alpha$s is twice the number of index loops in (\ref{e1}) while
the number of auxiliary variables is twice the number of momentum
loops in (\ref{auxferm}).

Now in order to soak up the fermionic $\pi$  integrations in
(\ref{mominteg}), we must expand (\ref{e1}) in powers and  extract
terms of the form \beq {\cal W}_{(p_1)}^2 {\cal W}_{(p_2)}^2
\ldots {\cal W}_{(p_l)}^2, \eeq where $\ell$ is the number of
momentum loops and all the $p_i$ are distinct. The range of $p$'s
are over $1, \ldots, h$, with $h$ the number of index loops. In
the integral over the anticommuting momenta, we have all $h$
${\cal W}_{(p)}$ appearing. However, one linear combination, which
is the `center of mass' of the ${\cal W}_{(p)}$, does not appear.
This can be seen from (\ref{e1}): if we add a constant to all
${\cal W}_{(p)}$ simultaneously, the propagators do not change.
Thus, without loss of generality, one can set the ${\cal W}_{(p)}$
corresponding to the outer loop in a planar diagram equal to zero.
Let us assume this variable is  ${\cal W}_{(h)}$ and later
reinstate it. All ${\cal W}_{(p)}$ corresponding to inner index
loops remain, leaving as many of these as there are momentum loops
in  a planar diagram. It is then straightforward to demonstrate
that the $\CW$ appearing in (\ref{auxferm}) in linear combinations
reproduce the relations between propagator momenta and loop
momenta.   In other words, in this ``gauge" where the $\CW$
corresponding to the outer loop is zero, we recover the
decomposition of $\CW_\alpha$ in terms of auxiliary fermions
associated to momentum loops that was used in \cite{DGLVZ} and
reviewed in (\ref{auxferm}) above.

We can  now reproduce the overall factors arising from the
fermionic integrations in the planar diagrams contributing to
(\ref{planferm}). The result from the $\pi$ integrations is some
constant times \beq \prod_{p=1}^\ell {\cal W}_{(p)}^2.
\label{res1} \eeq Reinstating ${\cal W}_{(h)}$ by undoing the
gauge choice, namely by shifting \beq {\cal W}_{(p)} \rightarrow
{\cal W}_{(p)} + {\cal W}_{(h)} \eeq for $p=1,\ldots, h-1$,
(\ref{res1}) becomes \beq \prod_{p=1}^\ell ({\cal W}_{(p)} + {\cal
W}_{(h)})^2. \eeq The terms on which each index loop there has
either zero or two ${\cal W}$ insertions are easily extracted:
\beq \sum_{k=1}^h \left(\prod_{p\neq k} {\cal W}_{(p)}^2\right).
\eeq In this final result we should replace each of the ${\cal
W}_{(p)}^2$ by $S$, and therefore the final result is of the form
\beq h S^{h-1}, \eeq as derived in \cite{DGLVZ} and reproduced in
(\ref{planferm}).

Having reproduced the result for single-trace theories we can
easily show that all non-planar and pinched contributions to the
multi-trace effective superpotential vanish.   Consider any
diagram with $\ell$ momentum loops and $h$ index loops. By the
same arguments as above, we attach some ${\cal W}_{(p)}$ to each
index loop, and again, the `center of mass' decouples due to the
commutator nature of the propagator. Therefore, in the momentum
integrals, only $h-1$ inequivalent ${\cal W}_{(p)}$ appear. By
doing $\ell$ momentum integrals, we generate a polynomial of order
$2\ell$ in the $h-1$ inequivalent ${\cal W}^{\alpha}_{(p)}$. Each
index loop can have zero or two $\CW$'s.  Therefore, we reach the
important conclusion that the total number of index loops must be
larger than the number of momentum loops
\begin{equation}
h > \ell
\end{equation}
while the naive selection rule (\ref{selectno}) says that it could
be larger or equal.

Consider pasting and pinching $k$ primitive diagrams together,
each with $h_i$ index loops and $\ell_i$ momentum loops.
According to the rules set out in the previous subsection, the
total number of index loops and the total number of momentum loops
are given by:
\begin{equation}
h = \sum_i h_i   ~~~;~~~  \ell \geq \sum_i \ell_i
\end{equation}
with equality only when all the primitive diagrams are pasted
together without additional momentum loops.  Now the total number
of independent $\CW$s that appear in full diagram is $\sum_i (h_i
- 1)$ since in each primitive diagram the ``center of mass'' $\CW$
will not appear.  So the full diagram is non-vanishing only when
\begin{equation}
\ell \leq \sum_i (h_i -1).
\end{equation}
This inequality is already saturated by the momenta appearing in
the primitive diagrams if they are planar.  So we can conclude two
things. First, only planar primitive diagrams appear in the full
diagram. Second, only pasted diagrams are non-vanishing, since
pinching introduces additional momentum loops which would violate
this inequality.

\paragraph{Summary: } The only diagrams that contribute to the
effective multi-trace superpotential are pastings of planar
primitive diagrams.  These are tree-like diagrams which string
together double-trace vertices with ``propagators'' and ``external
legs'' which are themselves primitive diagrams of the single-trace
theory.   Below we will explicitly evaluate such diagrams and
raise the question of whether there is a generating functional for
them.

%=
\subsection{Summing pasted diagrams}

In the previous subsection we generalized steps 5 and 6 of the the
single trace case in Sec. \ref{DVCDD} to the double-trace theory,
and found that the surviving diagrams consist of planar connected
primitive vacuum graphs pasted together with double-trace
vertices.   Because of momentum conservation, no momentum can flow
through the double-trace vertices in such graphs.  Consequently
the fermionic integrations and the proof of localization can be
carried out separately for each primitive graph, and the entire
diagram evaluates to a product of the primitive components times a
suitable power of $\widetilde{g}_2$, the double-trace coupling.

Let $G_i$, $i = 1, \ldots, k$ be the planar primitive graphs that
have been pasted together, each with $h_i$ index loops and $\ell_i
= h_i - 1$ momentum loops to make a double-trace diagram $G$.
Then, using the result (\ref{prefact}) for the single-trace case,
the Schwinger parameters in the bosonic and fermionic momentum
integrations cancel giving a factor
\begin{equation}
I_{boson} \cdot I_{fermion} = \prod_i (N h_i S^\ell_i) = N^k
S^{\sum_i (h_i - 1) } \prod_i h_i,
\end{equation}
where the last factor arises from the number of ways in which the
glueballs $S$ can be inserted into the propagators of each
primitive diagram.  Defining $C(G) = \prod_i h_i$ as the glueball
symmetry factor, $k(G)$ as the number of primitive components,
$h(G) = \sum_i h_i$ as the total number of index loops and
$\ell(G) = \sum_i \ell_i = h(G) - k(G)$ as the total number of
momentum loops, we get
\begin{equation}
I_{boson} \cdot I_{fermion} = \prod_i (N h_i S^\ell_i) = N^{h(G) -
\ell(G)} S^{l(G)} C(G).
\end{equation}
We can assemble this with the Veneziano-Yankielowicz contribution
contribution for pure gauge theory \cite{VY} to write the complete
glueball effective action as
\begin{equation}
W_{eff}=-NS(\log (S/\Lambda^2)-1)+\sum_{G} C(G){\cal
F}(G)N^{h(G)-\ell(G)}S^{\ell(G)}, \label{effact}
\end{equation}
where $\CF (G)$ is the combinatorial factor for generating the
graph $G$ from the Feynman diagrams of the double-trace theory. We
can define a free energy related to above diagrams as \beq
\label{free0} {\cal F}_0= \sum_{G}{\cal F}(G)S^{h(G)}. \eeq ${\cal
F}_0$ is a generating function for the diagrams that contribute to
the effective superpotential, but does not include the
combinatorial factors arising from the glueball insertions.   In
the single-trace case that combinatorial factor was simply $N
h(G)$ and so we could write $W_{eff} = N (\partial {\cal
F}_0/\partial S)$.   Here $C(G) = \prod h_i$ is a product rather
than a sum $h(G) = \sum h_i$, and so the effective superpotential
cannot be written as a derivative of the free energy.

Notice that if we rescale $\widetilde{g}_2$ to
$\widetilde{g}_2/N$, there will be a $N^{-(k(G)-1)}$ factor from
 $k(G)-1$ insertions of the double-trace vertex. This factor will
change the $N^{h(G)-l(G)}$ dependence in \eref{effact} to just $N$
for every diagram.  This implies that the matrix diagrams
contributing to the superpotential are exactly those that survive
the large $M$ limit of a bosonic $U(M)$ Matrix model with a
potential
\begin{equation} \label{DTMM}
V(\Phi) = g_2 \Tr(\Phi^2) + g_4 \Tr(\Phi^4) + {\widetilde{g}_2
\over M} \Tr(\Phi^2) \Tr(\Phi^2).
\end{equation}

\subsection{Perturbative calculations}

Thus equipped, let us begin our explicit perturbation
calculations. We shall tabulate all combinatoric data of the
pasting diagrams up to third order. Here $C(G)=\prod_i h_i$ and
${\cal F(G)}$ is obtained by counting the contractions of $\Phi$s.
For pure single-trace diagrams the values of ${\cal F(G)}$ have
been computed in Table 1 in \cite{Brezin}, so we can utilize their
results.

\subsubsection{First Order}
\begin{figure}
  \begin{center}
 \epsfysize=2cm
   \mbox{\epsfbox{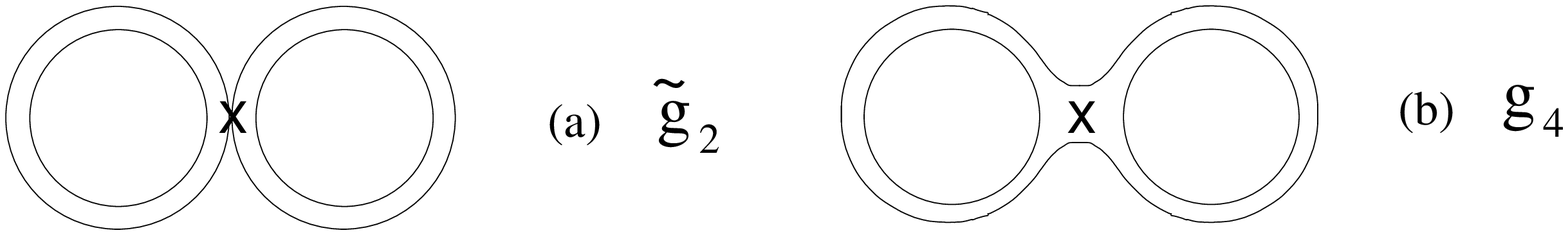}}
\end{center}
\caption{All two-loop primitive and pasting diagrams. The vertices
have been marked with a cross.} \label{g2-1}
\end{figure}

To first order in coupling constants, all primitive (diagram (b))
and pasting diagrams (diagram (a)) are presented in \fref{g2-1}.
Let us illustrate by showing the computations for (a). There is a
total of four index loops and hence $h=4$ for this diagram.
Moreover, since it is composed of the pasting of two primitive
diagrams each of which has $h=2$; thus, we have $C(G) = 2\times 2
= 4$. Finally, $\CF = \widetilde{g}_2$ because there is only one
contraction possible, {\em viz}, $\Tr (\overbracket{\Phi \Phi})
\Tr( \overbracket{\Phi \Phi})$.

In summary we have: \beq\label{two}\begin{tabular}{|c|c|c|}\hline
    diagram  &  $(a)$  &  $(b)$
\\ \hline
$h$ &  $4$ & $3$ \\  \hline
$C(G)$ & $4$ &  $3$ \\
\hline ${\cal F}(G)$ & $\widetilde{g}_2$ &  $2g_4$ \\
\hline
\end{tabular}
\eeq

\subsubsection{Second Order}
\begin{figure}
  \begin{center}
 \epsfysize=4cm
   \mbox{\epsfbox{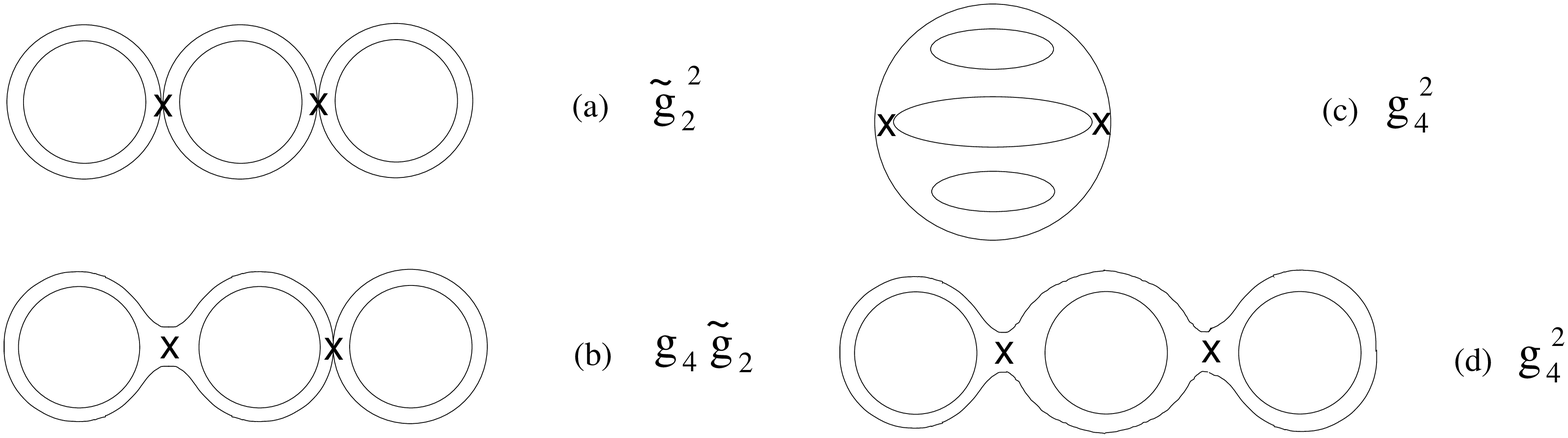}}
\end{center}
\caption{All three-loop primitive and pasting diagrams. The
vertices have been marked with a cross.} \label{g2-2}
\end{figure}

To second order  in the coupling all primitive ((c) and (d)) and
pasting diagrams ((a) and (b)) are drawn in \fref{g2-2} and the
combinatorics are summarized in table \eref{three}. Again, let us
do an illustrative example. Take diagram (b), there are five index
loops, so $h=5$; more precisely it is composed of pasting a left
primitive diagram with $h=3$ and a right primitive with $h=2$, so
$C(G) = 2 \times 3 = 6$. Now for $\CF(G)$, we need contractions of
the form $\Tr (\overbracket{\Phi \Phi} \; \overbracket{\Phi
\overbracket{\Phi) \Tr (\Phi} \Phi}) \Tr (\overbracket{\Phi \Phi})
$; there are $4 \times 2 \times 2 = 16$ ways of doing so.
Furthermore, for this even overall power in the coupling, we have
a minus sign when expanding out the exponent. Therefore $\CF(G) =
-16 \widetilde{g}_2 g_4$ for this diagram.

In summary, we have:
\beq\label{three}\begin{tabular}{|c|c|c|c|c|}\hline
    diagram  &  $(a)$  &  $(b)$ & $(c)$ & $(d)$
\\ \hline
$h$ &  $6$ & $5$ & $4$ &$4$  \\  \hline
$C(G)$ & $8$ &  $6$&  $4$ & $4$  \\
\hline ${\cal F}(G)$ & $-4\widetilde{g}_2^2$ &
$-16\widetilde{g}_2g_4$&
$-2g_4^2$ & $-16g_4^2$   \\
\hline
\end{tabular}
\eeq

\subsubsection{Third Order}
\begin{figure}
  \begin{center}
 \epsfysize=14cm
   \mbox{\epsfbox{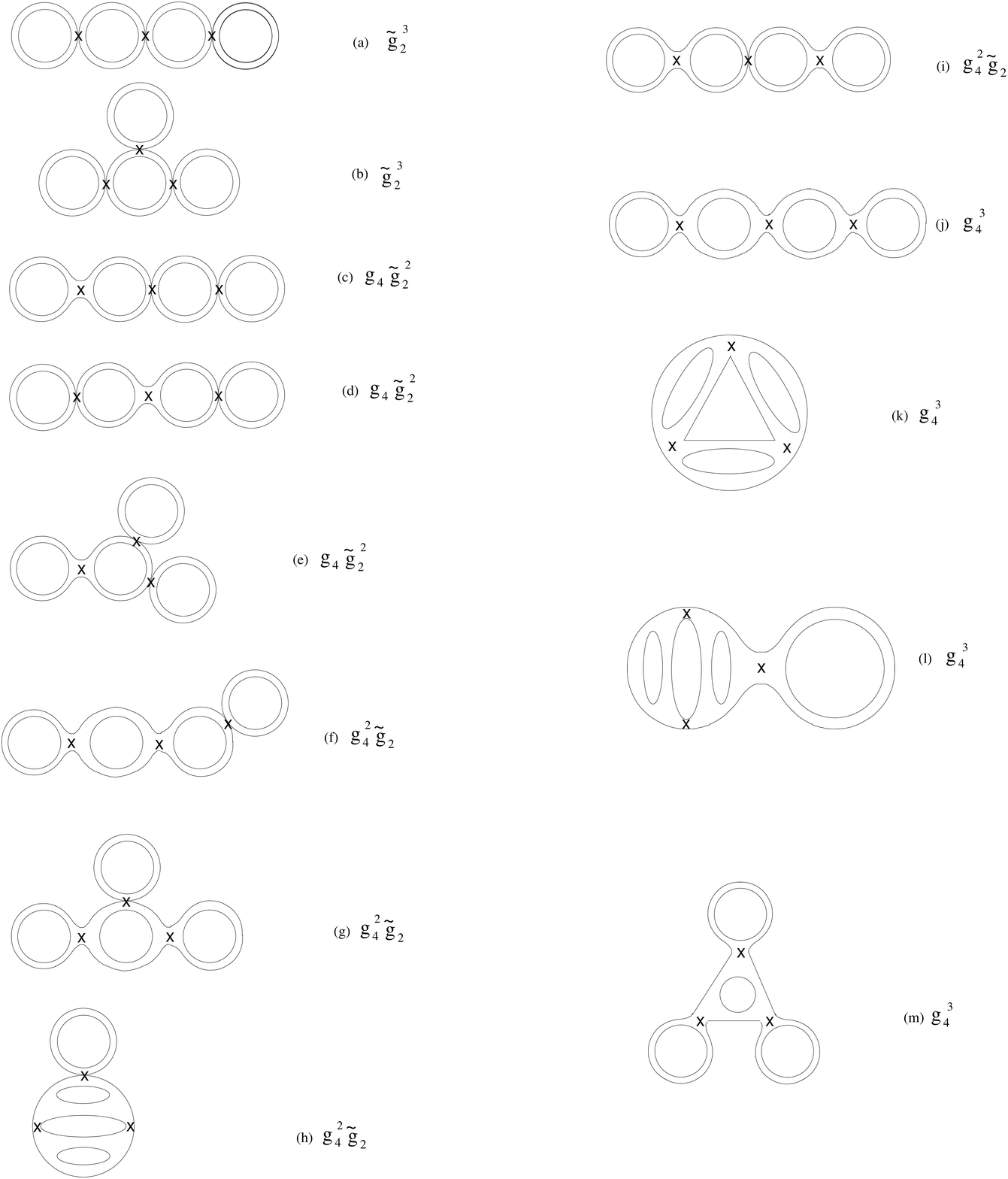}}
\end{center}
\caption{All four-loop primitive and pasting diagrams. The
vertices have been marked with a cross.} \label{g2-3}
\end{figure}

Finally, the third order diagrams are drawn in \fref{g2-3}. The
combinatorics are tabulated in \eref{four}.

Here the demonstrative example is diagram (b), which is composed
of pasting four diagrams, each with $h=2$, thus $h(G) = 4 \times 2
= 8$ and $C(G) = 2^4 = 16$. For $\CF(G)$, first we have a factor
$\frac{1}{3!}$ from the exponential. Next we have contractions of
the form $\Tr(\overbracket{\Phi \Phi})^3 \Tr(\overbracket{\Phi
\overbracket{\Phi) \Tr(\Phi} \; \overbracket{\Phi) \Tr(\Phi}
\Phi})$; there are $2^3 \times 4 \times 2$ ways of doing this.
Thus altogether we have $\CF(G) = \frac{32}{3}\widetilde{g}_2^3$
for this diagram.

In summary:
\beq\label{four}\begin{tabular}{|c|c|c|c|c|c|c|c|}\hline
    diagram  &  $(a)$  &  $(b)$ & $(c)$ & $(d)$ & $(e)$ & $(f)$ &
    $(g)$
\\ \hline
$h$ &  $8$ & $8$ & $7$ & $7$ & $7$ & $6$ & $6$  \\
\hline
$C(G)$ &  $16$ & $16$ & $12$ & $12$ & $12$ & $8$ & $8$  \\
\hline ${\cal F}(G)$ &  $16\widetilde{g}_2^3$ &
$\frac{32}{3}\widetilde{g}_2^3$ &
    $64\widetilde{g}_2^2g_4$ &
$32\widetilde{g}_2^2g_4$ & $64\widetilde{g}_2^2g_4$ &
$128\widetilde{g}_2g_4^2$ &
    $128\widetilde{g}_2g_4^2$
\\
\hline \hline diagram & $(h)$ & $(i)$ & $(j)$ & $(k)$ & $(l)$ &
$(m)$ &
\\ \hline $h$ & $6$ & $6$ & $5$ & $5$ & $5$ & $5$ & \\ \hline
$C(G)$ & $8$ & $9$ & $5$ & $5$ & $5$ & $5$ & \\ \hline ${\cal
F}(G)$ & $32\widetilde{g}_2g_4^2$ & $64\widetilde{g}_2g_4^2$ &
$128g_4^3$
& $\frac{32}{3}g_4^3$ & $64g_4^3$ & $\frac{256}{3}g_4^3$ &\\
\hline
\end{tabular}
\eeq

\subsubsection{Obtaining the Effective Action}
Now to the highlight of our calculation.
 From tables (\ref{two}), (\ref{three}), and (\ref{four})
we can readily compute the effective glueball superpotential and
free energy. We do so by summing the factors, with the appropriate
powers for $S$, in accordance with (\ref{effact},\ref{free0}).

We obtain, up to four-loop order,
\begin{eqnarray} \label{free1}
{\cal F}_0
&=& \sum_{G=all~diagrams}{\cal F}(G)S^{h(G)} \nonumber \\
&=&(2g_4+\widetilde{g}_2S)S^3-2(9g_4^2+8g_4\widetilde{g}_2S+2\widetilde{g}_2^2S^2)S^4
\nonumber \\
&+&\frac{16}{3}(54g_4^3+66g_4^2\widetilde{g}_2S+30g_4\widetilde{g}_2^2
S^2+5\widetilde{g}_2^3S^3)S^5+\cdots,
\end{eqnarray}
and subsequently,
\begin{eqnarray}
\label{glue1} W_{eff} &=& -NS(\log
(S/\Lambda^2)-1)+\sum_{G=all~diagrams}C(G){\cal
F}(G)N^{h(G)-l(G)}S^{l(G)} \nonumber \\
&=&-NS(\log (S/\Lambda^2)-1)+(6g_4+4\widetilde{g}_2N)NS^2-
(72g_4^2+96g_4\widetilde{g}_2N+32\widetilde{g}_2^2N^2)NS^3 \nonumber \\
&& + \frac{20}{3}(6g_4+4\widetilde{g}_2N)^3NS^4+\cdots.
\end{eqnarray}

In \cite{Balasubramanian} we check the free energy (\ref{free1})
are reproduced by the large $M$ limit of the double trace matrix
model (\ref{DTMM}). The double trace matrix model can be solved by
a mean field saddle point method \cite{Double}, and sum up exactly
the same ``planar pasted diagrams'' that we described above and
give the free energy defined by (\ref{free1}). Furthermore, we
show that the glueball superpotential (\ref{glue1}) is reproduced
by the analysis based on the Seiberg-Witten curve as in the single
trace case. However, unlike the single-trace case, the Matrix
model will not reproduce the combinatorial factors $C(G)$
appearing in (\ref{effact}). This subtlety can be thought of as
arising from the question of how to correctly identify the
glueball in the matrix model, and can be seen more clearly when
one linearize the double trace by an auxialiary singlet field.
Here we will not explore these points further. Interested readers
can consult \cite{Balasubramanian} for more details.

\section{Phases of ${\cal N}=1$ supersymmetric gauge theories}

Traditionally, there are two senses in which two quantum field
theories can be ``dual" to each other.   On the one hand, two
theories can be {\it equivalent} to each other in the sense that
all the correlation functions of one can be computed from the
other (and vice versa) by a suitable identification of dual
variables. Examples include the electric-magnetic duality of the
$\CN = 4$ super Yang-Mills theory, or the AdS/CFT correspondence.
Another kind of duality, described by Seiberg for $\CN = 1$
supersymmetric field theories occurs when two different
microscropic theories have identical macroscopic (or infra-red)
dynamics \cite{Seiberg:1994pq}.   For example, an $\CN = 1$
supersymmetric $SU(N_c)$ gauge theory with $3N_c > N_f > N_c+1$
fundamental flavours has the same long distance physics as an
$SU(N_f - N_c)$ gauge theory with $N_f$ flavours. Motivated by the
development of Dijkgraaf-Vafa conjecture , Cachazo, Seiberg and
Witten have proposed another notion of duality in certain $\CN =
1$ supersymmetric gauge theories \cite{CSW1}. It is found in
\cite{CSW1} that various vacua can be smoothly connected in the
phase diagrams. This is similar to the famous M-theory phase
diagrams, and one may hope to get more understanding of the
M-theory picture by studying these models. In this Section we
first review the case of pure gauge theory in \cite{CSW1}, then we
consider the case of theory with matter in the (anti)fundamental
representation \cite{CSW2, BFHN}. We will follow the approach in
\cite{BFHN}. For related observations and further generalization
other gauge groups see \cite{ferrari, Ann}.

\subsection{Phases of pure gauge theories}
We consider our familiar example, the pure $\CN =2$ $U(N)$ gauge
theory with one $\CN =2$ vector multiplet, and break the $\CN =2$
supersymmetry to $\CN=1$ by a tree level superpotential of the
adjoint chiral superfield $W(\Phi)$. For simplicity we consider a
cubic tree level superpotential. Classically, the gauge group is
broken to $U(N)\rightarrow U(N_1)\times U(N_2)$. The question
addressed by Cachazo et al \cite{CSW1}, is whether two different
classical limits $U(N)\rightarrow U(N_1)\times U(N_2)$ and
$U(N)\rightarrow U(\tilde{N}_1)\times U(\tilde{N}_2)$ can be
smoothly connected in the strongly coupled quantum mechanical
regime. It is found that such an interpolation is indeed possible.
Let us briefly review the results in \cite{CSW1}.

Quantum mechanically, once we turn on a cubic tree level cubic
superpotential, the points in the Coulomb moduli space that are
not lifted by the tree level superpotential are those points where
at least $N-2$ magnetic monopoles become massless and condense.
This is characterized by at least $N-2$ double roots in the
Seiberg-Witten curve. \footnote{The general solution to the
factorization problem of $U(N)$ gauge group is not known except at
the confining vacuum where $N-1$ monopoles condense (See
\cite{Janik} for results on some other cases.) . In \cite{CSW1}
some specific cases, namely from $U(2)$ to $U(6)$ gauge group, are
studied since the factorization can be solved explicitly.} Thus
the quantum moduli space is a two-dimensional sub-space of the
original Coulomb moduli space. Minimizing the tree level
superpotential on this sub-space we can find the discrete vacua of
the theory.  We can also reverse the problem by fixing a point in
the sub-space and solve for the cubic tree level superpotential
that produces vacuum at this given point. When the classical limit
$\Lambda\rightarrow 0$ is taken, we can see how the gauge group is
broken. It was found in \cite{CSW1} that the same branch of
solutions to the Seiberg-Witten curve factorization can have
different classical limits by taking different limits of the cubic
tree level superpotential.

In low energy there are two $U(1)$'s. The $SU(N_1)$ and $SU(N_2)$
confine, and give $N_1$ and $N_2$ discrete vacua according to
Witten index. So there are a total of $N_1N_2$ discrete vacua in
low energy. How can we predict which vacua are in the same branch
of the Seiberg-Witten curve factorization? In \cite{CSW1} an order
parameter, the {\it confinement index}, is found. Vacua that have
different confinement index can not be in the same branch, while
vacua that have the same confinement index may or may not be in
the same branch.

Let us discuss briefly the confinement index. Suppose $W$ is a
Wilson loop in the fundamental representation of the gauge group
$SU(N)$. To see whether a theory confines, we put in $r$-tensor
product of the Wilson loop $W^r$, and see whether it has an area
law. If the representation of the Wilson loop contains gauge
singlet, then it has no area law and is unconfined ; otherwise it
will have an area law and is said to be confined. It is obvious
that if two Wilson loops $W_1$ and $W_2$ are unconfined, then the
tensor product $W_1W_2$ is also unconfined. The confinement index
is defined as the smallest positive integer $t$ that the Wilson
loop $W^t$ that is unconfined. According to this definition, if
$t=1$, we say the vacuum is unconfined; if $t>1$, we say the
vacuum is confined. In pure gauge theory, the Wilson loop can be
combined with the center of the gauge group that represents the
gluons to make electric screening and 't Hooft loop to make
magnetic screening. For cubic tree level superpotentail, the gauge
group is broken to $U(N)\rightarrow U(N_1)\times U(N_2)$. There
are $N_1N_2$ discrete vacua that can be parameterized by a pair of
integers $(r_1, r_2)$, with $1\leq r_1\leq N_1$ and $1\leq r_2\leq
N_2$.  The electric screening makes $W^{N_1}$ and $W^{N_2}$
unconfined, while the magnetic screening makes $W^{r_1-r_2}$
unconfined. Taking into electric and magnetic screening it is
found in \cite{CSW1} that the confinement index should be the
greatest common divisor (GCD) of $N_1$, $N_2$, $r_1-r_2$.

\subsection{Phases of gauge theories with flavors}
It is interesting to understand how these above ideas extend to
the case where fundamental matter is included.  In particular, how
does Seiberg duality of $\CN = 1$ theories fit into the story of
\cite{CSW1}? Accordingly, in \cite{BFHN} we study $\CN=1$ $U(N)$
theory with an adjoint chiral multiplet $\Phi$, and $N_f$
fundamental and anti-fundamental chiral multiplets $\CQ_i$ and
$\tilde{\CQ^i}$ ($i =1 \dots N_f$). We also add a superpotential
$W= \Tr(W(\Phi))+ \sqrt{2} \widetilde{\CQ}_i \Phi \CQ^i+
\sqrt{2}m_i \widetilde{\CQ}_i  \CQ^i $. This model is an $\CN=2$
$U(N)$ gauge theory with $N_f$ fundamental hypermultiplets which
is deformed to $\CN=1$ by addition of the tree level
superpotential $\Tr W(\Phi)$. \footnote{The coupling $\sqrt{2}
\widetilde{\CQ}_i \Phi \CQ^i$ is required by $\CN =2$
supersymmetry, and the mass terms $\sqrt{2}m_i \widetilde{\CQ}_i
\CQ^i$ is allowed for $\CN =2$ supersymmetry.}

First we can consider confinement in this case. A Wilson loop $W$
in any representation can always combine with fundamental matter
to make gauge singlet, thus there is no area law. This is known as
``maximal screening'' or ``perfect screening''. It means there is
no true confinement once we have light fundamental matter. For
example, in QCD if we try to pull out of a single light quark by
high energy electrons, the quark will produce a cloud of gluons
with energy density of the $\Lambda^4$. This cloud then pair
produces quarks and antiquarks. Thus if quarks have a mass much
less than $\Lambda$, it will be screened maximally and we will not
see single isolated quarks. The usual referring of ``quark
confinement'' in QCD really means maximal screening, and should
not be confused with the true color confinement we discuss here.

Since there is no confinement once we have (anti)fundamental
matter, the order parameter of confinement index used in pure
gauge theory to distinguish various phases is not immediately
applicable here. In \cite{CSW2}, a general tree level
superpotential for the chiral superfield and mass matrix for the
quarks are considered. In this case the quarks can be thought of
as massive in classical limit $\Lambda\rightarrow 0$ and can be
integrated out. In low energy we are left with a few $U(1)$
factors. It is found that vacua with the same number of $U(1)$
factors can be smoothly connected to each other by changing the
superpotential and mass matrix.

In the following we summarize the results in \cite{BFHN} where
another situation is considered. The interested readers can
consult \cite{BFHN} for more details. We consider the case that
the quarks have the same mass $m$, and the cubic tree level always
has a minimal point at $-m$. In this case there will be massless
quarks in low energy even in classical limit $\Lambda\rightarrow
0$. Thus the moduli space of vacua can have different branches
with the same number of $U(1)$'s in low energy.

Classically, the D-term equation requires adjoint scalar to be
diagonal. Suppose the two roots of the $W'(\Phi)$ is $-m$ and
$-\alpha$, then diagonal elements of $\Phi$ have to be either $-m$
or $-\alpha$. So classically the gauge group is broken to two
pieces $U(N)\rightarrow U(N_1)\times U(N_2)$, with $-m$ as $N_1$
diagonal elements and $-\alpha$ as $N_2$ diagonal elements. The
$N_f$ fundamental flavors are massless in $U(N_1)$ piece, but
become massive in the $U(N_2)$ piece and can be integrated out in
low energy . So classically we have two decoupled pieces in low
energy: $U(N_1)$ with $N_f$ massless flavors and pure $U(N_2)$
theory. We denote such a classical vacua as $U(N)\rightarrow
\widehat{U(N_1)}\times U(N_2)$. We study whether we can go from
the same branch of strongly coupled regime to different classical
limits by taking different classical limits of the cubic tree
level suprpotential.

The classical picture is modified significantly in the quantum
theory. We can analyze the $\CN=1$ quantum theory in two ways by
strong coupling analysis and weak coupling analysis. The strong
coupling analysis is to consider this theory as a small
perturbation of a strongly coupled $\CN=2$ gauge theory with tree
level superpotential $W=0$. The moduli space of $\CN=2$ SQCD is
analyzed in details in \cite{Argyres:1996eh}. The Seiberg-Witten
curve encodes the low energy quantum dynamics of the $\CN=2$
theory on the Coulomb branch. Turning on a tree-level
superpotential lifts almost all points on the Coulomb branch,
except points in the Higgs branch roots where a certain number of
mutually local monopoles and become massless. Furthermore, on this
sub-manifold of the Coulomb branch, the tree-level superpotential
has to be minimized to find the $\CN=1$ vacua.  By varying the
parameters of the superpotential, these $\CN=1$ vacua can be moved
around on the Coulomb branch of $\CN = 2$ moduli space. In
particular, special corners in the parameter space will place
these vacua in regions where the $U(N)$ gauge symmetry breaking
scale is much greater than $\Lambda$, the $\CN=2$ dynamical scale.
Thus the gauge group is higgsed. These regions can then be
analyzed by weak coupling analysis.

In weak coupling analysis the description in terms of
non-interacting product gauge group factors is valid. More
concretely, we can integrate out (in each gauge group factor) the
massive adjoint chiral field, $\Phi$, which has a mass well above
$\Lambda$. The corresponding $\CN=1$ theory thus obtained, valid
below scales of order $\mu$ can be analyzed in various group
factors separately. In each factor, it will become strongly
coupled in the infra-red and will have vacua, details of which
will depend on the number of flavors charged under the group
factor. We can interpolate smoothly between vacua in the weak and
strong coupling regions by holomorphic variation of the parameters
in the superpotential $W(\Phi)$ because the theory has $\CN = 1$
supersymemtry. What is perhaps most interesting is the fact that
different weak coupling regions with different microscopic physics
can be reached smoothly from the same strongly coupled point.

\subsubsection{ An example}

The above discussion seems a little abstract. Let us discuss an
explicit example with a rather rich phase structure. We will
discuss the case of a cubic tree level superpotential
 \be \label{cubic-W} W(\Phi)= u_3+ (m+\alpha) u_2+
m \alpha u_1 \ee so that $W'(x)=(x+m)(x+\alpha)$.   In this case
the gauge group will break into two  factors in the semiclassical
limit $\Lambda\rightarrow 0$.  Suppose the Seiberg-Witten curve is
$y^2=P_{N_c}(x)^2-4\Lambda^{2N_c-N_f}(x+m)^{N_f}$. Then there are
various $r$'th branches in which  the SW  curve factorizes as
$P_{N_c}(x,u_k)=(x+m)^rP_{N_c-r}(x)$. In the $r>0$ branches we
have massless flavors quantumly.

We consider the case of $U(4)$ gauge theory with $N_f=4$ flavors
in $r=0$ non-baryonic branch. We have 2 double roots in the
Seiberg-Witten curve. Suppose the Seiberg-Witten curve factorize
as
\begin{equation}
y^2=P_4(x)^2-4\Lambda^4(x+m)^4=F_4(x)H_2(x)^2
\end{equation}
The 2 double roots can have various distributions in the
Seiberg-Witten curve. We discuss these various branches and
extrapolate to weak coupling regimes.

\paragraph{\underline{Non-baryonic $r=0$ branch in $(2,0)/(0,2)$
distributions:}} On this branch with the $(0,2)$ and $(2,0)$
distribution of roots, we have $P_4(x)-2 \eta \Lambda^2
(x+m)^2=(x+a_1)^2(x+a_2)^2$, ($\eta=\pm1$), then
\begin{eqnarray}
F_4(x) & = & (x+a_1)^2(x+a_2)^2+4\eta\Lambda^2 (x+m)^2\nonumber
\\& = &(x^2+(a_1+a_2)x+a_1a_2+2\eta\Lambda^2)^2+{\cal O}(x)
\end{eqnarray}
We can find the tree level superpotential that gives rise to the
vacuum by using a theorem in \cite{CIV} and generalized to the
case with flavors in \cite{BFHN}.\footnote{There is a subtle of of
additional contributions from the flavor when $N_f\geq 2N_c-2$.
See \cite{BFHN}. Here we will not need to worry about this for
$N_f=N_c=4$.} It is found there the tree level superpotential is
determined by
\begin{eqnarray}
F_{2n}(x)  = W^{\prime}(x)^2+{\cal O}(x^n)
\end{eqnarray}
So here we find
\begin{eqnarray}
W^{\prime}(x)&=& x^2+(a_1+a_2)x+a_1a_2+2\eta\Lambda^2  \nonumber \\
&=&(x+m)(x+\alpha)
\end{eqnarray}
There are two solutions for $m$ : $m=\frac{1}{2}(a_1+a_2\pm
\sqrt{(a_1-a_2)^2-8\eta\Lambda^2})$. In semiclassical limit, each
solution leads to the breaking pattern $U(4)\rightarrow
\widehat{U(2)}\times U(2)$.

To count vacua, we fix $m,\alpha$ and solve $a_1,a_2$: \be
m+\alpha= a_1+a_2,~~~~~~m\alpha=a_1a_2+2\eta \Lambda^2 \ee Since
$a_1$ and $a_2$ are symmetric, we have only
one solution for each $\eta=\pm 1$.

\paragraph{\underline{Non-baryonic $r=0$ branch in the $(1,1)$
distribution:}} By shifting $x$ by a constant, we can arrange the
two double roots to be at $x=a$ and  $x=-a$. The general case can
be recovered by shifting  by a constant $b$: $x\rightarrow x+b$,
$m\rightarrow m-b$. The factorization we need is
\begin{eqnarray}
P_4(x)+2\Lambda^2(x+m)^2&=
&(x-a)^2\Bigl((x+a)^2+\frac{\Lambda^2}{a^3}((m^2-a^2)x+2ma(m-a))\Bigr)
\nonumber \\
P_4(x)-2\Lambda^2(x+m)^2 &=
&(x+a)^2\Bigl((x-a)^2+\frac{\Lambda^2}{a^3}((m^2-a^2)x+2ma(-m-a))\Bigr)
\nonumber
\end{eqnarray}
and we find
\begin{eqnarray}
F_4(x) &=&
\Bigl((x+a)^2+\frac{\Lambda^2}{a^3}((m^2-a^2)x+2ma(m-a))\Bigr)
\Bigl((x-a)^2 \nonumber \\ && +\frac{\Lambda^2}{a^3}((m^2-a^2)x+2ma(-m-a))\Bigr) \nonumber \\
&=&
\Bigl(x^2+\frac{(m^2-a^2)\Lambda^2}{a^3}x-a^2(1+\frac{2m\Lambda^2}{a^3})\Bigr)^2+{\cal
O}(x)
\end{eqnarray}
So $W^{\prime}(x)$ is given by
\begin{equation} \label{W44}
W^{\prime}(x)=x^2+\frac{(m^2-a^2)\Lambda^2}{a^3}x-a^2(1+\frac{2m\Lambda^2}{a^3})
\end{equation}
$m$ has to satisfy the following equation:
 \begin{equation} \label{17}
m^3-\frac{a^3}{\Lambda^2}m^2+a^2m+\frac{a^5}{\Lambda^2}=0
\end{equation}
This has three solutions which we denote by $m_1$, $m_2$ and
$m_3$.  Notice that
 (\ref{17}) has the symmetry $m\rightarrow -m$ and $a\rightarrow -a$.
We now consider the different semiclassical limits.
\begin{enumerate}
\item $\Lambda\rightarrow 0$ and $a$ fixed and finite.
Then  from (\ref{17}),  find two solutions $m_1=a$ or $m_2=-a$
when  the second and fourth term in (\ref{17}) dominates, and the
third solution $m_3\sim \frac{a^3}{\Lambda^2}$ blows up in the
$\Lambda \rightarrow 0$ limit . We will ignore this solution. For
$m_1$ and $m_2$ we obtain $P_4(x)\rightarrow (x+a)^2(x-a)^2$ which
implies that the breaking pattern is  $U(4)\rightarrow
\widehat{U(2)}\times U(2)$.

\item  $\Lambda\rightarrow 0$, and $a\sim \Lambda^p$ with
$0<p\leq 1$. The asymptotic behavior of $m_{1,2,3}$ can be read
off from (\ref{17}). We find the solutions  $m_{1,2}\sim \pm a$
when second and fourth term in (\ref{17}) dominate and $m_3\sim
\frac{a^3}{\Lambda^2}$ when the first and second terms dominate.
Thus for $m_{1,2}$ we obtain $P_4(x)\rightarrow x^4$, which yields
a singular limit since there is only one gauge group factor. For
$m_3$ if $0<p<\frac{2}{3}$ the solution blows up and should be
discarded. If $\frac{2}{3}<p\leq 1$ we obtain $P_4(x)\rightarrow
x^4$, which is a singular limit. For $p=\frac{2}{3}$ we obtain a
smooth semiclassical limit. In this case
$m_3=\frac{a^3}{\Lambda^2}$, and we obtain $P_4(x)\rightarrow
x^3(x+m_3)$. Hence the breaking pattern is $U(4)\rightarrow
\widehat{U(1)}\times U(3)$.

\item  $\Lambda\rightarrow 0$, and $a\sim \Lambda^p$ with $p>1$.
The asymptotic behavior of $m_{1,2,3}$ can be again read off from
(\ref{17}). We find $m_{1,2}\sim \pm ia$ and $m_3\sim
-\frac{a^3}{\Lambda^2}$ For $m_{1,2}$ we get $P_4(x)\rightarrow
x^3(x-\frac{2\Lambda^2}{a})$, which is a singular limit unless
$p=2$, in which case the gauge group breaks into $U(4)\rightarrow
U(1)\times \widehat{U(3)}$. For $m_{3}$ we have $P_4(x)\rightarrow
x^3(x-\frac{\Lambda^2}{a})$, which is a singular limit unless
$p=2$, in which case the gauge group breaks into $U(4)\rightarrow
U(1)\times \widehat{U(3)}$.
\end{enumerate}
Something interesting has happened here. To determine classical
limits, we take a point on the factorization locus (parameterized
by $a$ and $m$) and then determine the superpotential which would
yield that point as its minimum. That leads to the consistency
condition (\ref{17}), which has three solutions $m_1, m_2$ and
$m_3$.  The three solutions lead to different branches and
different interpolation patterns: $m_{1,2}$ smoothly interpolate
between $U(4)\rightarrow \widehat{U(2)}\times U(2)$  and
$U(4)\rightarrow U(1)\times \widehat{U(3)}$.   $m_3$ smoothly
interpolates between
 $U(4)\rightarrow \widehat{U(1)}\times U(3)$ and
$U(4)\rightarrow U(1)\times \widehat{U(3)}$. We do not encounter
this phenomenon in our other examples. For example, in the $U(4)$
theory with two flavors, the three classical limits smoothly
inteplate between each other with the same choice of $m$. Notice
that for the $U(4)$ theory with four flavors, the limit
$\widehat{U(2)} \times U(2)$ is not smoothly connected with
$\widehat{U(1)} \times U(3)$.

To count the number of vacua, we need to first fix $m$ and
$\alpha$ and find the number of solutions for $a$ and the shifted
constant $b$. From (\ref{W44})  we obtain the equations
\begin{eqnarray}
\alpha-b+m-b&=& \frac{((m-b)^2-a^2)\Lambda^2}{a^3} \nonumber \\
(\alpha-b)(m-b)&=&-a^2(1+\frac{2(m-b)\Lambda^2}{a^3})
\end{eqnarray}
Eliminating
$b=\frac{a^4(m+\alpha)+2a^3\Lambda^2+2m\Lambda^4-ma\Lambda^2(m-\alpha)}
{2a^4+2\Lambda^4-a\Lambda^2(m-\alpha)}$, we obtain the following
equation for $a$:
\begin{eqnarray}
&&\hspace{-0.5in}4a^8-(m-\alpha)^2a^6+4\Lambda^4a^4+\Lambda^2(m-\alpha)^3a^3
-5\Lambda^4(m-\alpha)^2a^2 \nonumber \\
& &
~~~~~~~~~~~~~~~~~~~~~~~~~~~~~~~~~~~~~+8\Lambda^6(m-\alpha)a-4\Lambda^8=0
\label{a44}
\end{eqnarray}
Thus $a$ has eight solution.  We keep $m$ and $\alpha$ fixed and
find the asymptotic behavior of the eight roots when
$\Lambda\rightarrow 0$. First by setting $\Lambda=0$ in
(\ref{a44}) we find two roots at $a\sim\pm \frac{m-\alpha}{2}$ and
six others $a\rightarrow 0$. The two non-zero solutions (which
correspond to $m_{1,2}$ in case 1 above) lead to a semiclassical
limit $U(4)\rightarrow U(2)\times \widehat{U(2)}$ . We now analyze
the six solutions for $a$ which vanish in the $\Lambda\rightarrow
0 $ limit. Assume $a\sim \Lambda^p$.  Then from (\ref{a44}) we
find that $a\sim \Lambda^{\frac{2}{3}}$ or $a\sim \Lambda^2$. For
$a\sim \Lambda^{\frac{2}{3}}$ the dominant terms in (\ref{a44})
give $-(m-\alpha)^2a^6+\Lambda^2(m-\alpha)^3a^3=0$ and  we find
three roots with $a^3\sim (m-\alpha)\Lambda^2$. These solutions
(which correspond to  $m_3$ in case (2) above) lead to the
semiclassical limit $U(4)\rightarrow \widehat{U(1)}\times U(3)$.
 For $a\sim \Lambda^2$ the dominant terms (\ref{a44}) give
\begin{equation}
\Lambda^2(m-\alpha)^3a^3-5\Lambda^4(m-\alpha)^2a^2+8\Lambda^6(m-\alpha)a-4\Lambda^8=0
\end{equation}
This implies that
$((m-\alpha)\frac{a}{\Lambda^2}-1)((m-\alpha)\frac{a}{\Lambda^2}-2)^2=0$.
We obtain one solution (corresponding to $m_3$ in case 3 above)
with  $a\sim \frac{\Lambda^2}{m-\alpha}$ and two solutions
(corresponding to $m_{1,2}$ in case 3) with $a\sim
\frac{2\Lambda^2}{m-\alpha}$. These three solutions lead to the
semiclassical limit  $U(4)\rightarrow U(1)\times \widehat{U(3)}$.

We can match the number of these strong coupling vacua with the
number obtained in the weak coupling region. In $r=0$ branch,
$\widehat{U(2)}\times U(2)$ has four vacua where two from
confining $U(2)$ factor and two from the $U(2)$ factor with four
flavors. Two of these four vacua are in $(2,0)/(0,2)$ distribution
and two, $(1,1)$ distribution. $\widehat{U(1)}\times U(3)$ has
three vacua in the $(1,1)$ distribution while
 $U(1)\times \widehat{U(3)}$ has three vacua in  $(1,1)$ distribution.

\subsubsection{ What are the order parameters?}

On a given branch, the vacua are all in the same phase. What order
parameters (or indices) distinguish between vacua on different
branches? An obvious characterization of a branch is the global
symmetry group (which in this case will be a flavor symmetry). One
such index is $r$, which characterizes the meson VEV in various
vacua in the weak coupling region, and labels the root of the
$r$-th Higgs branch in the strong coupling region. The global
symmetry group must be the same on each branch. On the $r$-th
branch, the global flavor symmetry is broken as
$SU(N_f)\rightarrow SU(N_f-2r)\times U(1)^r$. Another fact which
distinguishes the different branches is if they are `baryonic' or
`non-baryonic'. These two types of branches differ from each other
in the number of condensed monopoles, and hence have different
number of $U(1)$s at low energies. Furthermore, a finer
distinction is possible on the non-baryonic $r$-th branches which
arise when $N_f>N_c$ and $r < N_f - N_c$. In these cases, there
are two types of non-baryonic branch.  In the strong coupling
region one arises from  a generic non-baryonic root, while the
other is special case arising when the non-baryonic root lies
inside the  baryonic root. These two kinds of strong coupling
vacua match up with two types weak coupling vacua in which the
meson matrix is degenerate and non-degenerate. However, the
various indices that are available to us are not refined enough to
provide sufficient conditions which determine the phase structure
completely. It would be interesting to find more order parameters
that can determine the phase structure.
    %insert chapter 1 file name

\chapter{Topics in the AdS/CFT Correspondence}  \label{chapter2}%continue adding chapter titles
\section{Introduction: Classical Supergravity on $AdS_5\times S^5$ and $\CN =4$ Super Yang-Mills}
It has been speculated long time ago by 't Hooft \cite{'tHooft}
that when the rank of the gauge group $N$ becomes large, the gauge
dynamics can be described by classical string theory. The original
motivation is to study strongly coupled gauge theory. When $N$ is
large, Feymann diagrams can be organized in the $1/N$ expansion by
the topology of the diagrams. Since string theory is also well
known to have perturbative expansion in topology of the Riemann
surface that represents string worldsheet, it was speculated by t'
Hooft the two descriptions are equivalent. In particular, when $N$
is large only planar diagrams contribute, the gauge theory is
described by genus zero string theory, i.e. free string theory.

Maldacena made a remarkable conjecture relate type IIB string
theory on $AdS_5\times S^5$ to $\CN=4$ $SU(N)$ Super Yang-Mills
theory \cite{Maldacena}. The conjecture is non-trivial since it
relate theories with gravity, such as string theory, to a field
theory with no gravity at all. Some problems that are very hard to
solve on one side of the duality may be easy on the other side.
For example, we can extract many information in strongly coupled
$\CN=4$ Yang-Mills theory using classical supergravity on
$AdS_5\times S^5$. In this duality the $\CN=4$ super Yang-Mills
lives on a 4-dimensional space-time at the boundary of $AdS_5$.
This is an example that realize the ``holography principle'' that
comes from the study of black hole information paradox.

The motivation of the conjecture comes from the study of near
horizon geometry of extremal black branes. Here we will not go
into the detail of the origins of the conjecture, but instead
directly states the conjecture and explains various regime that it
can be tested. The $AdS_5\times S^5$ is a maximally supersymmetric
background that preserves all 32 supercharges of the type IIB
superstring theory. It is the near horizon geometry of D3-brane.
The metric in Poincare coordinate is
\begin{equation}
ds^2=\frac{r^2}{R^2}(-dt^2+dx_1^2+dx_2^2+dx_3^2)+\frac{R^2}{r^2}dr^2+R^2d\Omega_5^2,
\end{equation}
Here $R$ is the radius of the $AdS_5$ and $S^5$. The solution also
have Ramond-Ramond five form flux on the $S^5$. The flux number
$N$ is quantized
\begin{equation} \label{fl}
\int_{S^5}F_5=N
\end{equation}
The radius is related to the string scale by the relation
\begin{equation}
R^4=4\pi g_sNl_s^4
\end{equation}

The $\CN=4$ super Yang-Mills theory in four dimensions  contains
one $\CN=1$ gauge multiplet and three $\CN=1$ adjoint chiral
multiplets $\Phi_1, \Phi_2, \Phi_3$, with a tree level
superpotential
\begin{equation}
W_{tree}=\Tr(\Phi_1[\Phi_2,\Phi_3])
\end{equation}
This tree level superpotential is required to preserve $\CN=4$
supersymmetry. The $\CN=4 $ super Yang-Mills theory is known to
have vanishing beta function and is a conformally invariant theory
(CFT). The R-symmetry of theory is $SU(4)$. Guage invariant
operators can be organized in representations of the $SU(4)$
R-symmetry and is said to carry various R-charges.

The AdS/CFT correspondence states that Type IIB string theory on
$AdS_5\times S^5$ background is equivalent to $\CN=4$ $SU(N)$
supersymmetric Yang-Mills theory in $1+3$ dimension. The flux
number in (\ref{fl}) is identified with the rank $N$ in the
$SU(N)$ gauge group. The Yang-Mills coupling constant is related
to the string coupling by the relation
\begin{equation}
\tau=\frac{4\pi
i}{g_{YM}^2}+\frac{\theta}{2\pi}=\frac{i}{g_s}+\frac{\chi}{2\pi}
\end{equation}
Here $\chi$ is expectation value of the Ramond-Ramond scalar.

In the strongest form, the conjecture would hold for all finite
$N$ and $g_s$. Thus the $\CN=4$ Yang-Mills provided a
non-perturbative definition of type IIB string theory on
$AdS_5\times S^5$ since the string coupling constant could be any
values. However, at finite $N$ and $g_s$ this is not a duality per
say, since we do not have any other non-perturbative definition of
type IIB string theory on $AdS_5\times S^5$. On the other hand, if
we take $N$ to be large, then something very fantastic happens. A
miraculous  but not quite well understood ``duality transition''
occur at large $N$, and we could have a dual description of string
theory such as classical supergravity. This transition is in some
sense like a geometric transition \cite{GV}.

Let us discuss in what regime the string theory can be
approximated by classical supergravity. Firstly, if we take $N$ to
be large, then only planar diagrams contribute. We have argued
planar diagrams are genus zero diagrams. So in this case the gauge
theory corresponds to free string theory, with no string loop
effect.\footnote{From next section we will consider a sub-sector,
known as BMN sector, of the gauge theory with very large R-charge,
then non-planar diagrams can contribute although $N$ is large.}
Secondly, we consider $\alpha^{\prime}=l_s^2$ correction. The mass
scale of stringy excitation is of $\frac{1}{l_s}$, where $l_s$ is
the string length scale.  On the other hand, the typical energy
scale we encounter in the $AdS_5\times S^5$ background is
$\frac{1}{R}$. So when $R$ is much large $l_s$, we will only see
zero modes of the full string theory, and the string theory on
$AdS_5\times S^5$ can be truncated to its supergravity zero modes.
Thus the condition that the string theory can be approximated by
classical supergravity is the following
\begin{eqnarray}
N\gg 1,~~~~~~~~\frac{R}{l_s}\gg 1
\end{eqnarray}
Since $R^4=g_{YM}^2 N l_s^4$, we see that $\frac{R}{l_s}\gg 1$ is
same as $g_{YM}^2N\gg 1$. The constant $\lambda=g_{YM}^2N$ is
known as the 't Hooft coupling constant. In large $N$ limit, the
effective gauge coupling constant is  the 't Hooft coupling
constant $\lambda=g_{YM}^2N$ instead of $g_{YM}^2$. Thus we have a
strongly coupled gauge theory corresponding to the classical
supergarvity on $AdS_5\times S^5$.

In \cite{Gubser, Witten} a holographic map is proposed that
relates physical observable on both sides of the duality. It is
proposed that gauge invariant operators in $\CN=4$ Yang-Mills
theory are in one to one correspondence to fields in the string
theory side. Suppose an operator $\CO(\vec{x})$ of the CFT
corresponds to a field $\phi(\vec{x},z)$ in classical supergravity
on $AdS_5$. Here $\vec{x}$ is a four vector represent space-time
at the boundary, and $z$ is the radius direction of the $AdS_5$.
In Poincare coordinate the boundary is at $z=0$. It is proposed
that the partition function of string theory with a fixed boundary
field is the same as the generating function of the CFT with the
corresponding operator coupled to the boundary field as source,
namely
\begin{equation} \label{prop}
\CZ_{string}[\phi(\vec{x},z)|_{z=0}=\phi_0(\vec{x})]=\langle
e^{\int d^4x\phi_0(\vec{x})\CO(\vec{x})}\rangle_{CFT}
\end{equation}
We can then vary the generating function with respect to the
source $\phi_0(\vec{x})$ to computed correlation functions of
corresponding operators.  According to (\ref{prop}), the
correlation function can also be computed from the string theory
side by varying the supergravity action with respect to boundary
fields.

The AdS/CFT correspondence is a strong-weak duality. In the regime
that classical supergravity is valid, the gauge theory is strongly
coupled and the corresponding physical quantities can not be
computed by gauge pertuabation theory. How can we test the
duality? First, we can match the global symmetry of the two
theories, which does not change with the coupling. Both theories
are known to have a global symmetry group of $SU(2,2|4)$ whose
bosonic subgroup is $SO(4,2)\times SU(4)$. Also both theories are
believed to have a S-duality group symmetry $SL(2,Z)$ acting on
their coupling constant $\tau$.

Gauge invariant operators of the CFT fit in the representation the
global symmetry group $SU(2,2|4)$. There are operators that lie in
some short representations of the $SU(2,2|4)$. Operators in short
representations are annihilated by some combination supercharges,
so have less multiplicity of primary operators than those of long
representation. These operators are known as BPS or chiral
operaors. Their conformal dimension does not change with gauge
coupling constant, due to non-renormalization theorem in
supersymmetric gauge theory. Many correlation functions of BPS
operators has also been computed and shown to be not dependent on
the gauge coupling. We can compare these quantities to the
computation from classical supergravity side. Many successful
tests have been done in these cases. For review see \cite{Aharony,
Freedman}.

\section{Plane Waves and BMN Operators} \label{sec4}
Until recently, most tests of the AdS/CFT correspondence have been
restricted to the classical supergravity regime. Can we go beyond
classical supergravity to study stringy modes of theory, or even
interacting string theory? The difficulty is that we need $N$
large to have a duality transition. However, when $N$ is large the
anomalous dimension of non-BPS operators grow like
$(g_{YM}^2N)^{\frac{1}{4}}$, so is not subjected to perturbation
theory study when the classical supergravity is valid. Recently,
it is realized that a sub-sector of the $\CN=4$ can be non-BPS yet
have small anomalous dimension, due to the large R-charge of these
operators. In the string theory side, The $AdS_5\times S^5$
geometry goes to a Penrose limit and becomes a background known as
pp-wave, or plane wave. There are many excellent reviews on this
subject, see e.g. \cite{review}.

\subsection{The plane wave background}
First we explain the plane wave geometry. Long time ago, Penrose
pointed out that if we zoom in a null geodesics of any metric, we
can take a limit, known as Penrose limit, then we will obtain a
metric that has the form of a ``pp-wave'' \cite{Penrose}. Let us
follow the approach in \cite {BMN, Blau} and see how this works
for $AdS_5\times S^5$ background. It is convenient to write
$AdS_5\times S^5$ in global coordinate
\begin{equation}
ds^2=R^2(-\cosh^2\rho dt^2+d\rho^2+\sinh^2\rho
d\Omega_3^2+\cos^2\theta d\psi^2 +d\theta^2+\sin^2\theta
d\Omega^{\prime 2}_3)
\end{equation}
There is a null geodesics parameterized by $\rho=0$, $\theta=0$
and $\psi=t$. This trajectory satisfy $ds^2=0$ so it is a null
geodesics. We can take the Penrose limit by zooming in this
trajectory. We introduce the coordinate
$\tilde{x}^{\pm}=\frac{t\pm\psi}{2}$ and a parameter $\mu$, and
perform the scaling
\begin{eqnarray} \label{coor}
x^{+}=\frac{\tilde{x}^{+}}{\mu},~~~~x^{-}=\mu R^2\tilde{x}^{-}, \\
\nonumber
\rho=\frac{r}{R},~~~~\theta=\frac{y}{R},~~~~~~R\rightarrow \infty
\end{eqnarray}
In this limit we see the $AdS_5\times S^5$ becomes the following
metric
\begin{equation} \label{pp-wave}
ds^2=-4dx^{+}x^{-}-\mu^2(\vec{r}^{~2}+\vec{y}^{~2})(dx^{+})^2+d\vec{r}^{~2}+d\vec{y}^{~2}
\end{equation}
here the $\vec{r}$ and $\vec{y}$ parameterize points in $R^4$ from
$AdS_5$ and $S^5$. And the five form Ramond-Ramond flux become
\begin{equation}
F_{+1234}=F_{+5678}=const\times \mu
\end{equation}
The metric (\ref{pp-wave}) is a plane wave metric. This limit of
$AdS_5\times S^5$ is known as pp-wave (plane wave) limit, or BMN
limit. While string spectrum on $AdS_5\times S^5$ has not been
solved. It turns out the free string spectrum in plane wave
(\ref{pp-wave}) is solvable. The solvability is largely due to the
light cone gauge in the metric. Since the plane wave is a
Ramond-Ramond background, we have to use Green-Schwarz formalism
instead of NS-R formalism. Th free string spectrum is obtained in
\cite{Metsaev}. The light cone Hamiltonian spectrum is
\begin{equation} \label{HLC}
H_{lc}=-p_{+}=2p^{-}=\sum_{n=-\infty}^{+\infty}N_n\sqrt{\mu^2+\frac{n^2}{(\alpha^{\prime}p^{+})^2}}
\end{equation}
Here $n$ is the Fourier string mode. We use the notation of
\cite{BMN} where $n>0$ label left movers and $n<0$ label right
movers. $N_n$ denote the total occupation number of that bosonic
and fermionic mode with Fourier number $n$. The light cone momenta
is defined as
\begin{eqnarray} \label{light}
2p^{-}=-p_{+}=i\partial_{x^+} \\ \nonumber
2p^{+}=-p_{-}=i\partial_{x^-}
\end{eqnarray}

We now consider how we can compare the string spectrum to dual
$\CN=4$ gauge theory. The energy in global coordinate $AdS$ is
given by $E=i\partial_t$ and the angular momentum in $\psi$
direction is given by $J=-i\partial_{\psi}$. In the dual gauge
theory the energy will correspond to conformal dimension of an
operator. We denote the conformal dimension as $\Delta= E$. The
angular momentum in $\psi$ direction correspond to the R-charge of
the $U(1)$ factor in the $SU(4)\sim SO(6)$ R-symmetry group that
corresponds to the $\psi$ direction. Use the coordinate
transformation (\ref{coor}) we can write the light cone momenta
(\ref{light}) as
\begin{eqnarray} \label{cone}
&&2p^{-}=-p_{+}=i\partial_{x^+}=i\mu\partial_{\tilde{x}^+}=i\mu(\partial_t
+\partial_\psi)=\mu(\Delta-J) \nonumber \\
&&2p^{+}=-p_{-}=-i\partial_{x^-}=\frac{1}{\mu
R^2}i\partial_{tilde{x}^-}=\frac{1}{\mu R^2}i
(\partial_t-\partial_\psi)=\frac{\Delta+J}{\mu R^2}
\end{eqnarray}
We should keep the light cone momenta $p^{\pm}$ finite while scale
$R\rightarrow \infty$. Thus we must scale the energy and angular
momentum as $\Delta\sim J\sim R^2$ while keep $\Delta-J$ finite.
The second equation in (\ref{cone}) becomes
\begin{equation} \label{cone1}
p^{+}=\frac{J}{\mu R^2}
\end{equation}
From (\ref{HLC}) (\ref{cone}) (\ref{cone1}), and using the
relation $R^4=4\pi g_sN\alpha^{\prime 2}$ we find that the
contribution to $\Delta -J$ for an oscillator mode $n$ is given by
\begin{equation} \label{3.16}
\omega_n=(\Delta-J)_n=-\frac{p_{+}}{\mu}=\sqrt{1+\frac{n^2}{(\mu\alpha^{\prime}p^{+})^2}}=\sqrt{1+\frac{4\pi
g_sNn^2}{J^2}}
\end{equation}
This result is very important and will be compared to perturbative
gauge theory calculations. We see when $\mu\alpha^{\prime}p^{+}\gg
1 $, the string modes are very light and almost degenerate. This
corresponds to strings in a very curved RR background. On the
other hand, the opposite limit $\mu\alpha^{\prime}p^{+}\ll 1 $
corresponds to strings in nearly flat background. The comparison
to dual $\CN=4$ gauge theory will be made in the highly curved
limit. The essential point is that the large R-charge in
(\ref{3.16}) keep the anomalous dimension from growing in large
$N$. These stringy modes are near-BPS modes that can be studied in
gauge perturbation theory.

\subsection{The BMN operators}
Berenstein, Maldacena and Nastase proposed some operators that
correspond to string states in the plane waves background
\cite{BMN}. These operators are known as BMN operators. We should
now review the construction of BMN operators.

The string states in the plane waves consist of a vacuum state and
string excitation modes. The vacuum state is denoted as
$|0,p^{+}\ket$. It has R-charge $J$ in one of the direction in
$SO(6)$, and zero light cone Hamiltonian $\Delta-J=0$. The six
real scalars in $\CN=4$ super Yang-Mills are usually written in 3
complex scalars
\begin{equation}
X=\frac{\phi^1+i\phi^2}{\sqrt{2}},~~Y=\frac{\phi^3+i\phi^4}{\sqrt{2}},
~~Z=\frac{\phi^5+i\phi^6}{\sqrt{2}},
\end{equation}
Suppose the $\psi$ direction that strings move on correspond to
the R-charge of complex scalar $Z$. There is an operator with the
correct R-charge and dimension corresponding to the vacuum state
\begin{equation} \label{vac}
\Tr(Z^J)~ \longleftrightarrow ~ |0,p^{+}\ket
\end{equation}
Here we have not fixed the normalization of the operator.

The string excited states are created by acting creation operators
on the vaccum state. There are eight bosonic modes and eight
fermionic modes corresponding to the eight transverse directions
to the light cone. The bosonic and fermionic creation operators
are denoted as $(a^{i}_n)^\dagger,~i=1,2,\cdots,8$, and
$(S^{b}_n)^\dagger,~b=1,2,\cdots, 8$. Here $n$ is the string
Fourier and $n>0$ denote left movers and $n<0$ denote right
movers. These operators and their complex conjugates are the same
as creation and annihilation operators in usual harmonic
oscillators, with the frequency given by (\ref{3.16}). Since we
are considering close string states, we need to impose the usual
level matching condition to cancel the world sheet momentum for
left movers and right movers
\begin{equation} \label{level}
\sum_{n=-\infty}^{+\infty}nN_n=0
\end{equation}
The level matching condition impose constrains on string states.
For example, if we have only one string mode acting on the vacuum,
the level matching condition (\ref{level}) implies that this
string mode must be zero mode. Thus for one string mode we can
only have states $(a^{i}_0)^\dagger|0,p^{+}\ket$ and
$(S^{b}_0)^\dagger|0,p^{+}\ket$, with $i, b=1,2,\cdots,8$. These
are BPS states corresponding to supergravity modes. On the other
hand, string states with two string modes can have opposite
non-zero modes, such as
$(a^{i}_n)^\dagger(a^{j}_{-n})^\dagger|0,p^{+}\ket$. There are
truly non-BPS stringy modes.

The operators corresponding to strings with one supergravity mode
have $\Delta-J=1$. We should consider insert some operators in the
``strings of Z'' in the vacuum operator (\ref{vac}) that satisfies
this condition. For bosonic modes, there are eight such operators.
We can insert the other four scalars $\phi^i, i=1,\cdots, 4$ and
the covariant derivative $D_i=\partial_i+[A_i, \cdot], i=1,\cdots,
4$. There operators have dimension one and no R-charge in the
``Z'' direction. For fermionic operators, there are 16 gaugino
components that have dimension $\frac{3}{2}$. Eight of the
components $\chi_{J=\frac{1}{2}}$ have R-charge $\frac{1}{2}$ and
the other eight components $\chi_{J=-\frac{1}{2}}$ have R-charge
$-\frac{1}{2}$. These eight components $\chi_{J=\frac{1}{2}}^{a},
a=1,\cdots, 8$, should be identified as the correct operators with
$\Delta-J=1$ to be inserted in the vacuum operator. It is argued
in \cite{BMN} that other operators such as $\bar{Z}$,
$\chi_{J=-\frac{1}{2}}$ have $\Delta-J>1$ and their anomalous
dimensions will grow in the BMN limit, so we will not consider
these operators further. In summary, it is argued in \cite{BMN}
that the operators corresponding to string states with one
supergravity mode are the following
\begin{eqnarray} \nonumber
\Tr(\phi^{i}Z^J)~ \longleftrightarrow ~
(a^{i}_0)^\dagger|0,p^{+}\ket,~~~~~ i=1,\cdots, 4 \\ \nonumber
\Tr((D_{i}Z)Z^{J-1})~ \longleftrightarrow
~ (a^{i+4}_0)^\dagger|0,p^{+}\ket,~~~~~ i=1,\cdots, 4 \\
\Tr(\chi^{b}_{J=\frac{1}{2}}Z^J)~ \longleftrightarrow ~
(S^{b}_0)^\dagger|0,p^{+}\ket,~~~~~ b=1,\cdots, 8
\end{eqnarray}

In the above discussion we consider only one supergarvity mode, so
it does not matter where we insert the operators in ``string of
Z's'' due to the cyclicity of the trace. However we should really
think of the inserted operator as summing over all positions in
the ``string of Z's''. This is required when we have more that one
string excitation modes. For supergravity modes, this
symmetrization will produce BPS operators. For string modes with
Fourier mode $n\neq 0$, BMN \cite{BMN} proposed to use a phase
factor $e^{\frac{2\pi inl}{J}}$ when we insert an operator, say
$\phi$, in position $l$ of ``string of Z's'', $\Tr(Z^l\phi
Z^{J-l})$. For $n\neq 0$ this will be a non-BPS operator. Let us
see how the level matching condition can be obtained by this
proposal. For string state with one string mode, the operator is
\begin{equation}
\sum_{l=0}^{J-1}\Tr(Z^l\phi Z^{J-l})e^{\frac{2\pi inl}{J}}
\end{equation}
By the cyclicity of the trace and the identity
$\sum_{l=0}^{J-1}e^{\frac{2\pi inl}{J}}=0, (n\neq 0$), we
immediately see this operator vanishes unless $n=0$.\footnote{In
some cases, such as the case if we take $l$ from $0$ to $J$, the
sum does not exactly vanish but is small compared to $n=0$ in
large $J$. These operators can also be neglected in BMN limit. See
\cite{Constable} for more discussion. } This is expected from the
level matching condition (\ref{level}) that string states with one
string mode must be zero mode. In general it is straightforward to
show the level matching condition is reproduced for arbitrary
number of string modes. The first non-BPS BMN operators we can
construct consist of two string modes,
$(a^{I_1}_{-n})^\dagger(a^{I_2}_{n})^\dagger|0,p^{+}\ket$. Suppose
$1\leq I_1,I_2, \leq 4$, then the corresponding operator is
\begin{equation}
\sum_{l_1,l_2=0}^{J}\Tr(Z^{l_1}\phi^{I_1}Z^{l_2-l_1}\phi^{I_2}Z^{J-l})e^{-\frac{2\pi
inl_1}{J}} e^{\frac{2\pi inl_2}{J}}
\end{equation}
It is easy to use the cyclicity to move $\phi^{I_1}$ to the first
position of the trace and eliminate one of the sum, we find
\begin{equation}
\sum_{l=0}^{J}\Tr(\phi^{I_1}Z^{l}\phi^{I_2}Z^{J-l}) e^{\frac{2\pi
inl}{J}}~ \longleftrightarrow ~
(a^{I_1}_{-n})^\dagger(a^{I_2}_{n})^\dagger|0,p^{+}\ket,~~~~~ I_1,
I_2=1,\cdots, 4
\end{equation}
If $I_1=I_2$, this operator is referred to as ``singleton''. We
will focus on the case of scalar modes $1\leq I_1\neq I_2\leq 4$
in our discussion.

\subsubsection{Computation of the planar anomalous dimension}

Now we compute the anomalous dimension of BMN operators and see
how the spectrum (\ref{3.16}) can be reproduced from gauge theory.
In conformal field theory the conformal dimension of an operator
$O$ can be found by computing the following two point function
\begin{equation}
\bra O(x)\bar{O}(0)\ket=\frac{C}{|x|^{2\Delta}}
\end{equation}
Here $C$ is a constant that is not dependent on $x$. We then read
off the conformal dimension from the two point function. To
compute the two point function we need the free field propogator
\begin{equation} \label{propagators}
\bra Z_i^j(x) \overline{Z}_k^l(0) \ket = \bra Y_i^j(x)
\overline{Y}_k^l(0) \ket =  \bra X_i^j(x) \overline{X}_k^l(0) \ket
= \delta^l_i \delta^j_k { g_s \over 2 \pi}{1 \over |x|^2},
\end{equation}
and the $\CN=4$ action
\begin{equation}
S={1 \over 2\pi g_s} \int d^4 x ~\tr\Bigl({1 \over 2}F_{\mu
\nu}F^{\mu \nu}+D_\mu Z D^\mu \overline{Z} +D_\mu Y D^\mu
\overline{Y}+D_\mu X D^\mu\overline{X} +V_D+V_F\Bigr)
\end{equation}
Where the D-term potential and the F-term potential are
\begin{equation}
V_D=\frac{1}{2}\tr|[X,\overline{X}]+[Y,\overline{Y}]+[Z,\overline{Z}]|^2
\end{equation}
\begin{equation}
V_F=2\tr(|[X,Y]|^2+|[X,Z]|^2+|[Y,Z]|^2)
\end{equation}

The two point function in an interacting gauge theory can be
computed by putting in the action $e^{-S}$ and doing perturbation
with free field contractions
\begin{eqnarray} \label{correlator}
\bra O(x)\bar{O}(0)\ket &=& \bra O(x)\bar{O}(0)e^{-S}\ket_{free}
\nonumber \\ &=&\bra O(x)\bar{O}(0)(1-S+\cdots) \ket_{free}
\end{eqnarray}

We consider the BMN operator
\begin{equation}
O=\sum_{l=0}^{J}\tr(XZ^{l}YZ^{J-l})e^{\frac{2\pi inl}{J}}
\end{equation}
Here we consider only planar diagrams. The calculation of
anomalous dimension was originally done for real scalar insertions
in \cite{BMN} . Here we have used complex scalars $X$ and $Y$
insertions because the calculation is simplified due to an
argument in Appendix B in \cite{Constable}, where it was shown for
a holomorphic operator consisting of $X,Y,Z$, the D-term and gluon
exchange cancel at one loop order (this is based on techniques in
previous papers \cite{DHoker, Skiba} ), so we only need to
consider the contributions from F-term. We need to compute the
free part and one-loop part in (\ref{correlator}). Using the
propagator (\ref{propagators}) it is easy to find the free part
\begin{equation} \label{fre}
\bra O(x)\bar{O}(0)\ket_{free}=JN^{J+2} ({g_s \over 2\pi
|x|^2})^{J+2}\equiv \frac{C}{|x|^{2(J+2)}}
\end{equation}
The one-loop part is
\begin{equation} \label{one}
\bra O(x)\bar{O}(0)\ket_{one-loop}=\bra
O(x)\bar{O}(0)(-\frac{1}{2\pi g_s}\int d^4yV_F(y))\ket_{free}
\end{equation}
We will need to the log divergence formulae
\begin{equation}
|x|^4 \int d^4 y ~{1 \over |y|^4|x-y|^4}=4\pi^2\log(|x|\Lambda)
\end{equation}
The $2\tr(|[X,Z]|^2)$ and $2\tr(|[Y,Z]|^2)$ give the same
contributions to the one-loop two point function, so we only need
to consider one of them. The four terms in $2\tr(|[Y,Z]|^2)$
contribute as follows
\begin{eqnarray}
&&
\sum_{l=0}^{J}\bra\tr(XZ^{l}YZ^{J-l})(x)\tr(\bar{Z}^{J-l+1}\bar{Y}\bar{Z}^{l-1}\bar{X})(0)\int
d^4y (-\frac{1}{2\pi g_s})(-2\tr(YZ\bar{Y}\bar{Z})(y)) \ket
e^{\frac{2\pi in}{J}} \nonumber \\
&=& \frac{g_s N}{\pi}e^{\frac{2\pi in}{J}}
\frac{C}{|x|^{2(J+2)}}\log(|x|\Lambda),
\nonumber \\
&&
\sum_{l=0}^{J}\bra\tr(XZ^{l}YZ^{J-l})(x)\tr(\bar{Z}^{J-l-1}\bar{Y}\bar{Z}^{l+1}\bar{X})(0)\int
d^4y (-\frac{1}{2\pi g_s})(-2\tr(ZY\bar{Z}\bar{Y})(y)) \ket
e^{-\frac{2\pi in}{J}} \nonumber \\
&=& \frac{g_s N}{\pi}e^{-\frac{2\pi in}{J}}
\frac{C}{|x|^{2(J+2)}}\log(|x|\Lambda),  \nonumber \\&&
\sum_{l=0}^{J}\bra\tr(XZ^{l}YZ^{J-l})(x)\tr(\bar{Z}^{J-l}\bar{Y}\bar{Z}^{l}\bar{X})(0)\int
d^4y (-\frac{1}{2\pi g_s})(2\tr(YZ\bar{Z}\bar{Y})(y)) \ket
\nonumber \\&=&
\sum_{l=0}^{J}\bra\tr(XZ^{l}YZ^{J-l})(x)\tr(\bar{Z}^{J-l}\bar{Y}\bar{Z}^{l}\bar{X})(0)\int
d^4y (-\frac{1}{2\pi g_s})(2\tr(ZY\bar{Y}\bar{Z})(y)) \ket \nonumber \\
&=& -\frac{g_s N}{\pi} \frac{C}{|x|^{2(J+2)}}\log(|x|\Lambda).
\end{eqnarray}
Adding up all contributions and keep leading terms in large $J$
limit we find
\begin{eqnarray}
\bra O(x)\bar{O}(0)\ket &=& \bra O(x)\bar{O}(0)\ket_{free}+\bra
O(x)\bar{O}(0)\ket_{one-loop} \nonumber \\
&=& \frac{C}{|x|^{2(J+2)}}(1+\frac{2g_s N}{\pi}(e^{\frac{2\pi
in}{J}}+e^{-\frac{2\pi in}{J}}-2)\log(|x|\Lambda))
\nonumber \\
&=& \frac{C}{|x|^{2(J+2)}}(1-\frac{8\pi
g_sNn^2}{J^2}\log(|x|\Lambda))
\end{eqnarray}
The contribution of $X$ and $Y$ insertion in the ``string of Z's''
are the same. It can be read off from the above equation
\begin{equation}
(\Delta-J)_n=1+\frac{2\pi g_s Nn^2}{J^2}
\end{equation}

Thus the one-loop calculation reproduced the first order expansion
in the square root in (\ref{3.16}).

It has also been shown in \cite{Gross} the higher loop
calculations also reproduce higher order terms of the square root
expansion in (\ref{3.16}). One can also study the sub-leading
order in ``$1/J$'' correction to the spectrum, see e.g.
\cite{Callan}. This is known as the ``near plane wave'', which
corresponds to small deformation the plane wave toward the AdS
geometry. Some connections to integrable structure have also been
studied, see e.g. \cite{Minahan}. One can also study operators
corresponding rotating strings, for a review see e.g.
\cite{Tseytlin}. Here we will not explore these very important
issues further.

\section{String Interactions in Plane Wave Backgrounds}
We have seen the  BMN limit is to scale the R-charge $J$
\begin{equation}
J\sim \sqrt{N}\sim +\infty,
\end{equation}
with $\frac{J^2}{N}$ fixed. There is another dimensionless
parameter $g_{YM}$ in the theory, so we have a total of two
dimensionless parameters. \footnote{The string scale
$\alpha^{\prime}$ is a dimensionful parameter, so it can be set at
any value, (or just to ``1'') and it will not affect our
discussion.} These two dimensionless parameters that determine the
theory are conventionally denoted as $\lambda^{\prime}$ and $g_2$
in the literature,  and are parameterized as
\begin{equation}
\lambda^{'}=\frac{g_{YM}^2 N}{J^2}=\frac{1}{(\mu p^{+}
\alpha^{'})^2}
\end {equation}
\begin{equation}
g_2=\frac{J^2}{N}=4\pi g_s (\mu p^{+} \alpha^{'})^2
\end{equation}
Here $\lambda^{\prime}$ is usually referred to as the effective 't
Hooft coupling constant. In previous section we have considered
planar diagrams of gauge perturbation theory. This is an expansion
in $\lambda^{\prime}$ while keeping $g_2=0$. In this case the free
string spectrum are reproduced by gauge interactions. To study
string interaction, we must consider non-planar diagrams and do
``$1/N$'' expansion. This means we will do expansion in the
parameter $g_2$.

One might then try to study the string spectrum involving
non-planar diagrams. However, the BMN operators appear to have
some kind of dangerous mixings and no long have well defined
anomalous conformal dimensions \cite{Constable, Kristjansen, BN}.
Further studies in this direction can be found in e.g. \cite{GMR,
Gomis, DPPRT}. One can also propose some Hamiltonians, such as a
string bit model, that could reproduce the spectrum
\cite{Verlinde}.

In the following we will not explore these important directions
further. Instead, we consider the situation where we set
$\lambda^{\prime}=0$ and do expansion solely in the parameter
$g_2$. In this case the Yang-Mills theory is free, so the
correlation functions of the BMN operators have the usual nice
form of spacetime dependence. We note this nice form of spacetime
dependence is present when one of the parameters
$\lambda^{\prime}$ or $g_2$ is zero, but is ruined when both
parameters are finite. By setting $\lambda^{\prime}=0$, we can
focus on the coefficients of the correlation functions and not
worry about the not quite well understood spacetime dependence
part. We also see from the spectrum formula (\ref{3.16}) that the
string spectra are degenerate in this highly curved background.
However, the interactions of strings does not vanish. The physical
string amplitudes are represented in dual free gauge theory by
correlation functions of BMN operators.

Suppose a Feymann diagram has genus $g$, number of holes (index
loops) $h$, propagators (edges) $E$ and vertices $V$, then it is
easy to see $h=E-V+2-2g$. This diagram will has a contribution
proportional to $N^{h}$. Since we are considering free field
theory, for a given correlation functions we can not add
interaction vertices. The number of edges and vertices $E$ and $V$
are fixed and determined by the given correlation functions we
want to compute. Thus we find the implementation of the original
t'Hooft idea of ``summing over all holes for a given genus'' is
pretty simple in the case of free field theory! For a given genus
we only need to sum over a finite number of diagrams with the same
number of holes $h=E-V+2-2g$, and they are all proportional
$N^{h}\sim N^{-2g}$. Furthermore, in large R-charge limit the
diagram will be also proportional to $J^{4g}$ due to many ways of
doing free field contractions. Thus we see the diagram is
proportional to $J^{4g}N^{-2g}=g_2^{2g}$. In string perturbation
theory we know that string loop expansion is organized in the
genus expansion of the Riemann surface of the string worldsheet.
{\it Thus it is natural in this case to identify the parameter
$g_2$ as the effective string coupling constant.}

As opposed to previous section, where the free string spectrum is
reproduced by gauge interaction, here we have a situation that
string interactions are described by free gauge theory. In this
section we will compare the planar three point function to light
cone string field theory vertex \cite{Huang1}, and make a proposal
on string loop diagrams that corresponds to non-planar correlation
functions \cite{Huang2}.

\subsection{Planar three point functions}
The definition of the vacuum operators and BMN operators with one
and two excitation modes are
\begin{equation}
O^{J}=\frac{1}{\sqrt{N^JJ}}TrZ^J
\end{equation}
\begin {equation}
O^{J_1}_{0}=\frac{1}{\sqrt{N^{J_1+1}}} Tr(\phi^{I_1} Z^{J_1})
\end {equation}
\begin {equation}
O^{J_2}_{0}=\frac{1}{\sqrt{N^{J_2+1}}} Tr(\phi^{I_2} Z^{J_2})
\end {equation}
\begin{equation}\label{bnnop}
O^J_{m,-m} = \frac1{\sqrt{JN^{J+2}}} \sum_{l=0}^Je^{2\pi iml/J}
Tr(\phi^{I_1} Z^l\phi^{I_2} Z^{J-l}).
\end{equation}
Here $\phi^{I_1}$ and $\phi^{I_2}$ represent excitations in two of
the four scalar transverse directions. The normalization is fixed
by choosing planar two point functions to be orthonormal.

The computation of free planar three point function are
straightforward, see e.g. \cite{Constable}. Some results are the
followings (Assuming $m\neq 0$ and $n\neq 0$)
\begin {eqnarray} \label{planar1}
\langle\bar{O}^JO^{J_1}O^{J_2}\rangle &=&
\frac{g_2}{\sqrt{J}}\sqrt{x(1-x)}
\nonumber \\
\langle\bar{O}^J_{0}O^{J_1}O^{J_2}_{0}\rangle &=&
\frac{g_2}{\sqrt{J}}x^{\frac{1}{2}}(1-x)
\nonumber \\
\langle\bar{O}^J_{00}O^{J_1}_{0}O^{J_2}_{0}\rangle &=&
\frac{g_2}{\sqrt{J}}x(1-x)
\nonumber \\
\langle\bar{O}^J_{m,-m }O^{J_1}_{0}O^{J_2}_{0}\rangle
&=&-\frac{g_2}{\sqrt{J}}\frac{\sin^2(\pi mx)}{\pi^2m^2}
\nonumber \\
\langle\bar{O}^J_{00}O^{J_1}_{00}O^{J_2}\rangle &
=&\frac{g_2}{\sqrt{J}}x^{\frac{3}{2}}\sqrt{1-x}
\nonumber \\
\langle\bar{O}^J_{m,-m}O^{J_1}_{n,-n}O^{J_2}\rangle &=&
\frac{g_2}{\sqrt{J}}x^{\frac{3}{2}}\sqrt{1-x}\frac{\sin^2(\pi
mx)}{\pi^2 (mx-n)^2} \nonumber \\
\langle\bar{O}^J_{00}O^{J_1}_{n,-n}O^{J_2}\rangle &=& 0
\end {eqnarray}
where $x=J_1/J$ and $J=J_1+J_2$. Note the spacetime dependences of
two point and three point functions in this case of free field
theory is quite simple. Here and elsewhere in this section we have
omitted the factors of spacetime dependence in the correlators.

Here we note that a three point function can be thought of as a
two point function of a single trace operator and a double trace
operator. In order to have non-vanishing correlation functions we
must have equal number of $Z$'s and $\bar{Z}$'s. Since in BMN
limit an operator always has a large number of $Z$'s but no
$\bar{Z}$'s, a non-vanishing correlation function can always be
thought of as a two point function of multi-trace operators. In
AdS/CFT a multi-trace operator represents multi-particle states,
so here we will still conventionally call these correlators
``three point functions''. We also caution the readers although
these planar three point function can be drawn on a plane, they
really increase the number of traces and thus have power of $1/N$
compared to the planar diagrams we considered in previous section.

\subsubsection{How to compare three point function to string theory
vertex}

In \cite{Constable} it is proposed that the matrix element for a
single string $|\Phi_3\rangle$ to split into a two-string state
$|\Phi_1\rangle|\Phi_2\rangle$ in the string field theory light
cone Hamiltonian is
\begin {equation}
\label{1}
(\Delta_3-\Delta_1-\Delta_2)\langle\bar{O}_3O_1O_2\rangle
\end {equation}
Where $O_i$'s are the properly normalized corresponding operators
in CFT and $\Delta_i$'s are their conformal dimensions. In light
cone string field theory the matrix element is calculated by
applying the three string states to a prefactor $\hat{h}_3$ and
the cubic interaction vertex state $|V\rangle$ in the three-string
Hilbert space ${\cal H}_3$. The $\hat{h}_3$ and $|V\rangle$ are
calculated in details in \cite{SV1, SV2}. It is conjectured in
\cite{Constable} that at large $\mu$ limit the dressing factor
$(\Delta_3-\Delta_1-\Delta_2)$ in equation (\ref{1}) comes from
the prefactor $\hat{h}_3$ and assuming discretization of the
string world sheet at large $\mu$, a heuristic proof is given
there that the delta functional overlap agrees exactly with the
planar 3-point function in field theory, i.e.

\begin{equation}
\langle\Phi_1|\langle\Phi_2|\langle\Phi_3|V\rangle \sim
\langle\bar{O}_3O_1O_2\rangle
\end {equation}

We will explicitly check this proposal \cite{Huang1}. Here we will
not calculate the overall normalization of the matrix element. To
make an explicit check of the PP-wave/Yang-Mills duality, we will
calculate the ratio with vacuum amplitude on both sides. We should
verify

\begin {equation}
\label{proposal}
\frac{\langle\Phi_1|\langle\Phi_2|\langle\Phi_3|V\rangle}{\langle
0_1|\langle 0_2|\langle 0_3|V\rangle}
=\frac{\langle\bar{O}_3O_1O_2\rangle}{\langle\bar{O}^J
O^{J_1}O^{J_2}\rangle}
\end {equation}
Here $J=J_1+J_2$, and $O^{J}=\frac{1}{\sqrt{N^JJ}}TrZ^J$ is the
corresponding operator of the vacuum state.

From (\ref{planar1}) we can read off the ratios
\begin {equation}\label{supergravityexample}
 \frac{\langle\bar{O}^J_{00}O^{J_1}_{0}O^{J_2}_{0}\rangle}{\langle
 \bar{O}^JO^{J_1}O^{J_2}\rangle}=\sqrt{x(1-x)}
 \end {equation}
\begin {equation}\label{stringexample1}
 \frac{\langle\bar{O}^J_{m,-m}O^{J_1}_{0}O^{J_2}_{0}\rangle}{\langle
 \bar{O}^JO^{J_1}O^{J_2}\rangle}=-\frac{1}{\sqrt{x(1-x)}}\frac{\sin^2(\pi mx)}{\pi^2m^2}
 \end {equation}

 \begin {equation} \label{stringexample2}
 \frac{\langle\bar{O}^J_{m,-m}O^{J_1}_{n,-n}O^{J_2}\rangle}{\langle
 \bar{O}^JO^{J_1}O^{J_2}\rangle}=x\frac{\sin^2(\pi
mx)}{\pi^2 (mx-n)^2}
 \end {equation}

We will calculate the ratios of the three point correlators
(\ref{supergravityexample}) (\ref{stringexample1})
(\ref{stringexample2}) from light cone string field theory in
pp-wave. We will find exact agreements with equation
(\ref{proposal}).

Some notation is the following. Following the notation of
\cite{SV1, Huang1} we denote $\alpha=\alpha^{'}p^{+}$. The strings
are labeled by $r=1,2,3$ and in light-cone gauge their widths are
$2 \pi|\alpha_{(r)}|$, with $\alpha_{(1)} + \alpha_{(2)} +
\alpha_{(3)} = 0$. We will take $\alpha_{(1)}$ and $\alpha_{(2)}$
positive for purposes of calculation. Also note that
\begin{equation}
x=\frac{J_1}{J}=\frac{|\alpha_{(1)}|}{|\alpha_{(3)}|}
\end{equation}

\subsubsection{ Computation of the bosonic Neumann matrices in large $\mu
p^{+} \alpha^{'}$ limit}

Light cone string field theory is an old subject dating back to
the 80's (see e.g. \cite{GreenTC, Green}). Recently the results
have been extend to plane wave backgrounds, see e.g. \cite{PS,
HSSV, Open}. For our purpose we will need to use the cubic
interaction vertex $|V\rangle$, which can be written as an element
in the 3-string Hilbert space. Roughly speaking, the interaction
amplitude of three strings is the inner product of the three
string state with the cubic interaction vertex.

The string modes interaction vertex is $|V\rangle=E_aE_b|0\rangle$
where $E_a$ and $E_b$ are bosonic and fermionic operators that are
calculated in details in \cite{SV1, SV2} . Here will not consider
the fermionic sector. Up to a overall factor, the bosonic operator
$E_a$ is

\begin{equation} \label{bos}
 E_a \sim \exp \left[  \frac{1}{2} \sum_{r,s = 1}^3\sum_{I =
 1}^8 \sum_{m,n=-\infty}^{\infty}
a_{m(r)}^{\dagger I} \bar{N}^{(rs)}_{(mn)} a_{n(s)}^{\dagger I}
\right]
\end{equation}
where $I$ denote the eight transverse directions, and
$\bar{N}^{(rs)}_{(mn)}$ are  the Neumann matrices.

Here we only need to consider the limit of large $\mu p^{+} \alpha
{'}$. we will show that the infinite dimensional Neumann matrices
turn out to simplify in large $\mu p^{+} \alpha^{'}$ limit. The
Neumann matrices $\bar{N}^{(rs)}_{mn}$ ($r,s=1\cdots 3,
m,n=-\infty \cdots +\infty$ ) is calculated in \cite{SV1}

\begin{equation}
 \overline{N}_{mn}^{(rs)} = \delta^{rs} \delta_{mn} - 2 \sqrt{
\omega_{m(r)} \omega_{n(s)} } (X^{(r) {\rm T}} \Gamma_a^{-1}
X^{(s)})_{mn}
\end{equation}

where $\omega_{m(r)}=\sqrt{m^2+(\mu \alpha_{(r)})^2}$, and

\begin{equation} \label{gamma}
(\Gamma_a)_{mn} = \sum_{r=1}^3 \sum_{p=-\infty}^\infty
\omega_{p(r)} X^{(r)}_{mp} X^{(r)}_{np}
\end{equation}

The definition of $X^{(r)}$ is the following. Consider for $m,n>0$
the matrices of \cite{GreenTC, SV1},
\begin{equation}
 A^{(1)}_{mn}  =(-1)^{n} {2 \sqrt{m n} \over \pi} {x \sin{m
\pi x}\over n^2-m^2 x^2},
\end {equation}
\begin{equation}
A^{(2)}_{mn} =- {2 \sqrt{m n} \over \pi} {(1-x) \sin{m \pi x}
\over n^2-m^2 (1-x)^2}
\end{equation}

\begin{equation}
C_{mn} = m \delta_{mn}
\end{equation}
 and the vector
\begin{equation}
 B_m = - {2 \over \pi} {\alpha_{(3)} \over
\alpha_{(1)} \alpha_{(2)}} m^{-3/2} \sin m \pi x
\end{equation}
 We define
$X^{(3)}_{mn} = \delta_{mn}$, while for $r=1,2$ we can express the
matrices $X^{(r)}$ as

\begin{eqnarray}
 X^{(r)}_{mn} &=& (C^{1/2} A^{(r)}
C^{-1/2})_{mn} \qquad\qquad\qquad{\rm if}~m,n>0,\nonumber \\& =&
{\alpha_{(3)} \over \alpha_{(r)}} (C^{-1/2} A^{(r)} C^{1/2})_{
-m,-n}, \qquad{\rm if}~m,n<0,\nonumber \\&= & - {1 \over \sqrt{2}}
\epsilon^{rs} \alpha_{(s)} (C^{1/2} B)_m\qquad \qquad~~~{\rm
if}~n=0~{\rm and}~m>0,\\
&=&1\qquad\qquad\qquad\qquad\qquad\qquad\qquad{\rm if}~m=n=0,
\nonumber \\&=&0\qquad\qquad\qquad\qquad\qquad\qquad\qquad{\rm
otherwise}. \nonumber
\end{eqnarray}

In the limit of large $\mu\alpha$, $\omega_{m(r)}=\sqrt{m^2+(\mu
\alpha_{(r)})^2}\approx \mu|\alpha_{(r)}|$. Using equation
(\ref{gamma}), we find that for $m,n> 0$,
\begin {eqnarray}
(\Gamma_a)_{mn}&=& |\alpha_{(3)}|\mu\frac{4mn}{\pi^2}\sin(m\pi
x)\sin(n \pi x)
[\sum_{l=1}^{+\infty}\frac{x^3}{(l^2-m^2x^2)(l^2-n^2x^2)}\\&&
+\sum_{l=1}^{+\infty}\frac{(1-x)^3}{(l^2-m^2(1-x)^2)(l^2-n^2(1-x)^2)}
 +\frac{1}{2m^2n^2x(1-x)}]+|\alpha_{(3)}|\mu\delta_{mn} \nonumber
\end {eqnarray}

\begin {eqnarray} (\Gamma_a)_{-m,-n}&=&
|\alpha_{(3)}|\mu\frac{4}{\pi^2}\sin(m\pi x)\sin(n \pi x)
[\sum_{l=1}^{+\infty}\frac{xl^2}{(l^2-m^2x^2)(l^2-n^2x^2)} \\&&+
\sum_{l=1}^{+\infty}\frac{(1-x)l^2}{(l^2-m^2(1-x)^2)(l^2-n^2(1-x)^2)}
 ]+|\alpha_{(3)}|\mu\delta_{mn} \nonumber
\end {eqnarray}
 and $(\Gamma_a)_{00}=2|\alpha_{(3)}|\mu$. All other components
 such as $(\Gamma_a)_{m0}$ are zero.

Using the summation formulae in the appendix D of \cite{GreenTC}
we find
\begin{equation}
(\Gamma_a)_{mn}=2|\alpha_{(3)}|\mu\delta_{mn}\qquad\qquad\qquad\qquad
for \qquad m,n=-\infty\cdots +\infty
\end {equation}
So the Neumann matrices in large $\mu|\alpha|$ limit is
\begin{equation} \label {Neumann}
\bar{N}^{(rs)}_{(mn)}=\delta^{rs}\delta_{mn}-\frac{\sqrt{|\alpha_{(r)}||\alpha_{(s)}|}}{|\alpha_{(3)}|}(X^{(r)T}X^{(s)})_{mn}
\end{equation}
We note $\bar{N}^{(rs)}_{(mn)}=\bar{N}^{(sr)}_{(nm)}$.

In \cite{SV1} the cubic coupling matrix of supergravity modes is
derived. It is

\begin{equation} \label{supergravityNeumann}
M^{rs} = \left(\matrix { 1-x &-\sqrt{x(1-x)} &-\sqrt{x} \cr
-\sqrt{x(1-x)} &x &-\sqrt{1-x} \cr -\sqrt{x} &-\sqrt{1-x} & 0 }
\right)
\end{equation}
One would be tempted to identify $M^{rs}$ as the zero-zero
component of the Neumann matrices (\ref{Neumann}). But this is
incorrect. Actually one can check $M^{rs}=\bar{N}^{(rs)}_{(00)}$
is true when $\mu p^{+} \alpha^{'}=0$, but at large $\mu p^{+}
\alpha^{'}$ limit $M^{rs}=\bar{N}^{(rs)}_{(00)}$ is true only when
$r=3$ or $s=3$. But since we will only use these components, it
will be same whether we use $M^{rs}$ or $\bar{N}^{(rs)}_{(00)}$.

\subsubsection{ Interaction of supergravity modes}

We consider the interaction of three supergravity modes
 $a^{\dagger I_1}_{0(1)}|0\rangle$,$a^{\dagger I_2}_{0(2)}|0\rangle$,$a^{\dagger I_1}_{0(3)}a^{\dagger
 I_2}_{0(3)}|0\rangle$. Here $I_1$ and $I_2$ are two different transverse directions. We want to compute
 the object
\begin {equation}
 \frac{\langle0|a^{I_1}_{0(1)}a^{I_2}_{0(2)}a^{I_1}_{0(3)}a^{I_2}_{0(3)}|V\rangle}{\langle0|V\rangle}
 \end {equation}
We will need the zero components of the Neumann matrices. From
equation (\ref{Neumann}) we find
$\bar{N}^{(13)}_{(00)}=\bar{N}^{(31)}_{(00)}=-\sqrt{x}$,
$\bar{N}^{(23)}_{(00)}=\bar{N}^{(32)}_{(00)}=-\sqrt{1-x}$.
(Actually this is true without taking the large $\mu p^{+}
\alpha^{'}$ limit.)

 From the bosonic operator (\ref{bos}) and the Baker-Hausdorff formula
\footnote{The Baker-Hausdorff formula is
$e^{A}Be^{-A}=B+[A,B]+\frac{1}{2!}[A,[A,B]]+\cdots$.} we know
\begin{equation}
 (E_a)^{-1}a^{I_1}_{0(1)}a^{I_1}_{0(3)}E_a=
 a^{I_1}_{0(1)}a^{I_1}_{0(3)}-\frac{1}{2}(\bar{N}^{(13)}_{(00)}+\bar{N}^{(31)}_{(00)})
 \end {equation}
 \begin{equation}
 (E_a)^{-1}a^{I_2}_{0(2)}a^{I_2}_{0(3)}E_a=
 a^{I_2}_{0(2)}a^{I_2}_{0(3)}-\frac{1}{2}(\bar{N}^{(23)}_{(00)}+\bar{N}^{(32)}_{(00)})
 \end {equation}
So
\begin {equation} \label{ppsupergravityexample}
 \frac{\langle0|a^{I_1}_{0(1)}a^{I_2}_{0(2)}a^{I_1}_{0(3)}a^{I_2}_{0(3)}|V\rangle}{\langle0|V\rangle}
 =\frac{1}{2}(\bar{N}^{(13)}_{(00)}+\bar{N}^{(31)}_{(00)})\frac{1}{2}(\bar{N}^{(23)}_{(00)}+\bar{N}^{(32)}_{(00)})=\sqrt{x(1-x)}
 \end {equation}
On the field theory side, the three modes $a^{\dagger
I_1}_{0(1)}|0\rangle$,$a^{\dagger
I_2}_{0(2)}|0\rangle$,$a^{\dagger I_1}_{0(3)}a^{\dagger
I_2}_{0(3)}|0\rangle$ correspond to chiral operators $O^{J_1}_0$,
$O^{J_2}_0$ and $O^{J}_{00}$ (Suppose $I_1$ and $I_2$ correspond
to scalar instead of the $D_\mu$ insertions in the string of
$Z$'s) . Thus we have found equation (\ref{ppsupergravityexample})
is in agreement with equation (\ref{supergravityexample}).

\subsubsection{ Interaction of string theory modes} \label{string}

\bigskip{\noindent{\it Example 1}}

We consider the interaction of three states
$a^{I_1(BMN)\dagger}_{0(1)}|0\rangle$,
$a^{I_2(BMN)\dagger}_{0(2)}|0\rangle$,
$a^{I_1(BMN)\dagger}_{m(3)}a^{I_2(BMN)\dagger}_{-m(3)}|0\rangle$,
which correspond to operators $O^{J_1}_{0}$, $ O^{J_2}_{0}$ and
$O^J_{m, -m}$. We caution the reader here the $a^{+}$ notation we
use
 is not the familiar string theory basis of BMN \cite {BMN}, but is the
same as in \cite{SV1}. These two basis are related by
\begin {equation}
a^{BMN}_n=\frac{1}{\sqrt{2}}(a_{|n|}-ie(n)a_{-|n|})
\end {equation}
where $e(n)$ is the sign of $n$ (For $n=0$, $a^{BMN}_0=a_0$).
Notice that the $a_{-n}$ mode contribution vanish since the
corresponding Neumann matrices elements are zero. The calculation
here follows similarly as in the case of supergravity modes, we
find
\begin{eqnarray}
&&\frac{\langle 0|a^{I_1(BMN)}_{0(1)} a^{I_2(BMN)}_{0(2)}
a^{I_1(BMN)}_{m(3)}a^{I_2(BMN)}_{-m(3)}|V\rangle}{\langle0|V\rangle}
\nonumber
\\&& =\frac{\langle 0|a^{I_1}_{0(1)} a^{I_2}_{0(2)}
a^{I_1}_{m(3)}a^{I_2}_{m(3)}|V\rangle}{2\langle0|V\rangle}\\&&
=\frac{1}{2}\bar{N}^{(31)}_{(m0)}\bar{N}^{(32)}_{(m0)} \nonumber
\end{eqnarray}
Using equation (\ref{Neumann}) we find
$\frac{1}{2}\bar{N}^{(31)}_{(m0)}\bar{N}^{(32)}_{(m0)}=-\frac{1}{\sqrt{x(1-x)}}\frac{\sin^2(\pi
mx)}{\pi^2m^2}$, in agreement with equation
(\ref{stringexample1}).

\bigskip{\noindent{\it Example 2}}

In this example we consider the interaction of three states
$a^{I_1(BMN)\dagger}_{n(1)}a^{I_2(BMN)\dagger}_{-n(1)}|0\rangle$,$|0\rangle$,
$a^{I_1(BMN)\dagger}_{m(3)}a^{I_2(BMN)\dagger}_{-m(3)}|0\rangle$,
which correspond to operators $O^{J_1}_{n, -n}$, $ O^{J_2}$ and
$O^J_{m, -m}$. Notice the Neumann matrix elements
$\bar{N}^{(rs)}_{(m ,-n)}$ and $\bar{N}^{(rs)}_{(-m,n)}$ vanish,
 so
\begin{eqnarray}
&&\frac{\langle 0|a^{I_1(BMN)}_{n(1)} a^{I_2(BMN)}_{-n(1)}
a^{I_1(BMN)}_{m(3)}a^{I_2(BMN)}_{-m(3)}|V\rangle}{\langle0|V\rangle}
\nonumber
\\&&
=\frac{1}{4}(\bar{N}^{(31)}_{(m,n)}-\bar{N}^{(31)}_{(-m,-n)})^2
\\&& =x\frac{\sin^2(\pi mx)}{\pi^2 (mx-n)^2}\nonumber
\end{eqnarray}
Again it agrees with equation (\ref{stringexample2}).

\subsection{Non-planar correlation functions}
\begin{figure}
  \begin{center}
 \epsfysize=2.5in
   \mbox{\epsfbox{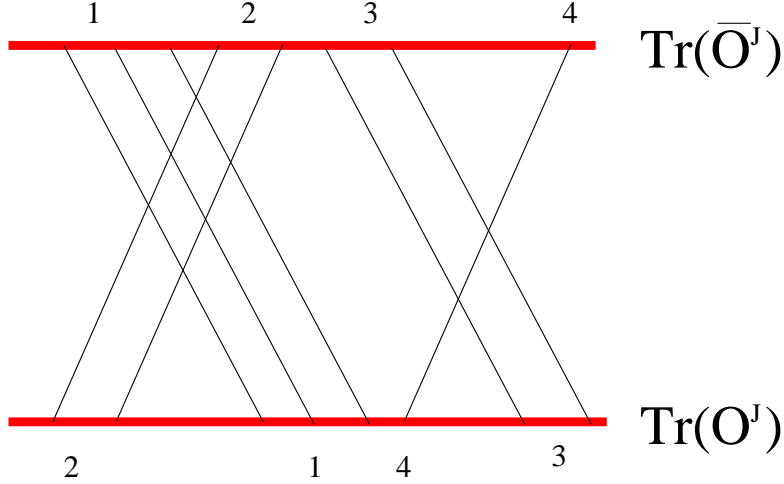}}
\end{center}
\caption{Feymann diagram of torus contraction of large $N$ gauge
indices. We are contracting non-planarly by dividing the string
into 4 segments.} \label{F3}
\end{figure}
We consider free torus two point function of BMN operators. The
calculation involves a gauge theory diagram as shown in Fig
\ref{F3}. The calculation is to divide the string of Z's into 4
segments. The scalar insertions can then be summed over all
positions with phases. The calculation was first done in
\cite{Constable, Kristjansen}. Here we simply quote the result

\begin{eqnarray}\label{torus}
&&\langle \bar{O}_{n,-n}^J O_{m,-m}^J \rangle_{torus}   \\
&& = \frac{g_2^2}{24}, \qquad\qquad\qquad\qquad\qquad\qquad\qquad\qquad\qquad\qquad\qquad\qquad m=n=0; \nonumber\\
&& = 0,  \qquad\qquad\qquad\qquad\qquad\qquad\qquad\qquad\qquad\qquad m=0, n\neq0~~or~~n=0, m\neq0; \nonumber\\
&& =g_2^2(\frac{1}{60} - \frac{1}{24 \pi^2 m^2} + \frac{7}{16 \pi^4 m^4}), \qquad\qquad\qquad\qquad\qquad\qquad\qquad m=n\neq0; \nonumber\\
&& =\frac{g_2^2}{16\pi^2m^2} ( \frac{1}{3}+\frac{35}{8\pi^2m^2}),
\qquad\qquad\qquad\qquad\qquad\qquad\qquad\qquad m=-n\neq0; \nonumber\\
&& =\frac{g_2^2}{4\pi ^{2}(m-n)^2} ( \frac{1}{3}+\frac{1}{\pi
^2n^2}+\frac{1}{\pi ^2m^2}-\frac{3}{2\pi ^2mn}-\frac{1}{2\pi
^2(m-n)^2}), ~~all~other~cases \nonumber
\end{eqnarray}
We should note that unlike planar two point functions, torus two
point functions do not vanish between different BMN operators. We
can not absorb the torus two point function by proper
normalization of the BMN operators. Therefore the torus two point
functions represent physical string propagation amplitudes. One
should try to reproduce them from a string theory calculation. We
will follow \cite{Huang2} and make a proposal as to what the
string theory calculation is.

\begin{figure}
  \begin{center}
 \epsfysize=2.5in
   \mbox{\epsfbox{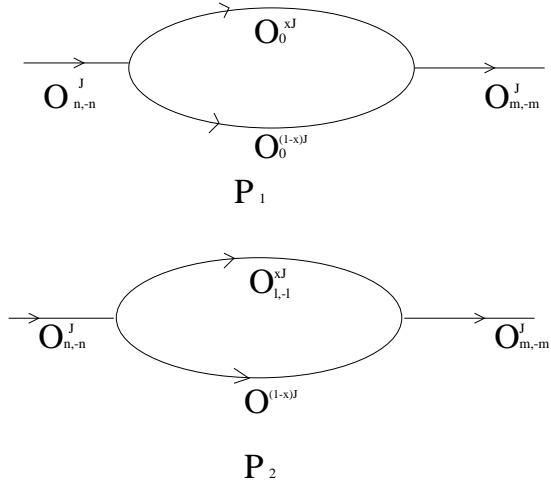}}
\end{center}
\caption{There are 2 diagrams contributing the one loop string
propagation. The BMN string $O^J_{n,-n}$ can split into two
strings $O^{J_1}_{l,-l}$, $O^{J_2}$ or $O^{J_1}_0$, $O^{J_2}_0$
and joining back into another string $O^J_{m,-m}$. We denote
contributions to these two diagrams $P_1$ and $P_2$.} \label{F2}
\end{figure}

\subsubsection{ A proposal on string theory loop diagrams}

\label{twopoint1} We consider a single string propagating in the
pp-wave background. We expect the one loop correction to the
string propagation to be the torus contribution to the two point
function of corresponding BMN operators. On the other hand, the
one loop amplitude can be calculated by summing over the
amplitudes of the string splitting into two strings and then
joining back into a single string. The cubic vertices of string
splitting and joining can be represented by free planar three
point functions.  There are two diagrams associated with this
process as shown in figure \ref{F2}. The BMN string $O^J_{n,-n}$
can split into two strings $O^{J_1}_{l,-l}$, $O^{J_2}$ or
$O^{J_1}_0$, $O^{J_2}_0$ and joining back into another string
$O^J_{m,-m}$. We denote the contributions from these two processes
by $P_1$ and $P_2$. Then

\begin{eqnarray}\label{P1}
P_1&=&\sum_{J_1=0}^{J}\langle \bar{O}^J_{n,-n} O^{J_1}_0 O^{J_2}_0
\rangle_{planar} \langle \bar{O}^{J_1}_0 \bar{O}^{J_2}_0
O^J_{m,-m} \rangle_{planar}\\&&\nonumber
=g_2^2\int_0^1dx\frac{\sin^2(m\pi x)}{m^2\pi^2}\frac{\sin^2(n\pi
x)}{n^2\pi^2}
\end{eqnarray}

\begin{eqnarray}\label{P2}
P_2&=&\sum_{J_1=0}^{J}\sum_{l=-\infty}^{+\infty} \langle
\bar{O}^J_{n,-n} O^{J_1}_{l,-l} O^{J_2} \rangle_{planar} \langle
\bar{O}^{J_1}_{l,-l} \bar{O}^{J_2} O^J_{m,-m}
\rangle_{planar}\\&&\nonumber =g_2^2\sum_{l=-\infty}^{+\infty}
\int_0^1dx~x^{3}(1-x)\frac{\sin^2(m\pi
x)}{\pi^2(mx-l)^2}\frac{\sin^2(n\pi x)}{\pi^2(nx-l)^2}
\end{eqnarray}

The string theory diagrams are computed by multiplying all
vertices and summing over all possible intermediate operators.
Here we do not use propagators in calculating the diagrams. In
large $J$ limit we can approximate the sum in $J_1$ by a integral
$\sum_{J_1=0}^{J}=J\int^1_0 dx$. It is straightforward to put
equations in (\ref{planar1}) into equations (\ref{P1}) (\ref{P2})
and explicitly compute the sum and integral. We find an agreement
with equation (\ref{torus}) in all 5 cases

\begin{equation} \label{a1}
\langle \bar{O}^J_{n,-n} O^J_{m,-m}
\rangle_{torus}=\frac{1}{2}(P_1+P_2)
\end{equation}
Here the $\frac{1}{2}$ can be thought of as the symmetry factor of
the string theory diagrams. The symmetry factor can be understood
from the example of free torus two point function of chiral
operators, which is computed in the field theory side as shown in
figure \ref{F3} \cite{Constable, Kristjansen}. The twistings in
the large $N$ gauge index contractions can be thought of
intuitively as string splitting and rejoining. Below we will give
an argument why we have overcounted by a factor of $2$ when we do
string theory diagrams. In more general cases of one loop cubic
interaction and two loop propagation diagrams the symmetry factors
will be determined by more complicated combinatorics and will
generally differ from the symmetry factors in usual Feymann
diagrams in quantum field theory.

One can also easily calculate the one loop string propagation
diagram for chiral operators $O^J$ and $O^J_0$. In both cases
there is only one diagram. The results are again agree with the
field theory calculations by the symmetry factor of $\frac{1}{2}$.

\subsubsection{ Derivation of symmetry factors} \label{count}

We propose a practical prescription for deriving symmetry factors
of string theory diagrams we computed. We denote a close string
with n segments by $(a_1a_2\cdots a_n)$, where the strings are
regarded as the same by cyclic rotation. For example,
$(a_1a_2\cdots a_n)$ and $(a_2a_3\cdots a_na_1)$ are the same
string. We denote the processes of string splitting and joining by
$(a_1a_2\cdots a_n)\rightarrow(a_1a_2\cdots a_i)(a_{i+1}\cdots
a_n)$ and $(a_1a_2\cdots a_i)(a_{i+1}\cdots
a_n)\rightarrow(a_1a_2\cdots a_n)$. Now imagine figure \ref{F3} as
a string of 4 segments goes from $(1234)$ to $(2143)$. How many
ways can we do this with our rules? A little counting reveals that
at one loop level there are only two processes as the following
\begin{eqnarray}
&&(1234)\rightarrow(12)(34)\rightarrow(2143) \nonumber
\\&& (1234)\rightarrow(23)(41)\rightarrow(2143) \nonumber
\end{eqnarray}
Here since $(12)$ and $(21)$, $(34)$ and (43) are the same, we can
join $(12)(34)$ in to $(2143)$.  These two processes are one loop
string propagation diagrams. Thus we conclude we have overcounted
by a factor of $2$ when we do string theory diagram calculations.
This explain the symmetry factor of $\frac{1}{2}$ in equation
(\ref{a1}).

\subsubsection{Further generalizations}

In flat space it is very hard to do string perturbation theory,
due to the difficulty in doing integration of the moduli space of
the string worldsheet. Most string calculations in flat space have
been restricted to less than two loops. On the contrary, here we
have found that the highly curved Ramond-Ramond plane wave
background with $\mu p^{+} \alpha^{\prime}=+\infty$ is an ideal
background for doing string perturbation theory. We proposed the
string loop amplitudes can be computed by a cubic string field
theory, whose diagrams are constructed by naively multiplying the
vertices. It would be interesting to directly derive this cubic
field theory from integration of the moduli of string worldsheet.
In this case it is the agreements with BMN correlation functions
in dual (free) Yang-Mill theory that give us confidences that
these are indeed the correct string amplitudes. In \cite{Huang2}
the comparison with Yang-Mills theory was further checked for free
torus three point functions and perfect agreements were found.  We
expect to this to work to all genera. A general derivation or
proof of this conjecture would increase our understanding of
string perturbation theory as well as give general lessons of
large $N$ duality.

In some cases of two-dimensional string theory or topological
string theory, the string amplitudes can also be computed to all
loops. In those cases the theory have much less degrees of freedom
than that of critical superstrings considered here, so the
amplitudes is usually solvable and can be summed up to all loops
by some kind of integrable structures. In our case, for the vacuum
operator, the free field correlation functions can be summed up to
all genera by a Gaussian matrix model \cite{Constable,
Kristjansen}. However, for general non-BPS BMN operators, we do
not know whether it is possible to find a formula to sum up all
genus (free field) correlation functions, although it is
straightforward to compute the correlation functions at any given
genus. If such a formula does exist, it would likely to be
provided by some matrix models or integrable structures.

\section{Giant Gravitons and Open Strings}
Historically, it was first pointed out in \cite{BBNS} that trace
operators with large R-charge will mix with each other and no
longer form a good orthogonal basis. This is telling us the
strings are strongly coupled and perturbation theory has broken
down. In strongly coupled regime we expect to encounter
non-perturbative objects such as D-branes. Indeed, it was shown in
\cite{McGreevy} that the size of a probe D-brane in $AdS_5\times
S^5$ expand if its angular momentum (R-charge of corresponding
operators ) becomes bigger. This is similar to Myers effect
\cite{Myers} where lower dimensional objects can blow up into
higher dimensional objects.

In large N limit, R-charge of operators in dual Yang-Mills theory
is closely related to string coupling. The string theory is free
when R-charge $J$ is less than $\sqrt{N}$. In the BMN limit
$J\sim\sqrt{N}$, in previous Section we claimed we can compute
string perturbation theory to all loops in a highly curved plane
wave background \cite{Huang2}. If the R-charge is larger,
non-planar diagrams dominate over planar diagrams, string
perturbation theory breaks down. Strings blow up into
non-perturbative objects known as giant gravitons. Giant gravitons
are D3-branes that can have open string excitations \cite{BHLN}.
Studies of these objects are likely to provide insights on
non-perturbative completions of string theory.

We have mentioned that when we stuck many D-branes together, a
mysterious ``duality transition'' will sometimes happen. Applying
this philosophy we can put a large number coincident giant
gravitons together. Then there is a new gauge symmetry emerging on
the giant gravitons. This gauge symmetry is different from the
original $SU(N)$ gauge symmetry we have, and is somehow encoded in
the open strings attached to the giant gravitons. There might be a
new duality transition in some decoupling limit analogous to
AdS/CFT correspondence. If this is true, we will have a very
interesting duality between two gauge theories with different
gauge groups. Some motivations and evidences of this conjecture
comes from the study of AdS black hole which is a condensate of
giant gravitons  \cite{Naqvi}.

Giant graviton states of spacetime are created in Yang-Mills
theory by determinant and sub-determniant operators as proposed
in~\cite{BBNS} and confirmed in~\cite{CJR}.  In this Section we
mainly follow the discussion in \cite{BHLN}. Other aspects of
giant gravitons have been studied in \cite{other}. \footnote{In
~\cite{openstrings, parklee} it was showed how open strings emerge
from gauge theories dual to string theories that have open strings
in the perturbative spectrum.  In these cases, the dual field
theory has quarks marking the endpoints of open string
worldsheets.  Here we are interested in situations in which open
strings emerge in pure supersymmetric Yang-Mills theory as
fluctuations around states dual to D-branes in spacetime. D-branes
in pp-wave backgrounds and open strings propagating on them have
been studied in e.g.~\cite{DP, kostasmarika}.} In Sec. \ref{sec2}
we show how Yang-Mills theory reproduces the spectrum of small
fluctuations of giant gravitons.    We discuss the emergence of
the $G^2$ degrees of freedom expected when $G$ giants nearly
coincide. In Sec. \ref{sec3} we display a Penrose limit in which
the open strings propagating on giants can be quantized simply.
Taking the corresponding large charge limit in Yang-Mills theory,
we reconstruct the open string worldsheets from field theory
degrees of freedom, and show that the one-loop field theory
calculation reproduces string spectra. The relevant operators are
generically not BPS, but nevertheless their dimensions do not grow
in the $N \rightarrow \infty$ limit.

Since we have a complete second-quantized formulation of ${\cal N}
= 4$ Yang-Mills, this theory  is supposed to give us a
non-perturbative description of strings.   If this is really so,
various subsectors of the theory should contain the holographic
duals to all possible string backgrounds. This is indeed the case.
A classical geometry $AdS_5\times S^5$ is seen as by small
R-charge BPS operators. As we go to BMN limit $J\sim \sqrt{N}$, we
encounter the geometry of plane wave and flat space. These are all
known maximally supersymmetric backgrounds of type IIB string
theory. We speculate the geometry seen by giant gravitons will be
very fuzzy and foamy. Further studies of this question would shed
light on the difficult problem of achieving a background
independence formulation of M-theory, see e.g. \cite{shape}.

\subsection{Spherical D3-branes and their fluctuations}
\label{sec2}

\paragraph{Scalar fluctuations:}
The best semiclassical description of a graviton with angular
momentum of order $N$  on the $\sph{5}$ of $\ads{5} \times
\sph{5}$ is in terms of a large D3-brane wrapping a 3-sphere and
moving with some velocity~\cite{McGreevy, MST}. This is the giant
graviton. In $\ads{5} \times \sph{5}$, the radius of the spherical
D3-brane is $\rho^2=lR^2/N$, where $l$ is the angular momentum on
the $\sph{5}$ of the state, $R$ is the radius of the sphere, and
$N$ the total 5-form flux through the 5-sphere. Since the radius
of the D3 brane giant graviton is bounded by the radius $R$ of the
$\sph{5}$, there is an upper bound on the angular momentum $ l
\leq N$.

The spectrum of small fluctuations of the giant graviton was
calculated in \cite{DJM}. When the giant graviton expands into an
$\sph{3}$ on the $\sph{5}$, it has six transverse scalar
fluctuations, of which four correspond to fluctuations into
$\ads{5}$ and two are fluctuations within $\sph{5}$ . These
vibration modes can be written as a superposition of  scalar
spherical harmonics $Y_k$ on the unit  $S^3$.    In~\cite{DJM} it
was found that the frequencies of the four modes corresponding to
fluctuations in $\ads{5}$ with wave-functions $Y_k$ are given by
\begin{equation}
\omega_k=\frac{k+1}{R} \label{frequency1}
\end{equation}
Similarly, the two vibration mode frequencies corresponding to
fluctuations in $\sph{5}$ are
\begin{equation}
\omega_k^-=\frac{k}{R},~~~~~\omega_k^+=\frac{k+2}{R}
\label{frequency2}
\end{equation}

\paragraph{Giants and their scalar fluctuations from CFT:}
In \cite{BBNS}, it was shown that  giant gravitons are dual to
states created by a family of subdeterminants:
\begin{equation}
O_l= {\rm subdet}{}_l Z \equiv \frac{1}{l!}\, \epsilon_{i_1 i_2
\cdots i_l a_1 a_2 \cdots a_{N-l}}\, \epsilon^{j_1 j_2 \cdots j_l
a_1 a_2 \cdots a_{N-l}}\,Z^{i_1}_{j_1}\, Z^{i_2}_{j_2}\,\cdots\,
Z^{i_l}_{j_l} \label{subdets}
\end{equation}
(So $O_{N}$ is the same as the determinant of $\Phi$.)  Here,
$Z=\frac{1}{\sqrt{2}}(\phi^5+i \phi^6)$ is a complex combination
of two of the six adjoint scalars in the $\CN=4$
theory.\footnote{The $\sph{5}$ in the bulk can be described by
$X_1^2+\dots X_6^2=R^2$. The operator $O_N$ in (\ref{subdets})
corresponds to a giant graviton moving in the $X^5, X^6$ plane.
The trajectory of such a giant will trace out a circle of radius
$(1-\frac{l}{N})R$ in this plane. Notice that the maximal giant
with $l=N$ is not really moving on the $\sph{5}$.   Its angular
momentum arises from the Chern-Simons interaction on its
worldvolume and the background flux.} These subdeterminants have a
bounded R-charge, with the full determinant saturating the bound.
The bound on the R-charge is the field theory explanation of the
angular momentum bound for giants.   A giant graviton is a 1/2 BPS
state of the CFT and breaks the SO(6) R-symmetry of the $\CN = 4$
theory down to U(1) $\times$ SO(4). The U(1) corresponds to the
plane of motion of the giant gravitons while the SO(4) corresponds
to the rotation group of the $\sph{3}$ worldvolume of the giants.
Under the U(1) $Z$ and $\bar{Z}$ have charges $\pm 1$ while the
other scalars $\phi^i$ ($i = 1 \cdots 4$) of the Yang-Mills theory
are neutral. The giant gravitons in (\ref{subdets}) therefore
carry a $U(1)$ charge $l$ and, being protected operators, their
conformal dimensions are $\Delta = l$. Under the SO(4), Z is
neutral, but the $\phi^i$ transform as a {\bf 4}.

To map the fluctuations of a giant graviton to the CFT, we can
replace $Z$ in (\ref{subdets}) by other operators, along lines
similar to~\cite{BHK} for the dibaryon in the theory of D3-branes
at a conifold singularity.  The resulting operator should carry
the same U(1) charge $l$ as the giant.    Therefore, their
conformal dimension in the free field limit should take the  value
\begin{equation}
\Delta = l + \omega \, R
\end{equation}
where $\omega$ is appropriate fluctuation frequency in
(\ref{frequency1}) or (\ref{frequency2}).\footnote{Recall that
giant gravitons in global AdS map onto states of Yang-Mills theory
on $\sph{3} \times R$ and that energy in spacetime maps to energy
$E$ in the field theory. Using the state-operator correspondence,
the energy of states on $\sph{3} \times R$ maps to the dimension
$\Delta = R \, E $ of operators on $R^4$, which we will typically
discuss.}     Finally, since the scalar vibrations of giants are
in  the $Y_k$ scalar spherical harmonics of $S^3$, i.e. the
symmetric traceless representation of $SO(4)$, it is natural to
use operators formed by the symmetric traceless products of the
four scalars $\phi^i$, ($i=1,2,3,4$).

Suppose $O^k$ is the $k$th symmetric traceless product of
$\phi^i$.  Consider the operators:
\begin{equation}
 \label{mode3}
\CO_{m}^k = \epsilon_{i_1\cdots i_la_1\cdots a_{n-l}}
\epsilon^{j_1\cdots j_la_1\cdots a_{n-l}}Z^{i_1}_{j_1}\cdots
Z^{i_{l-1}}_{j_{l-1}}(D_m Z \, O^k)^{i_{l}}_{j_{l}}
\end{equation}
$\CO^k_m$ are operators with U(1) charge $l$,  in the $k$th
symmetric traceless representation of SO(4), and have dimension
$\Delta = l+k + 1$.   The index $m=1 \cdots 4$ refers to the four
Cartesian directions of $R^4$ in radial quantization of $\sph{3}
\times R$.  Clearly, $\CO^k_m$ has the quantum numbers to match
the AdS polarized fluctuations with spectrum (\ref{frequency1}).
(Note that unlike (\ref{subdets}) we have not normalized these
operators to have unit two-point functions.)

Now consider
\begin{eqnarray}
\label{mode1} \CO^k_- &=& \epsilon_{i_1\cdots i_li_{l+1}a_1\cdots
a_{n-l-1}} \epsilon^{j_1\cdots j_lj_{l+1}a_1\cdots
a_{N-l-1}}Z^{i_1}_{j_1}\cdots
Z^{i_l}_{j_l}(O^k)^{i_{l+1}}_{j_{l+1}}
\\
\label{mode2} \CO^k_+ &=& \epsilon_{i_1\cdots i_{l+3}a_1\cdots
a_{n-l-3}} \epsilon^{j_1\cdots j_{l+3}a_1\cdots
a_{n-l-3}}Z^{i_1}_{j_1}\cdots
Z^{i_l}_{j_l}Z^{i_{l+1}}_{j_{l+1}}\overline{Z}^{i_{l+2}}_{j_{l+2}}(O^k)^{i_{l+3}}_{j_{l+3}}
\end{eqnarray}
These operators have U(1) charge $l$, are in $k$th symmetric
traceless representation of SO(4), and have conformal dimensions
$\Delta^- = l + k$ and $\Delta^+= l + k + 2$.   Clearly we have
found operators with quantum numbers matching the $\sph{5}$
polarized fluctuations whose spectrum is (\ref{frequency2}).
(Again, we have not chosen to normalize these operators to have
unit two-point functions.) Note that the operators
(\ref{mode1},\ref{mode2}) cannot be constructed for the maximal
giant graviton, i.e., when $l=N$.   The corresponding analysis of
fluctuations in~\cite{DJM} leads to a similar conclusion since the
relevant equations are ill-defined for the maximal giant.

In general, most fluctuations of giant gravitons are not
BPS~\cite{DJM} and so we expect anomalous dimensions to develop
quantum mechanically.  From the spacetime point of view these
would be studied by finding solutions to the open-string loop
corrected equations of motion of a D-brane.   (The DBI action used
in~\cite{DJM} included all $\alpha^\prime$ corrections at disk
order but not string loop corrections.)   Since these corrections
are hard to compute in spacetime it is interesting to examine them
in the field theory.  In \cite{BHLN} we show that the interactions
of $\CN = 4$ Yang-Mills do produce anomalous dimensions for
(\ref{mode3},\ref{mode1},\ref{mode2}), but, surprisingly, these
corrections do not grow with $N$.  Here we will not explore the
details, but give the result that the anomalous dimension is
$(J-1)g_s/\pi$. At weak coupling and large $N$, therefore, these
are non-BPS operators whose dimensions are protected from large
corrections.

\paragraph{Multiple giants from Yang-Mills:}
Consider a CFT states made by the product of $G$ identical giant
graviton operators.  This should represent $G$ giants of the same
size moving in concert on the $\sph{5}$.  Such a group of giants
should have $G^2$ strings stretching between them.   At low
energies these string should give rise to a $U(G)$ gauge theory
living on the worldvolume of the spherical D3-branes.   The
spectra that we described above we derived from fluctuations of a
single giant, and therefore apply to the $U(1)^G$ part of this
low-energy gauge theory.   Below we will display candidate
operators dual to the expected $6 G^2$ scalar fluctuations of the
branes.  In the quadratic limit relevant to small fluctuations all
of these will have the same spectrum as we see below.  (Again,
there are small quantum corrections to the spectrum that are
negligible in the large N limit.)

For simplicity consider two maximal giant gravitons, corresponding
to the product of two determinants in the CFT $\det_1Z \,
\det_2Z$.  Here we have introduced labels analogous to Chan-Paton
indices for each of the determinants representing a giant
graviton. Taking $x_1$ and $x_2$ to be coordinates on the
$\sph{3}$ on which the Yang-Mills theory is defined, we could
define the operator $\det Z(x_1) \, \det Z(x_2)$ so that it makes
sense to treat them as distinguishable in this way.   After
constructing the operators of interest to us as described in the
text we later let $x_1 \rightarrow x_2$.    For each of the
operators $\CO^k_{m,+,-}$ describing scalar fluctuations on a
single giant we might expect four operators here.  Two of these
would correspond to separate vibrations of each of the two giants
and should be given by $\CO^k_{11;m,+,-} \equiv \CO^k_{1;m,+,-} \,
\det_2 Z$ and  $\CO^k_{22;m,+,1} \equiv \det_1 Z \,
\CO^k_{2;m,+,-}$ where $\CO^k_{i;m,+,-}$ is a fluctuation on giant
$i$ as described above. Two of the fluctuations would arise from
off-diagonal components of the U(2) matrices, which in turn arise
from strings going between the two branes.   It is natural that
these vibrations should arise from operators that intertwine the
gauge indices in the two determinants:
\begin{eqnarray}
\CO^k_{12;-} &=& \epsilon_{i_1\cdots i_N} \epsilon^{j_1\cdots
j_N}Z^{i_1}_{j_1}\cdots Z^{i_{N-1}}_{j_{N-1}}(O^{k})^{i_N}_{l_N}
\, \, \, \epsilon_{k_1\cdots k_N} \epsilon^{l_1\cdots
l_N}Z^{k_1}_{l_1}\cdots Z^{k_N}_{j_N} \label{inter1}
\\
\CO^k_{21;-} &=& \epsilon_{i_1\cdots i_N} \epsilon^{j_1\cdots
j_N}Z^{i_1}_{j_1}\cdots Z^{i_{N}}_{l_{N}} \, \, \,
\epsilon_{k_1\cdots k_N} \epsilon^{l_1\cdots
l_N}Z^{k_1}_{l_1}\cdots Z^{k_{N-1}}_{l_{N-1}}(O^k)^{k_N}_{j_N}
\label{inter2}
\end{eqnarray}
Here the $O^k$ are the symmetric traceless products of scalars
mentioned earlier.  The other operators $\CO^k_{ij;+,m}$ are
written similarly.  The indices that we introduced on the two
determinants have played the  role of Chan-Paton factors of open
strings stretched between two giants.  It appears that the four
operators we find have a natural interpretation as the expected
operators from the adjoint representation of $U(2)$ group.
However, there is a subtlety.  When $x_1 \rightarrow x_2$,
$O_{12,-}^k = O_{21,-}^k$ and  $O_{11,-}^k = O_{22,-}^k$ by an
exchange of dummy indices.  What is more, it can be shown that in
this limit the operators $O_{11,-}^k = N \, O_{12,-}^k$ also.
This is surprising at first because $O_{12}$ intertwines indices
between two determinants, but this fact can be shown as follows.
First, observe that (\ref{inter1}) is zero when $l_N \neq j_N$.
In this case there exists an $l_x = j_n$ where $1 \leq x \leq
N-1$, and so the sum overs the permutations of $k_1,\cdots k_N$
gives zero.  We are left with $l_N = j_N$ in which case the
operator is unchanged by switching these two indices.  Thus
$O_{11;-}^{k}$ is proportional to $O_{12;-}^{k}$. The
proportionality factor between these operators is $N$ because the
former has $N^2$ choices of $j_N, l_N$,  while requiring $l_N=j_N$
leaves $N$ choices.

In fact this is exactly what we should expect since coincident
D-branes are identical and there is no difference between the
strings running between different pairs of branes; the change of
dummy variables relating (\ref{inter1}) and (\ref{inter2}) when
$x_1 \rightarrow x_2$ is an example of this.   To display the four
strings running between two branes we have to separate the branes
from each each other.     The generalization of this discussion to
$G$ giants and the associateed $G^2$ degrees of freedom from the
adjoint of $U(G)$ is obvious.

There is some issue as to whether multiple less-than-maximal giant
gravitons are described by a product of subdeterminant operators
or by a sudeterminant of products (see~\cite{BBNS,CJR,BHK}). Above
we restricted ourself to the largest giants for which this issue
does not arise since the determinant of products is the product of
determinants.    It would be interesting to test our proposal that
fluctuations of strings between branes are described by
intertwined operators, by taking one of them to be a maximal giant
and another to be a smaller one.    Strings running between such
branes are stretched and should have a corresponding gap in their
spectra.  It would be interesting to test this by trying to match
the vibrational energies of strings stretched between the maximal
and next-to-maximal sized giants.

\subsection{Open strings in plane wave limit} \label{sec3}

\subsubsection{Penrose limits for open strings} We will see that
for open strings moving with a large angular momentum on the giant
graviton D3 brane, we can construct the open string world sheet in
the $\CN=4$ $SU(N)$ Yang-Mills theory. To that end, we start by
looking at the geometry seen by such an open string.   The
$\ads{5} \times \sph{5}$ metric is:
\begin{eqnarray*}
ds^2&=&R^2[ -dt^2 \cosh^2 \rho + d \rho^2 + \sinh^2 \rho ~ d\Omega_3^2+d \psi^2
 \cos^2 \theta + d\theta^2 + \sin^2 \theta d {\Omega_3^\prime}^2] \\
d{\Omega_3^\prime}^2& =& d \varphi^2 + \cos^2 \varphi d \eta^2 +
\sin^2 \varphi d \xi^2
\end{eqnarray*}
where $R = (4\pi \alpha^{\prime2} g_s N)^{1/4}$ is the AdS scale.
Consider the near maximal giant graviton at $\theta \sim \pi/2$
which is moving in the $\psi$ direction.\footnote{Strictly
speaking, the maximal giant is at rest and all its angular
momentum comes from the five-form flux.}  The world volume of the
giant spans $(t,\varphi, \eta, \xi)$ and the giant graviton is at
$\rho=0$ and $\theta={\pi \over 2}$. We want to find the geometry
seen by an open string ending on the giant graviton, moving
rapidly in the $ \eta $ direction at the equator of $\sph{3}$
given by $\varphi=0$. We define light cone coordinates
${\tilde{x}}^{\pm}={ {t\pm\eta} \over 2}$, a new coordinate $\chi
= {\pi \over 2}- \theta$, and focus on the region near
$\rho=\chi=\varphi=0$ by rescaling:
\begin{equation}
x^+ = \frac{\tilde{x}^+}{\mu}, ~~~ x^-=\mu R^2 \tilde{x}^-, ~~~
\rho={r \over R}, ~~~ \chi={y \over R}, ~~~ \varphi={u \over R},
~~~R \rightarrow \infty .
\end{equation}
In this limit, the metric becomes,
\begin{equation}
ds^2=-4 dx^+
dx^--\mu^2(\vec{r}^2+\vec{y}^2+\vec{u}^2)(dx^+)^2+d\vec{y}^2+d\vec{u}^2+d\vec{r}^2\
\label{pp}
\end{equation}
where $\vec{u}$ and $\vec{y}$ parameterize points on two $R^2$s
and $\vec{r}$ parameterizes points on $R^4$.   The 5-form flux
that supports the $\ads{5} \times \sph{5}$ background becomes
\begin{equation}
F_{+1234} = F_{+5678} = {\rm Const} \times \mu
\end{equation}
in this limit, thereby breaking the $SO(8)$ isometry of the metric
(\ref{pp}) to SO(4) $\times$ SO(4).   We find that the open string
sees the standard pp wave geometry. The light cone action becomes
\cite{Metsaev},
\begin{equation}
S=\frac{1}{4\pi \alpha'} \int ~dt \int^{2\pi \alpha' p^+}_0 d
\sigma \Bigl[  {\partial_\tau X^I}{\partial_\tau X_I}  -
{\partial_\sigma X^I}{\partial_\sigma X_I} - \mu^2 \, X^I X_I  +
2i \bar{S} (\dslash+\mu \Pi)S \Bigr] \label{action}
\end{equation}
where $\Pi=\Gamma^{1234}$ and $S$ is a Majorana spinor on the
worldsheet and a positive chirality spinor ${\bf 8_s}$  under
SO(8) which is the group of rotations in the eight transverse
directions. The $X^I$ tranform as ${\bf 8_v}$ under this group.
The fermionic term in the action breaks the SO(8) symmetry that is
otherwise present to SO(4) $\times$ SO(4). The open strings we are
interested in have Neumann boundary conditions in the light cone
directions $x^{\pm}$ and $\vec{u}$ and Dirichlet boundary
conditions in the six transverse directions parameterized by
$\vec{r}$ and $\vec{y}$.
\begin{equation}
\partial_\sigma {X^\alpha}
=\partial_\tau {X^i}  =  0 \label{bcond}
\end{equation}
where $\alpha=7,8$. The coordinates used in (\ref{pp}),
$\vec{u}=(x^7,x^8)$, $\vec{r}=(x^1,x^2,x^3,x^4)$ and
$\vec{y}=(x^5,x^6)$. Such open strings were quantized by Dabholkar
and Parvizi in \cite{DP}. Here, we quote their results. The
spectrum of the light cone Hamiltonian ($H \equiv -p_+$) is
\begin{eqnarray}
H& = & E_0+E_\CN \nonumber \\
E_0&=&\mu \Bigl( \sum_{\alpha=7,8} \bar{a}_0^\alpha a_0^\alpha -2
i S_0 \Gamma^{56}
S_0 +e_0 \Bigr) \nonumber \\
 E_\CN& = &  \Bigl({1 \over 2} \sum_{n \neq 0} \omega_n~a_n^i a_{-n}^i + i \sum_{n \neq 0}
\omega_n ~S_n S_{-n} \Bigr).  \label{lightham}
\end{eqnarray}
where we have defined the bosonic and fermionic creation and
annihilation operators $a_n^i$ and $S_n$  as in~\cite{DP}.  Here
$e_0=1$ is the zero point energy for the D3 brane and
\begin{equation}
\omega_n=\mathrm{sign}(n)\sqrt{\Bigl({n \over {2 \alpha^\prime
p^+}}\Bigr)^2+\mu^2} \, . \label{freq}
\end{equation}
There are only two bosonic zero modes coming from two directions
in light cone gauge which have Neumann boundary conditions (these
would have been momentum modes but in the pp wave background, the
zero mode is also a harmonic oscillator). The fermionic zero mode
is $S_0$ which transforms in ${\bf 8_s}$ of SO(8).

The D3 brane occupies $x^+,x^-, x^7,x^8$ and has six transverse
coordinates $x^1 \cdots x^6$. In the light cone, only an
SO(2)${}_U$ subgroup of the SO(1,3) symmetry of the D3 brane world
volume is visible. In addition, the  SO(6) group transverse to the
D3-brane is broken down to SO(2)${}_Z \times $ SO(4)  by the
5-form background flux.   Hence we have the embedding
\begin{equation}
SO(8) \supset SO(2)_Z \times SO(2)_U  \times SO(4) \labell{symm1}
\end{equation}
The spinor ${\bf 8_s}$ decomposes as
\begin{equation}
{\bf 8_s} \rightarrow ({\bf 2},{\bf 1})^{(\half,\half)}\oplus
({\bf \bar{ 2}},{\bf 1})^{(-\half,-\half)} \oplus ({\bf 1},{\bf
2})^{(\half,-\half)}\oplus ({\bf 1},{\bf
\bar{2}})^{(-\half,\half)}
\end{equation}
where the superscripts denote SO(2)${}_Z \times$ SO(2)${}_U$, and
we have written SO(4) representations as representations of SU(2)
$\times$ SU(2). The fermionic zero modes $S_0$ can be arranged
into fermionic creation and annihilation operators:
\begin{eqnarray}
\bar{\lambda}_\alpha  \equiv S_{0\alpha}^{(\half,\half)} &,&
{\lambda}_\alpha  \equiv S_{0\alpha}^{(-\half,-\half)}  \nonumber \\
\bar{\lambda}_{\dot{\alpha}}  \equiv
S_{0{\dot{\alpha}}}^{(-\half,\half)} &,& {\lambda}_{\dot{\alpha}}
\equiv S_{0{\dot{\alpha}}}^{(\half,-\half)} \label{lambdas}
\end{eqnarray}
The commutation relations are
\begin{equation}
\{\bar{\lambda}_\alpha,\lambda^\beta \}= \delta^\beta_\alpha
~~~~,~~~~~
\{{\lambda}_{\dot{\alpha}},{\bar{\lambda}}^{\dot{\beta}} \}=
\delta^{\dot{\beta}}_{\dot{\alpha}}.
\end{equation}
The energy contribution from the zero mode oscillators is given by
\begin{eqnarray}
E_0&=& m(\bar{a}_0^7a_0^7+\bar{a}_0^8a_0^8+\bar{\lambda}_\alpha
\lambda^\alpha
-\bar{\lambda}_{\dot{\alpha}}\lambda^{\dot{\alpha}} + 1)\\
& = &  m(\bar{a}_0^7a_0^7+\bar{a}_0^8a_0^8+\bar{\lambda}_\alpha
\lambda^\alpha
+{\lambda}_{\dot{\alpha}}\bar{\lambda}^{\dot{\alpha}} -1)
\end{eqnarray}
as in~\cite{DP}.   We choose $\bar{\lambda}_\alpha$ and
$\lambda_{\dot{\alpha}}$ as creation operators.\footnote{This
convention differs from~\cite{DP} but is convenient for us.}   The
vacuum state  is invariant under SO(4) $\times$ SO(2)${}_U$ and
carries charge $-1$ under SO(2)${}_Z$:
\begin{equation}
a_0^r| - 1 , 0 \ket = 0 , ~~~~\lambda^\alpha | -1,0 \ket = 0, ~~~~
{\bar {\lambda}}^{\dot{\alpha}} |-1, 0 \ket = 0 \label{grounddef}
\end{equation}
Other modes with zero worldsheet momentum ($n=0$) are contructed
by acting with creation operators $\bar{\lambda}_\alpha$,
$\lambda_{\dot{\alpha}}$ and $\bar{a}_0^{r}$ . The vacuum state $|
-1,0 \ket$ has $E_0=-1$ and carries $-1$ units of angular momentum
in the $x^5 x^6$ direction. Since the maximal giant graviton that
we are considering carries angular momentum $N$ in this direction,
we see that  $N-1$ is the total angular momentum of the giant and
the ground state of its open strings in our Penrose limit.
Likewise the fermionic contribution to the energy ground state
lowers it to $N-1$ from the value $N$ for the maximal giant.
Below we will present the complete perturbative spectrum of the
string quantized in this way and map all the states to operators
of $\CN =4$ Yang-Mills theory.

\subsubsection{Open string world sheet in SYM} String theory in
global $\ads{5} \times \sph{5}$ is dual to $\CN = 4$ Yang-Mills
theory on $\sph{3} \times R$.   States in spacetime map to states
of the field theory, and by the state-operator correspondence for
conformal theories, are related to operators on $R^4$.    The
global symmetry of the theory is SO(4) $\times$ SO(6) where SO(4)
is the rotation group of $R^4$ corresponding to the SO(4)
appearing in (\ref{symm1}).   SO(6) is the R-symmetry group,
corresponding to the rotation group of $\sph{5}$ in the bulk
spacetime.  The Yang-Mills theory has six adjoint scalar fields
$\phi^1 \cdots \phi^6$ which transform as the fundamental of
SO(6).   The  complex combinations $Z=\frac{1}{2}(\phi^5+i
\phi^6)$, $Y=\frac{1}{2}(\phi^3+i \phi^4)$, $U=\frac{1}{2}(\phi^1
+ i\phi^2)$ are charged under three different SO(2) subgroups of
SO(6), SO(2)$_{Z,Y,U}$, which correspond to rotations in three
independent planes of the bulk $\sph{5}$.   We will denote charges
under these SO(2) groups as $J_{Z,Y,U}$.   As we have discussed,
giant gravitons carry a charge of order $N$ under SO(2)$_Z$, and
are created by subdeterminant operators.  The Penrose limit of
open strings on giants corresponds to strings moving in the
spacetime direction corresponding to $Y$.  We will propose a field
theory description of the worldsheets of open strings on the
maximal giant graviton.

We will denote the conformal dimenion of the operators dual to
such strings by
\begin{equation}
{\tilde\Delta} = N + \Delta
\end{equation}
where the additive $N$ arises because the background giant has
this dimension. Mapping the data of the Penrose limit to the field
theory we find that the conformal dimension $\Delta$ of the
excitation above the giant and $SO(2)_Y$ charges of these states
are related to the lightcone Hamiltonian (\ref{lightham}) of
strings as~\cite{BMN}
\begin{equation}
H = -p_+ = 2p^- = \Delta - J_Y \, = {\tilde\Delta} - N - J_Y.
\end{equation}
We are going to consider states of fixed $p_+$ so that $\Delta
\approx J_Y$. Likewise the other lightcone momentum maps as
\begin{equation}
-p_- = 2p^+ = {\Delta + J_Y \over R^2} \, ,
\end{equation}
where $R$ is the AdS scale.   Note that to have a fixed non-zero
value of $p_-$, $J_Y$ must be of order $\sqrt{N}$.  The
contribution to Hamiltonian from higher oscillator modes of the
lightcone string (\ref{freq}) then translates into \footnote{We
have set $\mu =1$ to make the comparison with field theory.}
\begin{equation}
( \tilde\Delta - N - J_Y)_n = (\Delta - J_Y)_n = \omega_n =
\sqrt{1 + {\pi g_s N n^2 \over J_Y^2}} \label{spectrum}
\end{equation}
in the Yang-Mills theory up to small corrections that vanish at
large $N$.

\paragraph{The ground state: }   From the previous section, the ground state
of strings on maximal giants carries SO(2)$_Z$ charge $-1$ and
lightcone energy $-1$.  So the overall state including the giant
carries an SO(2)$_Z$ charge $N - 1$ and $\Delta - J_Y = -1$.
Furthermore, we achieve the Penrose limit by considering states
with SO(2)$_Y$ charge $J$ of order $\sqrt{N}$.   To describe the
ground state of the string we therefore seek an operator that is a
modification of the $\det Z$ creating the maximal giant which has
the charges just listed. A suitable candidate is:
\begin{equation}
\epsilon_{i_1\cdots i_N} \epsilon^{j_1\cdots
j_N}Z^{i_1}_{j_1}\cdots Z^{i_{N-1}}_{j_{N-1}} (Y Y \cdots
Y)^{i_N}_{j_N} ~~~~ \leftrightarrow ~~~~ | G_N ; -1, 0 \ket
\label{opdef}
\end{equation}
where we have inserted a product of $J$ Ys in place of one Z. (We
have chosen not to normalize this operator to have a unit
two-point function.)  The notation $ |G_N ; -1, 0 \ket$ indicates
a single maximal giant graviton with an open string in its ground
state as defined in (\ref{grounddef}).  The Zs create the D3 brane
giant graviton and, as we show below, the string of Ys explicitly
reconstructs the worldsheet of open string propagating on a giant
in a Penrose limit.The absence of a trace on the indices of the
product of Ys will be responsible for making this an open string
worldsheet, and in the presence of multiple giants, Chan-Paton
factors will emerge from the ability to intertwine these indices
between different giant operators.

The operator (\ref{opdef}), just like the small fluctuations
(\ref{mode3},\ref{mode1},\ref{mode2}), is not BPS and therefore
receives quantum corrections to its dimension.  Surprisingly, as
shown in \cite{BHLN}, these corrections do not grow with $N$ and
lead to an anomalous dimension of $(J-1)g_s/\pi$.  This extra
piece is very small compared to the BMN anomalous dimension
$\frac{g_sN}{J^2}$ if we take $g_2=\frac{J^2}{N}$ small to
suppress the non-planar contributions anomalous dimensions . This
shift, being of $1/N$ order, suggests a loop open-string effect.

\paragraph{Rest of the zero modes: } The remainder of the zero modes on the string
worldsheet arose from the ground state (\ref{grounddef}) by the
action of the creation operators $\bar{\lambda}_\alpha$,
$\lambda_{\dot{\alpha}}$ and $\bar{a}_0^r$.   In Yang-Mills theory
we can construct operators corresponding to these states by
inserting into the string of the Ys the two scalars $\phi^{1,2}$
and four of the 16 gaugino components of $\CN =4$  Yang-Mills
which have SO(2)$_Y$ charge $J_Y = 1/2$ and SO$_Z$ charge $J=1/2$.
These gaugino components in which we are interested transform as
$({\bf 2,1 })$ and $({\bf 1, 2})$ under the SO(4) = SU(2) $\times$
SU(2) rotation group of four dimensional Yang-Mills and so we will
collect into two spinors $\psi_\alpha$ and $\psi_{\dot\alpha}$.
This matches the charges carried by the four creation operators
$\bar\lambda_\alpha$ and $\lambda_{\dot\alpha}$ identified on the
lightcone string worldsheet in (\ref{lambdas}).    So we identify
the string worldsheet zero mode creation operators with operator
insertions into (\ref{opdef}) as follows:
\begin{eqnarray}
\bar{a}_0^{7,8}  ~~~~ &\leftrightarrow& ~~~~ \phi^{1,2} \nonumber
\\
\bar\lambda_\alpha , \lambda_{\dot\alpha} ~~~~ &\leftrightarrow&
~~~~ \psi_\alpha, \psi_{\dot\alpha} \label{corresp1}
\end{eqnarray}
For example,
\begin{equation}
\bar\lambda_\alpha |G_N ; -1,0\ket ~~~~ \leftrightarrow ~~~~
\epsilon_{i_1\cdots i_N} \epsilon^{j_1\cdots
j_N}Z^{i_1}_{j_1}\cdots Z^{i_{N-1}}_{j_{N-1}} \left(
\sum_{l=0}^{J_Y} \, Y^l \, \psi_\alpha Y^{J_Y - l}
\right)^{i_N}_{j_N}
\end{equation}
Each action of a zero mode operator on the lightcone string vacuum
adds a similar sum to the dual field theory operator.  It is
interesting to see a detailed match between  field theory
operators and the quantum numbers for states created by acting by
worldsheet fermionic zero modes.   Each of these operators takes
the form $\frac{1}{N!}\epsilon_{i_1\cdots i_N} \epsilon^{j_1\cdots
j_N}Z^{i_1}_{j_1}\cdots Z^{i_{N-1}}_{j_{N-1}} \, \CV^{i_N}_{j_N}$
with $\CV$ given as below:
\begin{center}
\begin{tabular}{|r|l|c|l|}
\hline
State & Rep. & $H = \Delta - J_Y$ & $\CV$ \\
\hline
$|-1,0\ket$ & $({\bf 1},{\bf 1})^{(-1,0)}$ & $- 1$& $Y^{J_Y}$\\
$\bar\lambda_\alpha| -1,0\ket$& $({\bf 2},{\bf
1})^{(-\half,\half)}$ & $0$ & $\sum_{l=0}^{J_Y} \, Y^l \,
\psi_\alpha Y^{J_Y - l} $
\\
$\lambda_{\dot{\alpha}}|-1,0\ket$ &$({\bf 1},{\bf
2})^{(-\half,-\half)}$ &0 &$
 \sum_{l=0}^{J_Y} \, Y^l \, \psi_{\dot{\alpha}} Y^{J_Y - l} $ \\
$\bar\lambda_\alpha\bar\lambda_\beta| -1,0\ket$& $({\bf 1},{\bf
1})^{(0,1)}$ & $1$
& $ \sum_{l_1,l_2=0}^{J_Y} \, Y^{l_1} \, \psi_\alpha Y^{l_2}\psi_{\beta}Y^{J_Y-l_1-l_2} $\\
$\bar\lambda_\alpha \lambda_{\dot{\alpha}}| -1,0\ket$& $({\bf
2},{\bf 2})^{(0,0)}$ & $1$
& $ \sum_{l_1,l_2=0}^{J_Y} \, Y^{l_1} \, \psi_\alpha Y^{l_2}\psi_{\dot{\alpha}}Y^{J_Y-l_1-l_2} $\\
 $\lambda_{\dot{\alpha}}\lambda_{\dot{\beta}}| -1,0\ket$&
$({\bf 1},{\bf 1})^{(-1,-1)}$ & $1$
& $ \sum_{l_1,l_2=0}^{J_Y} \, Y^{l_1} \, \psi_{\dot{\alpha}} Y^{l_2}\psi_{\dot{\beta}}Y^{J_Y-l_1-l_2} $\\
$\bar\lambda_\alpha\lambda_{\dot{\alpha}}\lambda_{\dot{\beta}}|
-1,0\ket$& $({\bf 2},{\bf 1})^{(-\half,-\half)}$ & $2$
& $ \sum_{l_i=0}^{J_Y} \, Y^{l_1} \,\psi_\alpha  Y^{l_2}\ \psi_{\dot{\alpha}} Y^{l_3} \psi_{\dot{\beta}}
Y^{J_Y-\sum {l_i}} $\\
$\bar\lambda_\alpha\bar\lambda_{{\beta}}\lambda_{\dot{\beta}}|
-1,0\ket$& $({\bf 1},{\bf 2})^{(\half,\half)}$ & $2$
& $ \sum_{l_i=0}^{J_Y} \, Y^{l_1} \,\psi_\alpha  Y^{l_2}\ \psi_{{\beta}} Y^{l_3} \psi_{\dot{\beta}}
Y^{J_Y-\sum l_i} $\\
$\bar\lambda_\alpha\bar\lambda_{{\beta}}\lambda_{\dot{\alpha}}\lambda_{\dot{\beta}}|
-1,0\ket$& $({\bf 1},{\bf 1})^{(1,0)}$ & $3$
& $\sum_{l_i=0}^{J_Y}  Y^{l_1} \,\psi_\alpha  Y^{l_2}\ \psi_{{\beta}} Y^{l_3} \psi_{\dot{\alpha}}
Y^{l_4}\psi_{\dot{\beta}}Y^{J_Y-\sum{l_i}} $\\
\hline
\end{tabular}
\end{center}
The superscripts denote charges under SO(2)$_Z \times $SO(2)$_U$
and we have indicated the energy of fluctuations above the giant
graviton which itself has energy $N$.

\paragraph{Higher oscillators and string spectrum from $\CN=4$ theory: }  To construct the
higher oscillator states of open strings in analogy
with~\cite{BMN} we can insert operators representing string
fluctuations into the worldsheet represented by the string of Ys
in (\ref{opdef}). (Again we choose not to  normalize these
operators here to have a unit two point function.)  A phase
depending on the position of insertion into the string of Ys
represents the oscillator level. In effect, the phases reconstruct
the Fourier representation of a momentum state on the string
worldsheet in position space along the string.   The operators we
can insert include those in (\ref{corresp1}) corresponding to the
directions in which the open string has Neumann boundary
conditions. For example:
\begin{equation}
a^7_{-n}|G_N;-1,0 \ket \leftrightarrow \epsilon_{i_1\cdots i_N}
\epsilon^{j_1\cdots j_N}Z^{i_1}_{j_1}\cdots Z^{i_{N-1}}_{j_{N-1}}
\left( \sum_{l=0}^{J_Y} \, Y^l \, \phi^1 Y^{J_Y - l} \cos({{\pi
n l \over J_Y}})\right)^{i_N}_{j_N} \label{highstate}
\end{equation}
In addition, although there are no zero modes in directions with
Dirichlet boundary conditions for the open string,  there are
higher oscillator excitations. These will correspond to insertions
of
\begin{eqnarray}
\bar{a}^{i} ~~~~ &\leftrightarrow& ~~~~ D_i Y ~~~~ i=1\cdots4  \nonumber \\
\bar{a}^{5,6} ~~~~ &\leftrightarrow& ~~~~ \phi^{5,6}
\end{eqnarray}
with a position dependent phase $\sin({n\pi l \over J_Y})$:
\begin{equation}
a^5_{-n}|G_N;-1,0 \ket \leftrightarrow \epsilon_{i_1\cdots i_N}
\epsilon^{j_1\cdots j_N}Z^{i_1}_{j_1}\cdots Z^{i_{N-1}}_{j_{N-1}}
\left( \sum_{l=0}^{J_Y} \, Y^l \, \phi^5 Y^{J_Y - l} \sin({{\pi  n
l \over J_Y}})\right)^{i_N}_{j_N} \label{highstate1}
\end{equation}
The higher fermionic oscillators correspond to similar insertions
of $\psi_\alpha$ and $\psi_{\dot{\alpha}}$ with similar phases.
Note that the phase ${{\pi  n l \over J_Y}}$ is half of the phase
appearing in the closed string construction of \cite{BMN}. This is
necessary to correctly reproduce the open string spectrum from
field theory.

All operators constructed as in (\ref{highstate},\ref{highstate1})
carry charge $J_Y$ under SO(2)$_Y$ and so to compare the string
spectrum (\ref{spectrum})  to the field theory we need only
compute the conformal dimension of the operator.

Although the interactions between Zs and the operators within the
string of Ys continue to be suppressed as for the zero modes, the
presence of phases in (\ref{highstate},\ref{highstate1}) leads to
anomalous dimensions that we must compute in order to match the
spectrum (\ref{spectrum})~\cite{BMN}.   Below we will work with
the example (\ref{highstate}) but an identical story applies to
all the other operators.  (We leave out most of the details of the
calculation since it is exactly parallel to the work
in~\cite{BMN}.)

To start it is useful to expand the energies (\ref{spectrum}) in a
power series in $g_sN/J_Y^2$:
\begin{equation}
\omega_n = (\Delta - J_Y)_n = 1 + {\pi g_s N n^2 \over 2 J_Y^2} +
\cdots \label{expansion}
\end{equation}
The classical dimension of (\ref{highstate}) is $\tilde\Delta = N
-1 + J_Y + 1 = N + J_Y$.   In the interacting theory anomalous
dimensions will develop.   To study this we have to compute
correlation functions of (\ref{highstate}).   Even in the free
limit, there are many non-planar diagrams in these correlators
which are not suppressed even at large $N$ because the operator
itself has dimension comparable to $N$~\cite{BBNS}.    However,
within any diagram the interactions between the Ys and themselves
is dominated by planar sub-diagrams because $J_Y \sim \sqrt{N}$
and because when $N$ is large nonplanarity only becomes important
when more that $\sim N^{2/3}$ fields are involved~\cite{BBNS}.
The free contractions between Zs and themselves and the Ys and
themselves give rise to the classical dimension of the operators.

If we introduce an additional operator $\phi^I$ within the string
of Ys as in (\ref{highstate}). There are interactions between
$\phi^I$ and Z, which lead to small corrections that are
suppressed  in the large $N$ limit. There will be further
interactions between $\phi^I$ and Y which we will discuss here.
The diagrams connecting $\phi^I$ and $Y$  arise  because of the 4
point vertex in the $\CN=4$ theory.  Summing all such diagrams is
a computation almost identical to what we did for BMN operators as
in Sec.\ref{sec4}. The only difference arises from the different
position dependent phase in relating the higher oscillators to
operators as in (\ref{highstate}). The result is:
\begin{equation}
(\Delta-J_Y)_n=1+{\pi g_s N n^2 \over 2J_Y^2}
\end{equation}
This correctly reproduces the first order correction to the energy
in (\ref{expansion}).

 In Sec. \ref{sec2} we discussed how the $G^2$ low energy fluctuations of $G$ coinciding
 giants can arise from states with mutiple determinants with
intertwined indices (see the discussion around (\ref{inter1}) and
(\ref{inter2})).    A similar construction in the Penrose limit of
G coinciding D3-branes yields strings stretched between each pair
of branes.  The spectra of each of these strings is identical and
is reproduced as above.  The presence of a Chan-Paton factor
labelling the string endpoints is confirmed by point-splitting the
location of the determinant operators in Yang-Mills theory.   At
low energies these strings must give rise to a new $U(G)$ gauge
theory.  Note, however, that only gauge-invariant operators built
from this theory will be visible unless the $U(G)$ is broken by
separating the branes.  This is related to the observation in Sec.
\ref{sec2} that when the branes coincide, all the $G^2$ operators
describing fluctuations of the multiple giants become identical.
    %and files

\end{document}